\documentclass[12pt,oneside,openany,a4paper,afrikaans,english,PhD,goldenblock]{usthesis}
\usepackage[afrikaans, english]{babel}
\usepackage[utf8x]{inputenc}
\usepackage[T1]{fontenc}
\usepackage{geometry}
\usepackage{graphicx}
\usepackage{svg} % To be able to add SVG images
\usepackage{usbib}
\setcitestyle{square}
\bibpunct{[}{]}{,}{n}{,}{,}
\usepackage{caption}
\usepackage{subcaption}
\usepackage{mathtools}

\newcommand{\be}{\begin{equation}}
\newcommand{\ee}{\end{equation}}

\usepackage{amsmath}
\usepackage{amssymb}
\usepackage{float}
\usepackage{upgreek}
\usepackage{bbm}
\usepackage{xcolor}

\newcommand{\bs}{\boldsymbol}
\newcommand{\tn}{\textnormal}

\newcommand{\g}{\bs{g}}
\newcommand{\po}{\partial \Omega}
\newcommand{\om}{\Omega}

\newcommand{\norm}[1]{\left\lVert #1 \right\rVert}
\newcommand{\x}{\bs{x}}

\newcommand{\del}{\partial}
\newcommand{\avg}[1]{\left\langle#1\right\rangle}

\title{\AorE{Dinamiese Groot Afwykings van Diffusie Prosesse \\ \normalfont\small\itshape(‘‘Dynamical Large Deviations of Diffusions’’)}{Dynamical Large Deviations of Diffusions}}
\author{J.P.\ du Buisson}{Johannes Petrus du Buisson}
\faculty{Faculty of Science}
\degree{PhD (Theoretical Physics)}{Doctor of Science (Theoretical Physics)}
\supervisor{Prof.\ Hugo Touchette}
\cosupervisor{Prof.\ Kristian M\"{u}ller-Nedebock}
\setdate{1}{2023}
\SetCopyrightHolder

\begin{document}
\frontmatter
\TitlePage

\DeclarationDate{January, 2023}
\DeclarationPage

\address{\AorE{Departement Fisika,\\Universiteit van Stellenbosch,\\Stellenbosch, Suid-Afrika.}{Department of Physics,\\University of Stellenbosch,\\Stellenbosch, South Africa.}}

\begin{abstract}
We solve two problems related to the fluctuations of time-integrated functionals of Markov diffusions, used in physics to model nonequilibrium systems. In the first we derive and illustrate the appropriate boundary conditions on the spectral problem used to obtain the large deviations of current-type observables for reflected diffusions. For the second problem we study linear diffusions and obtain exact results for the generating function associated with linear additive, quadratic additive and linear current-type observables by using the Feynman-Kac formula. We investigate the long-time behavior of the generating function for each of these observables to determine both the so-called rate function and the form of the effective process responsible for manifesting the fluctuations of the associated observable. It is found that for each of these observables, the effective process is again a linear diffusion. We apply our general results for a variety of linear diffusions in $\mathbb{R}^2$, with particular emphasis on investigating the manner in which the density and current of the original process are modified in order to create fluctuations. 

\end{abstract}
\begin{abstract}[afrikaans]
Ons los twee probleme op wat verband hou met die fluktuasies van tyd-ge\"{i}ntegreerde funksionale van Markov-diffusies, wat in fisika gebruik word om sisteme buite termiese ewewig te modelleer. Vir die eerste probleem lei ons die toepaslike randvoorwaardes af vir die spektrale probleem wat gebruik word om die groot afwykings van stroom-tipe waarneembares van gereflekteerde diffusies te bekom en illustreer ook ons resultate. Vir die tweede probleem bestudeer ons line\^{e}re diffusies en verkry eksakte resultate vir die genererende funksie wat met line\^{e}re digtheids-, kwadratiese digtheids- en line\^{e}re stroom-tipe waarneembares geassosieer word deur van die Feynman-Kac formule gebruik te maak. Ons ondersoek die groot-tyd gedrag van die genererende funskie vir elk van hierdie waarneembares en verkry die sogenoemde tempo-funksie asook die vorm van die effektiewe proses wat verantwoordelik is vir die manifestering van fluktuasies van die geassosieerde waarneembare. Dit word bevind dat vir elk van hierdie waarneembares is die effektiewe proses weereens 'n line\^{e}re diffusie. Ons pas ons algemene resultate toe vir 'n verskeidenheid line\^{e}re diffusies in $\mathbb{R}^2$, met spesifieke klem op die ondersoek van die wyse waarop die digtheid en stroom van die oorspronklike proses aangepas word in orde om fluksuasies te skep. 

\end{abstract}

\chapter{Acknowledgements}
This work is based on research supported in part by the National Research Foundation of South Africa (Grant Number 134044).

Thank you to my supervisor, Hugo Touchette, for your support and friendship during the PhD. Working with you has been a privilege and a joy. 
\par To my parents, Annalise and Dubbies, thank you for your continuous support over the course of my life, without which none of this would have been possible. I am grateful for everything you have done for me.
\par To my sister, Lise, thank you for inspiring me to pursue greater intellectual honesty and curiosity. I miss you and hope to see you more often. 
\par Finally, thank you to my wife, Lia, for your support, love and generosity, for all the joy and laughter you bring into my life, for coffee in the morning and walks in the woods. You make me a happier person. I love you and am excited to embark on our next adventure. 

\chapter{Dedications}
Dedicated to my wife, Lia, and to my family: Lise, Annalise and Dubbies. 
\tableofcontents

\listoffigures

\mainmatter

\chapter{Introduction} \label{chap:chap1}

\section{Diffusion processes}
\begin{figure}[t] 
        \centering
        \includegraphics[width=10cm]{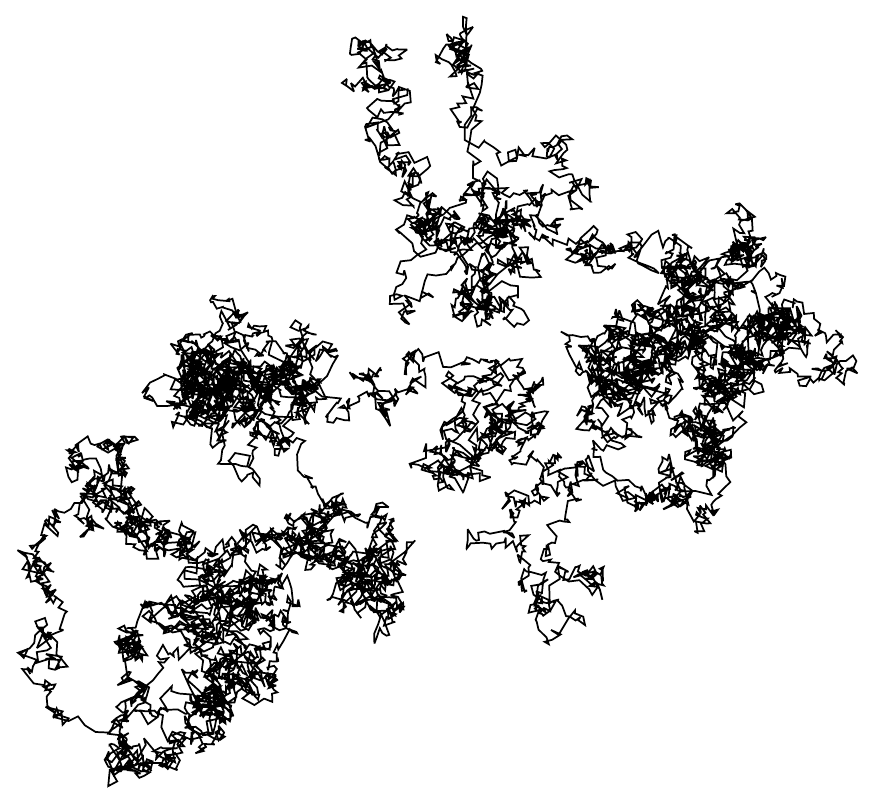}
        \caption{Sample path of Brownian motion in two dimensions.}
        \label{fig:BM2Da}
    \end{figure}

The physical phenomenon underlying the diffusion processes studied in this dissertation is Brownian motion, so named after Robert Brown who, in 1827, observed the irregular motion of pollen grains suspended in water. A trajectory of a process undergoing Brownian motion is shown in Fig.~\ref{fig:BM2Da}. Einstein~\cite{einstein1905molekularkinetischen} initiated the probabilistic study of this type of motion in order to understand diffusion processes in physics. His work allowed for the determination of Avogadro's number via the experimental measurement, made by Perrin, of the mean square displacement of particles suspended in a fluid and was a key piece of evidence in favor of the existence of atoms~\cite{newburgh2006einstein}. Independent of Einstein, the study of Brownian motion as a Markov process was undertaken by Smoluchowski~\cite{von1906kinetischen}.
\par An alternative formulation of Brownian motion as a system obeying Newton's equation of motion in the presence of a random force was constructed by Langevin~\cite{langevin1908theorie}. This formulation formed the basis for the modern theory of stochastic differential equations, later developed by It\^{o}~\cite{kunita2010ito}. 
\par A stochastic differential equation can be written informally as 
\begin{equation}
    \dot{\bs{X}}(t) = \bs{F}(\bs{X}(t)) + \bs{\xi}(t),
\end{equation}
where $\bs{F}$ is a deterministic force and the so-called Gaussian white noise $\bs{\xi}(t)$ represents the random force present in the system and accounts for uncertainty in the evolution of the system. The trajectories of a process satisfying a stochastic differential equation are everywhere continuous but nowhere differentiable and represent in this manner the irregular motion that is the hallmark of Brownian motion. A process that is governed by a stochastic differential equation is also known as a diffusion.
\par Diffusion processes appear in a variety of contexts, including: 
\begin{itemize}
    \item Nonequilibrium systems such as biomolecular motors or nano-machines in biophysics~\cite{seifert2012stochastic,julicher1997modeling}. 
    \item Mesoscopic systems suspended in a fluid and manipulated via laser tweezers~\cite{PhysRevX.7.021051}. 
    \item In finance, where stock prices are modelled by a stochastic differential equation known as the geometric Brownian motion~\cite{shreve2004stochastic}. 
    \item As the continuous limit of systems with a discrete state space, for example in population dynamics or queuing theory~\cite{jacobs2010stochastic,meyn2008control}. 
\end{itemize}
\par In this dissertation we will study separately two particular classes of diffusion processes: those evolving in a bounded domain and undergoing reflection at the boundary of that domain, and the class of linear diffusions for which the deterministic force $\bs{F}$ present in the stochastic differential equation is linear in the state $\bs{X}(t)$ of the diffusion. 
\par While diffusion processes are often considered as taking place in an unbounded domain $\mathbb{R}^n$, it is clearly also important to consider processes evolving in some subset of $\mathbb{R}^n$~\cite{gardiner1985handbook}. Examples of such processes are a diffusion taking place inside a biological cell~\cite{schuss2013,bressloff2014stochastic} or a process which has a state which is inherently constrained as a result of its definition, for example the price of a stock which cannot be negative. For such processes both the geometry of the boundary and the behavior of the process at the boundary are important and must be specified.
\par The study of bounded diffusions is an active area in probability theory and in statistical physics and was initiated in large part by Feller, who provided a complete classification of all possible types of boundary behavior in one dimension~\cite{feller1952parabolic,feller1954diffusion,peskir2015boundary}. We will consider in this dissertation only the case where the boundaries present in the system are reflecting in nature. A construction of reflected diffusions at the level of stochastic differential equations to account for reflection at the boundary was provided by Skorokhod~\cite{skorokhod1961stochastic, skorokhod1962stochastic} via the inclusion of a so-called local time term in the stochastic differential equation. A survey of results on reflected diffusions is provided in Grebenkov~\cite{grebenkov2007nmr}. 
\par Linear diffusions, the second type of diffusion that we shall consider, constitute an important class of stochastic differential equations, as they are simple enough to be analytically studied while being rich enough to exhibit interesting behavior. As such, these systems have been studied extensively in both mathematics and physics~\cite{bryc1997large,bercu1997large,bercu2002sharp,chatterjee2010exact,noh2014fluctuations,kwon2011nonequilibrium,saha2015work,volpe2006torque}. Examples of linear diffusions include colloidal particles trapped by optical tweezers~\cite{wang2002experimental,van2003extension,van2004extended}, RC electrical circuits~\cite{gonzalez2019experimental} and bead-spring models used to simulate polymers~\cite{liu2008effective}. Particularly interesting to us is the fact that systems of this sort can manifest both equilibrium behavior, for which probability currents are absent, and nonequilibrium behavior, characterized by the existence of probability currents as a result of non-conservative forces or the presence of multiple heat baths at different temperatures~\cite{noh2014fluctuations,kwon2011nonequilibrium,dotsenko2013two,falasco2015energy,saha2015work}. Linear diffusions have also been examined using a framework known as stochastic thermodynamics~\cite{seifert2012stochastic,sekimoto1998langevin,sekimoto2010stochastic}, which extends the notions of traditional thermodynamics such as heat, work and entropy production, to individual trajectories of a stochastic process. General results such as thermodynamic uncertainty relations \cite{horowitz2020thermodynamic,gomez2010steady} and fluctuation relations~\cite{gallavotti1995dynamical,lebowitz1999gallavotti,seifert2012stochastic}, which constrain the probability distributions of quantities such as heat and work, have been illustrated in the context of linear diffusions \cite{gupta2020thermodynamic,noh2014fluctuations,kwon2011nonequilibrium}. 

\section{Dynamical large deviations}
Statistical information regarding the state $\bs{X}(t)$ of a diffusion is contained in the probability density $p_t(\x) = P(\bs{X}(t) = \x)$ describing the probability of the diffusion having a particular value at a given time. In this dissertation, rather than studying the process $\bs{X}(t)$ directly, we will instead study time integrated functionals
\begin{equation}
    A_T = \frac{1}{T} \int_0^T f(\bs{X}(t)) dt + \frac{1}{T} \int_0^T \bs{g}(\bs{X}(t))\circ d\bs{X}(t),
\end{equation}
that depends on the entire trajectory of the diffusion $\bs{X}(t)$ over the course of the interval $t\in [0,T]$. These functionals are called in physics dynamical observables~\cite{touchette2018introduction}. Two broad classes of dynamical observable can be distinguished, namely additive observables for which $\bs{g} = \bs{0}$ and current-type observable for which $f = 0$. Examples of dynamical observables include: 
\begin{itemize}
    \item The empirical density $\rho_T(\bs{y})$ representing the fraction of time (out of a total time $T$) that the process spends at $\bs{y}$, for which $f(\x) = \delta(\x - \bs{y})$ and $\bs{g}=\bs{0}.$
    \item The empirical current $\bs{J}_T(\bs{y})$ representing the number of passes (counted with sign) through $\bs{y}$ per unit time over the interval $t \in [0,T]$. This vector observable is defined with $f = 0$ and $\bs{g} = \delta(\x - \bs{y})$.
    \item A variety of quantities of interest in stochastic thermodynamics, including~\cite{seifert2012stochastic,sekimoto2010stochastic} the nonequilibrium work done on the system, the heat exchanged between the system and its environment, and the entropy produced over the course of the evolution of the system. These are all examples of current-type observables. The entropy production, for instance, is given explicitly by $\bs{g}(\x) = 2 (D^{-1}\bs{F}(\x))$, where $D$ is the diffusion matrix associated with a diffusion~\cite{touchette2018introduction}. 
\end{itemize}
\par In order to understand the statistical properties of a dynamical observable we must obtain the density $P(A_T = a)$ associated with it, which describes both the typical values of the observable as well as the fluctuations away from these typical values. In practice obtaining this density directly is difficult, even for relatively simple systems and observables. However, following the theory of large deviations~\cite{dembo1998}, as developed by Varadhan, we expect that the density $P(A_T = a)$ often has the asymptotic form 
\begin{equation}
    P(A_T = a) \approx e^{-T I(a)}
\end{equation}
for large $T$. An observable for which this approximation applies is said to satisfy a large deviation principle, with the rate function $I(a)$ characterizing the typical values of the observable and the likelihood of fluctuations in the long-time limit. 
\par The theory of large deviations as applied to dynamical observables is concerned not only with obtaining the rate function, which describes the probability of fluctuations occurring, but also with describing the manner in which these fluctuations are created dynamically in time~\cite{touchette2018introduction}. This information is provided~\cite{chetrite2013nonequilibrium,chetrite2015nonequilibrium,chetrite2015variational} via an effective process which models, in the long-time limit, the behavior of the process $\bs{X}(t)$ leading to a particular fluctuation. In this dissertation we will apply the theory of dynamical large deviations to the study of both reflected diffusions and linear diffusions in order to understand the fluctuations of dynamical observables of these processes in the long-time limit. 
\section{Goals and previous works}
In practice the rate function $I(a)$ characterizing the probability of a particular fluctuation occurring as well as the effective process describing the behavior of the process associated with that fluctuation is given by the long-time behavior of the generating function of $A_T$, or equivalently by the dominant eigenvalue and the associated eigenfunction of a related spectral problem. 
\par For reflected diffusions, the appropriate boundary conditions for the spectral problem associated with the large deviations of an additive observable $A_T$, for which $\bs{g} = \bs{0}$ in the notation of the previous section, were obtained by Du Buisson and Touchette~\cite{dubuissonmasters,Buisson2020}. The argument used to obtain these boundary conditions does not extend to the case where the dynamical observable $A_T$ is of current-type, having $f = 0$ and $\bs{g} \neq \bs{0}$. The first problem that we will consider in the dissertation is to present a new argument, developed in collaboration with Mallmin~\cite{Mallmin2021} and which allows us to obtain the boundary conditions for the spectral problem associated with current-type observables. We show that the boundary conditions for these observables differ in a non-trivial manner from those obtained for the case of additive observables. 
\par Large deviations have been studied before for reflected diffusions, in particular, by Grebenkov~\cite{grebenkov2007residence}, Forde~\cite{forde2015large}, and Fatalov~\cite{fatalov2017}, who obtain the rate function of various functionals of reflected Brownian motion, including its area and the residence time at a reflecting point. The large deviations of bounded diffusions have also been studied previously using the so-called level 2 large deviations involving the empirical density, by Pinsky~\cite{pinsky1985function, pinsky1985classification} and Budhiraja and Dupuis~\cite{budhiraja2003large}. Finally, studies of large deviations of bounded diffusions in the low-noise limit (as opposed to the long-time limit which we consider) include Ignatyuk~\cite{ignatyuk1994boundary}, Sheu~\cite{sheu1998large}, Dupuis~\cite{dupuis1987large}, Bo and Zhang~\cite{bo2009large} and Sheu~\cite{sheu1998large}. In these studies the observable of interest typically involve the state $\bs{X}(t)$ of the diffusion at a fixed or random time rather than the entire history of the diffusion, as considered here. 
\par The second problem that we consider is to obtain general results for the large deviations associated with a variety of observables for linear diffusions. The classes of observables considered in this dissertation are two additive observables, which are linear and quadratic, respectively, in the state of the diffusion $\bs{X}(t)$, and a current-type observable which is linear in the state $\bs{X}(t)$. Linear current-type observables are particularly interesting and include many quantitites of physical interest, such as the nonequilibrium work, entropy production and heat transferred to and from the system~\cite{saha2015work,noh2014fluctuations,kwon2011nonequilibrium}. The rate function and effective process are here obtained by explicitly calculating the generating function for these observables and studying its long-time behavior. 
\par A variety of results have been obtained for the types of observables we study in this dissertation. In particular, the rate function for a variety of quadratic additive observables has been obtained by Bryc and Dembo~\cite{bryc1997large} and Bercu~\cite{bercu1997large,bercu2002sharp}. Linear current-type observables have also been studied extensively. Exact results for the generating function were obtained for the nonequilibrium work by Kwon et al.~\cite{kwon2011nonequilibrium}. This work was extended further by Noh~\cite{noh2014fluctuations} to the study of the heat transfer and energy change in linear systems, obtaining the rate function for these observables for a specific system. Similarly, Saha and Mukherji~\cite{saha2015work} obtained the probability density associated with the work for a linear system having a non-conservative force. An exact solution for the heat probability density for a trapped Brownian oscillator was obtained by Chatterjee and Cherayil~\cite{chatterjee2010exact}. As such, many exact or large deviation results exist for the probability densities associated with the observables we shall study. However, the effective process describing the manner in which fluctuations of these observables are manifested has not been obtained previously. One of our main goals is to obtain this effective process in order to gain an understanding of how the probability density and the probability current of a linear process are modified to create the fluctuations of a particular observable. The creation of currents to realize nonequilibrium fluctuations of equilibrium processes and the modification of currents in processes which are nonequilibrium to begin with are particularly interesting topics. 
\par The work most similar to that done in this dissertation pertaining to linear diffusions is that of Kwon et al.~\cite{kwon2011nonequilibrium} and Noh~\cite{noh2014fluctuations}, who obtained via path-integral methods the generating function associated with particular instances of linear current-type observables including the nonequilibrium work, heat and energy change. Our results are obtained instead via the Feynman-Kac formula and serve as a generalization of these works to all linear additive, quadratic additive, and linear current-type observables. Furthermore, long-time results for the generating function were obtained~\cite{kwon2011nonequilibrium,noh2014fluctuations} either via numerical methods or analytically for only particular systems, but did not explore the general features of the long-time solution. We study this aspect of the problem in more depth, both at a general level and for a variety of particular systems, including one equilibrium system and two nonequilibrium systems, and for a variety of observables, including the stochastic area, which has recently attracted some attention~\cite{ghanta2017fluctuation,gonzalez2019experimental}. We also obtain for the first time the large deviations and effective process associated with this observable for linear diffusions.  

\section{Outline} 
The dissertation is structured as follows. In Chap.~\ref{chap:chap2} we introduce those elements of the theory of diffusion processes and dynamical large deviations that we will need throughout this dissertation. Chapter~\ref{chap:chap3} is concerned with reflected diffusions. We first summarize recent work~\cite{dubuissonmasters, Buisson2020} on obtaining the boundary conditions on the spectral problem associated with the large deviations of additive observables for these processes and then explain why these arguments fail for current-type observables. We then show the novel argument, published in~\cite{Mallmin2021}, to obtain these boundary conditions and illustrate these results for a heterogeneous single file diffusion system, also studied in~\cite{Mallmin2021}. 
\par Chapter~\ref{chap:chap4} contains a derivation of the generating function for the three classes of observables mentioned previously for linear diffusions. Particular emphasis is placed on the long-time form of this solution which allows us to obtain both the rate function and the effective process for these observables. We also obtain explicit expressions for the asymptotic mean and variance for all the observables considered. 
\par The general results for linear diffusions are illustrated in Chap.~\ref{chap:chap5} for three linear systems: an equilibrium process, a nonequilibrium process having a nonconservative force, and another nonequilibrium process having a conservative force but in contact with multiple heat baths at different temperatures. We apply the formalism developed in Chap.~\ref{chap:chap4} for a quadratic additive observable and for a variety of linear current-type observables, including the nonequilibrium work, the entropy production, as well as the stochastic area, which is an observable that is of great interest currently as a metric of irreversibility in diffusions.
\par Finally, in Chap.~\ref{chap:chap6} we provide a summary of the results obtained in the dissertation and discuss open problems and possible directions for further study. 
\chapter{Mathematical preliminaries} \label{chap:chap2}
In this dissertation we will be concerned with the application and extension of the theory of dynamical large deviations as applied to Markov diffusions. We introduce in this chapter those elements of the theory of Markov diffusions and large deviations that will be needed for the work presented in this dissertation. For a general introduction to probability theory and stochastic processes the reader is referred to Grimmett and Stirzaker~\cite{grimmett2001probability}. For a text with a greater emphasis on diffusion processes, see Pavliotis~\cite{pavliotis2014stochastic}. Our introduction of the theory of dynamical large deviations mirrors that found in~\cite{touchette2018introduction} and also in the MSc thesis of Du Buisson~\cite{dubuissonmasters}. For a more comprehensive introduction to large deviation theory, see Dembo and Zeitouni~\cite{dembo1998}. 

\section{Markov diffusions and observables} \label{sec:sec21}
We consider in this dissertation $n$-dimensional Markov diffusions $\bs{X}(t)$ evolving in a subset $\om$ of $\mathbb{R}^n$ according to a \textit{stochastic differential equation} (SDE) having the form
\begin{equation} \label{SDE}
    d\bs{X}(t) = \bs{F}(\bs{X}(t)) \, dt + \sigma \, d\bs{W}(t),
\end{equation}
where 
\begin{itemize}
    \item $\bs{X}(t) \in \om$ is a random vector representing the state of the system at time $t \in \mathbb{R}^{+}$.
    \item The \textit{drift} $\bs{F}: \mathbb{R}^n \rightarrow \mathbb{R}^n$ is a vector field that describes the deterministic force acting on the system. 
    \item $\bs{W}(t)\in \mathbb{R}^m$ is a vector of independently distributed Wiener motions, with each component of the \textit{Wiener increment} $d\bs{W}(t)$ being Gaussian distributed with mean $0$ and variance $dt$. Note that the dimension of $\bs{W}(t)$ need not be the same as the dimension of the state $\bs{X}(t)$. 
    \item The \textit{noise matrix} $\sigma$ is an $n \times m$ matrix, which describes the strength and type of noise present in the system, with $m$ matching the dimensions of $\bs{v}(t)$. 
    
\end{itemize}
The process so defined is a Markov process because at every point in time the evolution of $\bs{X}(t)$, determined by its increment $d\bs{X}(t)$, depends only on the current state of $\bs{X}(t)$. Note that it is possible to consider a more general type of SDE for which the drift depends explicitly on the time $t$ and the noise depends explicitly on the state $\bs{X}(t)$ or the time $t$, or both. We will not consider these cases here. 
\par In this dissertation we will be concerned with studying random variables known as \textit{time-integrated functionals} or \textit{dynamical observables} having the form 
\begin{equation} \label{obs}
    A_T = \frac{1}{T} \int_0^T f(\bs{X}(t)) dt + \frac{1}{T} \int_0^T \g(\bs{X}(t)) \circ d\bs{X}(t),
\end{equation}
which represent physical quantities that depend on the entire history of the state $\bs{X}(t)$ from $t = 0$ to $t = T$. Since the trajectory $\bs{X}(t)$ is random, so is the value of the observable $A_T$. We will be interested in studying the probability distribution of such observables as $T$ becomes large. 
\par In order for the above observable to be well-defined, we must specify what is meant by the \textit{stochastic integral}
\begin{equation}
    \int_0^T \g(\bs{X}(t)) \circ d\bs{X}(t).
\end{equation}
This integral differs from the Riemann integral in that its value depends on the integration convention used, that is, in the manner in which values of the integrand are chosen in each discretisation interval when defining the integral as the limit of a discrete sum~\cite{jacobs2010stochastic}. The symbol $\circ$ indicates that we employ the middle-point convention, also known as the \textit{Stratonovich convention}, given explicitly as 
\begin{equation} \label{strato1}
    \int_0^T \g(\bs{X}(t)) \circ d\bs{X}(t) = \lim_{\Delta t \rightarrow 0} \sum_i \g\left(\frac{\bs{X}(t_i) + \bs{X}(t_i + \Delta t)}{2} \right) \cdot \left( \bs{X}(t_i+\Delta t) - \bs{X}(t_i)\right).
\end{equation}
A useful property of the Stratonovich convention is that it preserves the usual rules of calculus. In particular, we have that 
\begin{equation} \label{strato2}
    \int_0^T \bs{\nabla}h(\bs{X}(t)) \circ d\bs{X}(t) = h(\bs{X}(T)) - h(\bs{X}(0)). 
\end{equation}
Another commonly used integration convention is the \textit{It\^{o} convention}, where the left-most point in each discretization interval is chosen~\cite{jacobs2010stochastic}. 
\par Given that the formal solution to the SDE (\ref{SDE}) is given by 
\begin{equation}
    \bs{X}(t) = \bs{X}(0) + \int_0^T \bs{F}(\bs{X}(t)) dt + \int_0^T \sigma d\bs{W}(t),
\end{equation}
we must also specify~\cite{jacobs2010stochastic} an integration convention for the SDE in the event that the noise matrix $\sigma$ depends on the state $\bs{X}(t)$. In the case where the noise matrix $\sigma$ does not depend on the state $\bs{X}(t)$ all integration conventions produce the same results. Thus we do not need to discuss the transformations taking us from one integration convention to another, given that we will only ever consider state-independent diffusion matrices. Moreover for dynamical observables we will always be interested only in the Stratonovich convention. 
\par We reiterate that an observable $A_T$ having the form 
\begin{equation} \label{additive}
    A_T = \frac{1}{T} \int_0^T f\left(\bs{X}(t) \right)\, dt
\end{equation}
so that $\bs{g} = \bs{0}$ is known as an \textit{additive} or \textit{occupation-type} observable, while an observable of the form 
\begin{equation} \label{currentobs}
    A_T = \frac{1}{T} \int_0^T \bs{g}\left(\bs{X}(t)\right) \circ d\bs{X}(t)
\end{equation}
is known as a \textit{current-type} observable. 

\section{The Fokker-Planck equation} \label{sec:sec22}
The probability density $p_t(\x) = P(\bs{X}(t) = \x)$ associated with a diffusion $\bs{X}(t)$ satisfies a partial differential equation, known as the \textit{Fokker-Planck equation}, given for a diffusion evolving according to the SDE (\ref{SDE}) by
\begin{equation} \label{FP}
    \partial_t p_t(\x) = -\bs{\nabla}\cdot \left(\bs{F}(\x) p_t(\x) \right) + \frac{1}{2} \bs{\nabla}\cdot D \bs{\nabla} p_t(\x),
\end{equation}
with initial condition $p_0(\x) = P(\bs{X}(0) = \x)$, and where we have introduced the \textit{diffusion matrix} $D=\sigma\sigma^{\mathsf{T}}$. The interested reader is referred to Risken~\cite{risken1996fokker} for more information regarding the Fokker-Planck equation. Equation (\ref{FP}) can be rewritten as a linear equation
\begin{equation} \label{FPeq2}
    \partial_t p_t(\x)= \mathcal{L}^{\dagger} p_t(\x),
\end{equation}
in terms of a second-order differential operator 
\begin{equation} \label{FPO}
    \mathcal{L}^{\dagger} = -\bs{\nabla} \cdot \bs{F} + \frac{1}{2} \bs{\nabla}\cdot D \bs{\nabla},
\end{equation}
known as the \textit{Fokker-Planck operator}. This operator acts on an appropriate space of normalizeable, twice-differentiable densities. The requirement that densities are normalized amounts to the statement that 
\begin{equation}
    \int_{\om} p_t(\x) d\x = 1 
\end{equation}
for all times $t$. We note that in order for the Fokker-Planck operator to be completely specified we must specify not only its form as a differential operator (as in (\ref{FPO})) but also the domain of functions $\mathcal{D}(\mathcal{L}^{\dagger})$ on which it acts. We will discuss this in more detail in the next chapter.
\par The Fokker-Planck equation can also be expressed in the form of a conservation equation as
\begin{equation} \label{conservationeq}
    \partial_t p_t(\x) + \bs{\nabla} \cdot \bs{J}_{\bs{F}, p_t}(\x,t) = 0,
\end{equation}
featuring the time-dependent \textit{probability current} $\bs{J}_{\bs{F},p_t}$ which is a vector field describing the spatial flow of probability at a given point in time $t$, and which is given explicitly by 
\begin{equation} \label{current}
    \bs{J}_{\bs{F},p_t}(\x,t) = \bs{F}(\x) p_t(\x) - \frac{1}{2} D\bs{\nabla}p_t(\x).
\end{equation}
The subscripts in the notation $\bs{J}_{\bs{F},p}$ indicate that we are considering the probability current associated with a particular drift and density. This notation will prove useful later when discussing the current associated with various drifts and densities in describing the manner in which fluctuations of physical observables are manifested. We do not include the diffusion matrix $D$ as a subscript given that the diffusion matrix will always be clear from the context. 
\par In this dissertation we will mostly be interested in \textit{ergodic} processes for which there exists a unique stationary density $p^*(\x)$ satisfying 
\begin{equation} \label{stationarydef}
    \mathcal{L}^{\dagger}p^* = 0. 
\end{equation}
Associated with the stationary density is the stationary current $\bs{J}_{\bs{F},p^*}$, given explicitly as 
\begin{equation} \label{statcur}
    \bs{J}_{\bs{F},p^*}(\x) = \bs{F}(\x)p^*(\x) - \frac{1}{2}D \bs{\nabla}p^*(\x).
\end{equation}
It can be checked, from (\ref{conservationeq}) and (\ref{stationarydef}), that this current satisfies the divergence-free condition 
\begin{equation}
    \bs{\nabla}\cdot \bs{J}_{\bs{F},p^*} = 0.
\end{equation}
\par For an ergodic process, time-averaged quantities converge by the ergodic theorem to an expectation value with respect to the stationary density or stationary current~\cite{pavliotis2014stochastic}. This convergence is convergence in probability, meaning that for an observable $A_T$ having the form in (\ref{obs}) that 
\begin{equation} \label{ergodicdensity}
    \lim_{T\rightarrow\infty} P(|A_T -a^*| < \epsilon) = 1
\end{equation}
for all $\epsilon > 0$, where 
\begin{equation} \label{ergodicexp}
    a^* = \int_{\om} f(\x)p^*(\x)d\x + \int_{\om} \g(\x)\cdot \bs{J}_{\bs{F},p^*}d\x
\end{equation}
is the stationary expectation. 

\section{Infinitesimal generator}\label{sec:sec23}
In the previous section we discussed the time evolution of the probability density $p_t(\x)$ associated with a diffusion process $\bs{X}(t)$. We now turn our attention to the time evolution of expectation values. 
\par The expectation value $\mathbb{E}[h(\bs{X}(t))]$ of a `test' function $h$ with respect to the state $\bs{X}(t)$ at time $t$ is defined as
\begin{equation} \label{expvalue}
    \mathbb{E}\left[h(\bs{X}(t)) \right] = \int_{\om} h(\x) p_t(\x)d\x.
\end{equation}
Introducing the $L^2$ inner product 
\begin{equation} \label{innerproduct}
    \langle p, h \rangle = \int_{\om} p(\x) h(\x) d\x,
\end{equation}
with $p$ a normalized density and $h$ a suitable test function such that the inner product is finite, we can write the expectation value (\ref{expvalue}) as 
\begin{equation}
    \mathbb{E}\left[h(\bs{X}(t)) \right] = \langle p_t, h \rangle.
\end{equation}
The time evolution of this expectation value is given by 
\begin{equation}
    \partial_t \mathbb{E}\left[h(\bs{X}(t)) \right] = \int_{\om} h(\x) \partial_t p_t(\x) d\x = \int_{\om} h(\x) \mathcal{L}^{\dagger}p_t(\x) d\x = \langle \mathcal{L}^{\dagger} p_t, h\rangle, 
\end{equation}
where we have used (\ref{FPeq2}) and the definition of the inner product (\ref{innerproduct}). 
\par From this result, it is natural to introduce an \textit{adjoint} operator $\mathcal{L}$ that satisfies 
\begin{equation} \label{adjoint}
    \langle \mathcal{L}^{\dagger} p,h \rangle = \langle p, \mathcal{L} h\rangle ,
\end{equation}
where 
\begin{equation}
    \langle p,\mathcal{L} h\rangle = \int_{\om} (\mathcal{L}h)(\x) p(\x)d\x
\end{equation}
for all suitable densities and test functions. We then have that 
\begin{equation}
    \partial_t \mathbb{E}\left[h(\bs{X}(t)) \right] = \langle \mathcal{L}^{\dagger} p_t, h\rangle = \langle p_t,\mathcal{L} h\rangle,
\end{equation}
and therefore it follows that
\begin{equation}
    \partial_t \mathbb{E}\left[h(\bs{X}(t)) \right] = \mathbb{E}\left[\left(\mathcal{L}h \right)(\bs{X}(t))\right],
\end{equation}
so that the adjoint operator $\mathcal{L}$ governs the time evolution of expectation values. For this reason, $\mathcal{L}$ is called the \textit{infinitesimal generator} of the diffusion $\bs{X}(t)$. \par In order to obtain the explicit form of $\mathcal{L}$, we must consider the duality relation (\ref{adjoint}). We have that 
\begin{equation}
    \langle \mathcal{L}^{\dagger}p,h\rangle = \int_{\om} \left[-\bs{\nabla} \cdot \left(\bs{F}(\x)p(\x)\right) + \frac{1}{2} \bs{\nabla}\cdot D \bs{\nabla}p(\x)\right]h(\x) d\x.
\end{equation}
Using integration by parts in $\mathbb{R}^n$, we can shift the action of the derivatives from the density $p$ to the test function $h$, producing boundary terms in the process, so as to obtain 
\begin{align}
    \langle p,\mathcal{L} h\rangle&= \int_{\om}p(\x) \left[\bs{F}(\x)\cdot \bs{\nabla} + \frac{1}{2}\bs{\nabla}\cdot D\bs{\nabla}\right]h(\x) d\x  \nonumber \\ &\quad + \, \tn{boundary/surface terms involving $p$ and $h$}.
\end{align}
By referring to the definition (\ref{adjoint}) we see that in order for the adjoint $\mathcal{L}$ to exist and to be defined independently of a particular density $p$ and test function $h$ the boundary terms must be made to vanish by imposing appropriate restrictions on the domains of the operators $\mathcal{L}$ and $\mathcal{L}^{\dagger}$. Provided these conditions are satisfied, the infinitesimal generator must then have the form 
\begin{equation} \label{IG}
    \mathcal{L} = \bs{F}\cdot \bs{\nabla} +\frac{1}{2} \bs{\nabla} \cdot  D\bs{\nabla}.
\end{equation}
\par In general there are many possible ways of restricting the domains of $\mathcal{L}$ and $\mathcal{L}^{\dagger}$ which will make the boundary terms arising in the integration by parts procedure vanish. A particular choice of restriction that accomplishes this is known as a boundary condition. We will discuss issues related to the domains of the Markov operators $\mathcal{L}$ and $\mathcal{L}^{\dagger}$ in more depth in Chap.~\ref{chap:chap3}. We note here simply that the expressions (\ref{FPO}) and (\ref{IG}) for the form of $\mathcal{L}^{\dagger}$ and $\mathcal{L}$ as differential operators are always valid for an SDE having the form (\ref{SDE}), with only the domains of these operators differing depending on the particular physical situation under consideration. For more information regarding the generator and its adjoint, see Pavliotis~\cite{pavliotis2014stochastic}.
\section{Linear diffusions}\label{sec:sec24}
An important class of diffusions, which we shall study in depth in Chaps~\ref{chap:chap4} and~\ref{chap:chap5}, is the class of linear diffusions in $\mathbb{R}^n$ defined by the SDE 
\begin{equation} \label{eqSDE}
    d\bs{X}(t) = - M \bs{X}(t) \, dt + \sigma d\bs{W}(t),
\end{equation}
involving the drift 
\begin{equation}
    \bs{F}(\x) = - M \x
\end{equation} 
which is linear in the state $\x\in\mathbb{R}^n$. For reasons that will become clear, we will always assume that the drift matrix $M$ and the symmetric diffusion matrix $D = \sigma\sigma^{\mathsf{T}}$ are positive definite. In the mathematical literature, the definition of a positive definite matrix typically assumes that the matrix in question is symmetric and then requires that matrix to have positive eigenvalues. We instead say that a general (not necessarily symmetric) square matrix is positive definite if all its eigenvalues have positive real part. This latter definition will be used throughout this dissertation and reduces to the former definition when a matrix is real and symmetric. 
\par The infinitesimal generator $\mathcal{L}$ associated with the SDE (\ref{eqSDE}) is given from (\ref{IG}) by 
\begin{equation}
    \mathcal{L} = -M \x \cdot \bs{\nabla} + \frac{1}{2} \bs{\nabla} \cdot D \bs{\nabla},
\end{equation}
while the Fokker-Planck operator $\mathcal{L}^{\dagger}$ is found as the $L^2$ adjoint of $\mathcal{L}$ and is given from (\ref{FPO}) as 
\begin{align}
    \mathcal{L}^{\dagger} &= -\bs{\nabla}\cdot \left(M\x \right) + \frac{1}{2} \bs{\nabla} \cdot D \bs{\nabla} \nonumber \\ &=\tn{Tr}(M) + M\x \cdot \bs{\nabla} + \frac{1}{2} \bs{\nabla} \cdot D \bs{\nabla}.
\end{align}
Assuming the initial condition $\bs{X}(0) = \x_0$, the density $p(\x,t) = P(\bs{X}(t) = \x| \bs{X}(0) = \x_0)$ evolves according to the Fokker-Planck equation (\ref{FP}) and has an exact solution which is known to be a Gaussian density for all times $t$. The explicit form of this solution is given~\cite{pavliotis2014stochastic} as 
\begin{equation}
    p_t(\x) = \sqrt{\frac{1}{(2 \pi)^n \tn{det}C_t}} \exp\left(-\frac{1}{2}\langle \x - \bs{m}_t, C_t^{-1} (\x - \bs{m}_t)\rangle \right),
\end{equation}
where we now use $\langle \cdot, \cdot \rangle$ to indicate the standard vector inner product on $\mathbb{R}^n$ (not to be confused with the inner product (\ref{innerproduct})). In this expression $\bs{m}_t$ is the mean of $\bs{X}(t)$  
\begin{equation}
    \bs{m}_t = \mathbb{E}_{\bs{x}_0}\left[\bs{X}_t \right] = e^{-t M} \bs{x}_0,
\end{equation}
while $C_t$ is the covariance matrix at time $t$ defined as
\begin{equation}
    (C_t)_{ij} = \mathbb{E}_{\bs{x}_0}\left[(\bs{X}_t - \bs{m}_t)_i(\bs{X}_t - \bs{m}_t)_j  \right],
\end{equation}
which satisfies the differential equation
\begin{equation}
    \dot{C_t} = -MC_t - C_t M^{\mathsf{T}} + D.
\end{equation}
\par In the event that $M$ and $D$ are positive definite the above has a unique positive definite stationary solution $C = C_{\infty}$ satisfying a so-called \textit{Lyapanov equation}
\begin{equation} \label{ricccov}
    D = MC + CM^{\mathsf{T}}. 
\end{equation}
The requirement that $M$ and $D$ are positive definite is therefore sufficient in order for the Fokker-Planck equation to have a stationary solution and for the process to be ergodic~\cite{pavliotis2014stochastic}. The invariant density $p^*$ satisfying $\mathcal{L} p^* =0$ is then given by 
\begin{equation} \label{statden}
    p^*(\x) = \sqrt{\frac{1}{(2 \pi)^n \tn{det}C}} \exp\left(-\frac{1}{2}\langle \x, C^{-1} \x\rangle \right),
\end{equation}
while the stationary probability current $\bs{J}_{\bs{F},p^*}$ associated with this density is found to be 
\begin{equation} \label{statcurlin}
    \bs{J}_{\bs{F},p^*}(\x) = \left(\frac{D}{2} C^{-1} - M \right)\x p^*(\x),
\end{equation}
where we have used (\ref{statcur}) and the expression (\ref{statden}). 
\par A stationary density $p^*$ which is such that $\bs{J}_{\bs{F},p^*}(\x) = \bs{0}$ for all $\x$ is said to satisfy \textit{detailed balance}, and is an equilibrium steady state, while a non-zero stationary probability current is associated with violation of detailed balance and hence with a nonequilibrium steady state. For linear diffusions we can distinguish two distinct sources of nonequilibrium behavior: a non-symmetric drift matrix $M$ (and hence a non-gradient drift $\bs{F}$) and a diffusion matrix $D$ not proportional to the identity matrix $\mathbb{I}$. Note however that there are certain specific cases in which non-symmetric $M$ and a diffusion matrix $D$ not proportional to the identity satisfy 
\begin{equation}
    \frac{D C^{-1}}{2} - M = 0,
\end{equation}
in which case the process nonetheless has an equilibrium steady state. 
\par We consider next three examples of linear diffusions evolving in $\mathbb{R}^2$, one equilibrium and two nonequilibrium, which are important in physics. The two nonequilibrium systems illustrate the two sources of nonequilibirum behavior just described. We will study these systems again in Chap.~\ref{chap:chap5}.
\subsection{Gradient diffusion}\label{sec:sec241}
The first linear diffusion that we consider is a simple gradient diffusion $\bs{X}(t) = (X_1(t), X_2(t))^{\mathsf{T}}$ evolving in $\mathbb{R}^2$ according to the SDE
\begin{equation} \label{gradsde}
    d\bs{X}(t) = - \begin{pmatrix}\gamma && 0\\0 && \gamma \end{pmatrix}\bs{X}(t) \, dt + \epsilon d\bs{W}(t),
\end{equation}
with $\gamma > 0$ the friction coefficient and $\epsilon > 0$ the noise strength. For this system the drift matrix $M$ and diffusion matrix $D$ are given by $M = \gamma \mathbb{I}$ and $D = \epsilon^2 \mathbb{I}$, respectively, where $\mathbb{I}$ indicates the identity matrix. The process is an example of a gradient diffusion since the drift $\bs{F}(\x) = - M\x$, with $\x = (x_1,x_2)^{\mathsf{T}}$, can be written as the negative of the gradient of the function $U(\x) = \gamma \norm{\x}^2/2$ and the diffusion matrix is proportional to the identity matrix. Systems of this type are often used~\cite{wang2002experimental,van2003extension,van2004extended} to model colloidal particles trapped by optical tweezers, with the harmonic potential $U(\x)$ related to the friction coefficient $\gamma$ representing the optical trap. 
\par The stationary density $p^*$ for this process is given from (\ref{statden}) with the stationary covariance matrix $C$ satisfying the algebraic equation (\ref{ricccov}), which gives here
\begin{equation}
    C = \frac{\epsilon^2}{2\gamma}\mathbb{I}.
\end{equation}
As a result, we have 
\begin{equation} \label{statdengrad}
    p^*(\x) = \frac{\gamma}{\pi\epsilon^2}\exp\left(-\frac{\gamma}{\epsilon^2}\norm{\x}^2 \right).
\end{equation}
From the expression (\ref{statcurlin}) for the stationary current associated with this process we observe that, since 
\begin{equation}
    \frac{DC^{-1}}{2} - M = \frac{1}{2} \epsilon^2 \frac{2\gamma}{\epsilon^2}\mathbb{I} - \gamma \mathbb{I} = 0,
\end{equation}
this process has zero stationary current. As a result the stationary state attained by the process is an equilibrium stationary state, as expected for a gradient diffusion. 
\subsection{Transverse diffusion}\label{sec:sec242}
The second system that we consider is a linear diffusion in $\mathbb{R}^2$ having a antisymmetric component in the drift matrix. In particular, we consider the so-called \textit{transverse} process $\bs{X}(t) \in \mathbb{R}^2$ satisfying the SDE 
\begin{equation} \label{sdetrans}
    d\bs{X}(t) = - \begin{pmatrix} \gamma && \xi \\ -\xi && \gamma \end{pmatrix} \bs{X}(t) + \epsilon \, d\bs{W}(t), 
\end{equation}
with $\epsilon > 0$ and $\gamma > 0$, so that
\begin{equation}
    M = \begin{pmatrix} \gamma && \xi \\ -\xi && \gamma \end{pmatrix} \quad \tn{and} \quad D = \begin{pmatrix} \epsilon^2 && 0 \\ 0 && \epsilon^2 \end{pmatrix}.
\end{equation}
This process has a drift featuring both a symmetric (gradient) part associated with the friction parameter $\gamma$, as well as an antisymmetric part associated with the parameter $\xi$. The antisymmetric part of the drift is associated, for $\xi > 0$, with an anti-clockwise circular rotation of the system around the origin. For $\xi < 0$ the circular rotation is clockwise. \par
The solution of the Lyapunov equation (\ref{ricccov}) for this system yields the stationary covariant matrix 
\begin{equation}\label{transcov}
    C = \frac{\epsilon^2}{2\gamma},
\end{equation}
which is the same as that obtained in the case of the simple gradient diffusion considered in Sec.~\ref{sec:sec241}. As a result the process has a stationary density $p^*(\x)$ which is identical to that of (\ref{statdengrad}) and is given explicitly by \begin{equation}\label{rhotrans}
    p^*(\x) = \frac{\gamma}{\pi \epsilon^2} \exp \left(-\frac{\gamma}{\epsilon^2}\left(x_1^2 + x_2^2\right) \right).
\end{equation}
However, the process now has a non-zero current, as can be checked from (\ref{statcurlin}):
\begin{equation} \label{Jtrans}
    \bs{J}_{\bs{F},p^*}(\x) = \xi \begin{pmatrix} -x_2 \\ x_1 \end{pmatrix}p^*(\x).
\end{equation}
The stationary state for the transverse system is therefore a nonequilibrium steady state which violates detailed balance. 
\par
\begin{figure}[t]
    \centering
    \begin{minipage}{.45\textwidth}
    \includegraphics[width=6cm]{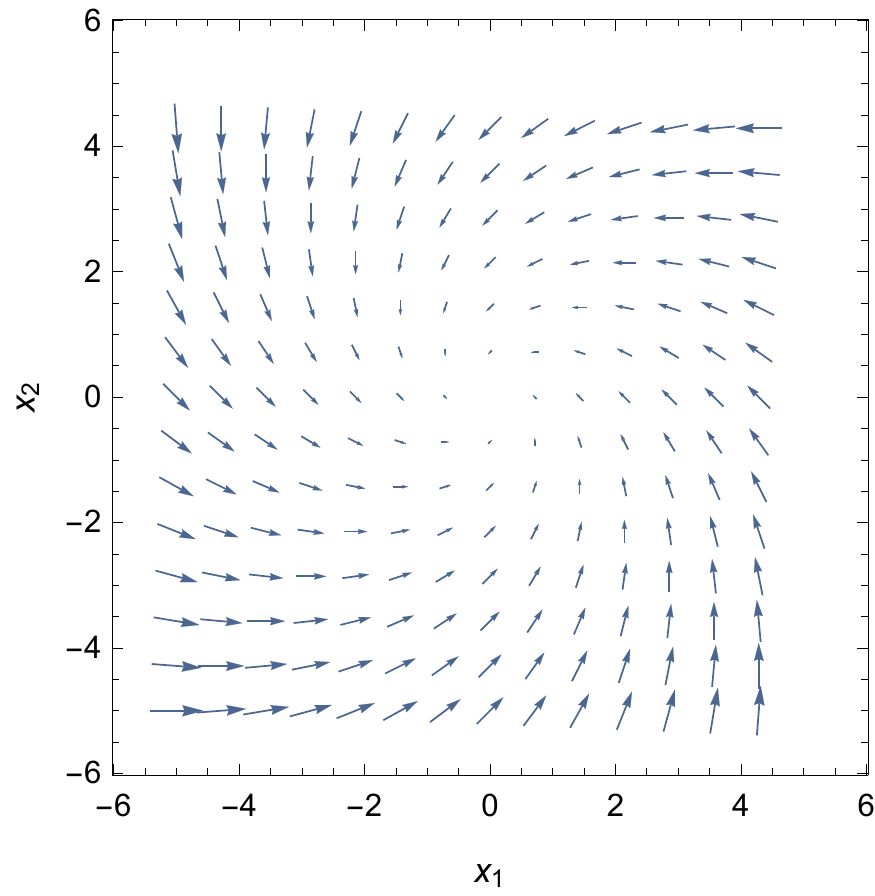}
    \end{minipage}
    \begin{minipage}{.45\textwidth}
    \includegraphics[width=6cm]{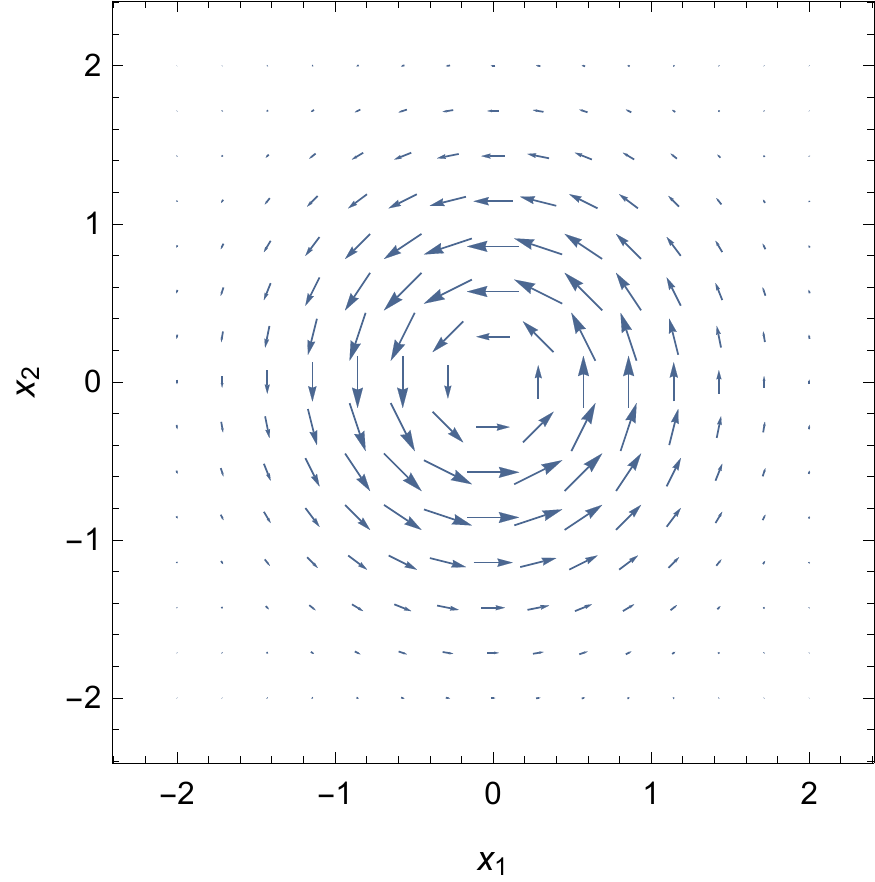}
    \end{minipage}
    \caption{Drift (left) and stationary current (right) for the transverse system for the parameter values $\gamma = 1,\xi=1$, and $\epsilon=1$.}
    \label{fig:Jtrans}
\end{figure}

 It is clear that for this process the nonequilibrium behavior is due to the presence of an antisymmetric component in the drift matrix, with $\xi$ the associated nonequilibrium parameter. In the event that $\xi = 0$, the force is gradient and the stationary current vanishes, as can be seen in the expression (\ref{Jtrans}). Furthermore, it can be seen from (\ref{Jtrans}) that the current will be anti-clockwise for $\xi > 0$ and clockwise for $\xi < 0$. The drift and stationary current for this process are shown in Fig.~\ref{fig:Jtrans} for a particular set of values for the parameters $\epsilon, \gamma$ and $\xi >0$. The anti-clockwise circular nature of the probability current for $\xi > 0$ can be clearly seen in Fig.~\ref{fig:Jtrans}. 
\subsection{Two temperature system with spring coupling}\label{sec:sec243}
\begin{figure}[t]
\centering 
\includegraphics[width = 6cm]{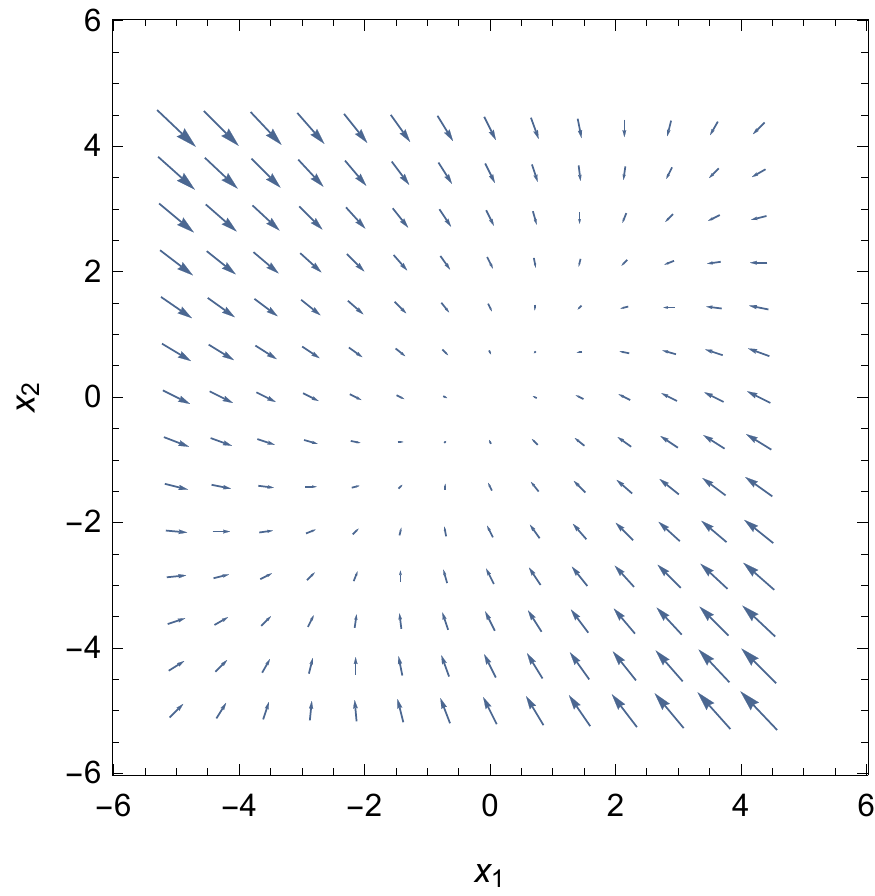}
\caption{Vector plot for the drift of the spring coupled system for the parameters $\gamma = 1, \kappa = 1$.}
\label{fig:Fspring}
\end{figure}

The third example that we consider consider is the linear diffusion in $\mathbb{R}^2$ given by the SDE 
\begin{equation} \label{sdespring}
    d\bs{X}(t) = - \begin{pmatrix}\gamma + \kappa && -\kappa \\ -\kappa && \gamma + \kappa \end{pmatrix}\bs{X}(t) + \begin{pmatrix}\epsilon_1 && 0 \\ 0 && \epsilon_2 \end{pmatrix} d\bs{W}(t). 
\end{equation}
The drift in this system represents a harmonic potential related to the friction parameter $\gamma$ applied to both $X_1(t)$ and $X_2(t)$, with a linear spring force with coupling $\kappa$ existing between $X_1(t)$ and $X_2(t)$. The presence of two separate noise strengths $\epsilon_1$ and $\epsilon_2$ indicates that $X_1(t)$ and $X_2(t)$ are coupled to two different heat baths having (in general) non-identical temperatures $T_{1,2} = \epsilon_{1,2}^2/2$. The drift for the process is shown in Fig.~\ref{fig:Fspring}. 
\par Systems of this type have been studied extensively~\cite{li2019quantifying, falasco2015energy, chun2015hidden, rieder1967properties, bonetto2004fourier} in statistical physics, being essentially identical to the spring bead system~\cite{li2019quantifying, liu2008effective}, and similar to the Feynman ratchet~\cite{feynman1965feynman}. A further example~\cite{ciliberto2013heat} of this model includes a system of two resistors kept at different temperatures and with a coupling capacitance allowing for the exchange of energy between the two resistors. 
\par The stationary density and current can be calculated exactly for this process, but are however too large to display here.
Instead we note only that the stationary covariance matrix $C$ is given from (\ref{ricccov}) by 
\begin{equation}
    C = \frac{1}{4\gamma(\gamma + 2\kappa)} \begin{pmatrix}\frac{2\gamma \epsilon_1^2 (\gamma + 2\kappa) + (\epsilon_1^2 + \epsilon_2^2)\kappa^2}{\gamma + \kappa} && (\epsilon_1^2 + \epsilon_2^2) \kappa \\ (\epsilon_1^2 + \epsilon_2^2) \kappa && \frac{2\gamma \epsilon_2^2 (\gamma + 2\kappa) + (\epsilon_1^2 + \epsilon_2^2)\kappa^2}{\gamma + \kappa} \end{pmatrix},
\end{equation}
from which the stationary density and current can easily be found via (\ref{statden}) and (\ref{statcurlin}), respectively. For $\kappa = 0$ it can be seen that the stationary covariance matrix becomes diagonal, representing the decoupling of the coordinates $X_1(t)$ and $X_2(t)$. \par
\begin{figure*}[t]
    \centering
            \begin{subfigure}[b]{.45\textwidth}
                \centering
                \includegraphics[width=6cm]{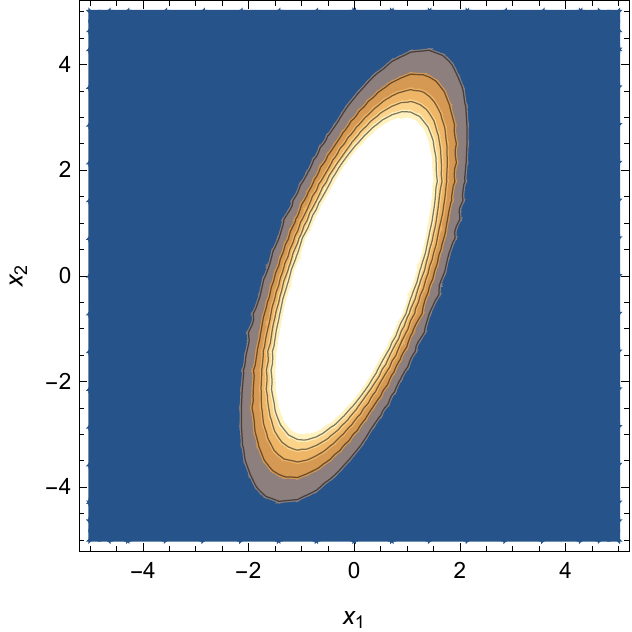}
                \caption{$\epsilon_1 = 1, \epsilon_2 = 3$}
                
            \end{subfigure}
            \begin{subfigure}[b]{.45\textwidth}
            \centering
                \includegraphics[width=6cm]{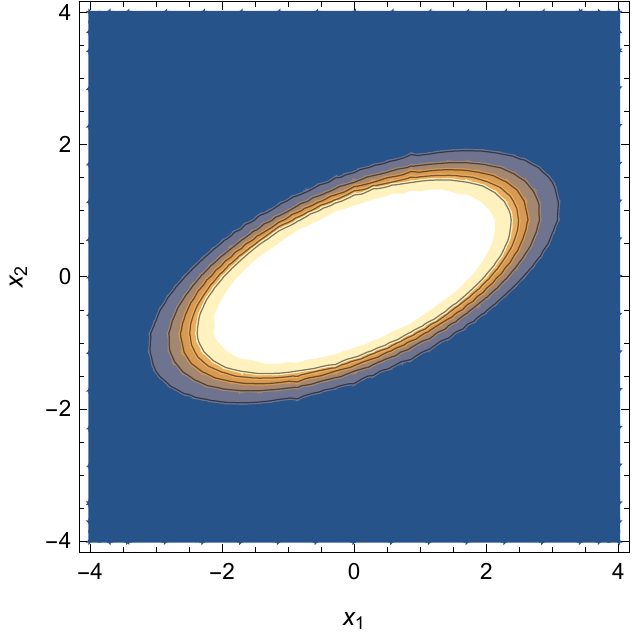}
                \caption{$\epsilon_1 = 2, \epsilon_2 = 1$}
            \end{subfigure}
        \caption{Contour plots of the stationary density for the two temperature spring system for different sets of noise strengths $\epsilon_1$ and $\epsilon_2$. The friction parameter and coupling parameters are $\gamma =1$ and $\kappa =1.$}
        \label{fig:rhospring}
\end{figure*}

 The contour lines of the stationary density, shown in Fig.~\ref{fig:rhospring}, are ellipses with the angle of the major axis with respect to the origin controlled by the ratio $\epsilon_1/\epsilon_2$, as is the eccentricity of the ellipse: the greater the disparity between the noise strengths, the more elongated the ellipse becomes. 
\par The stationary current has the form 
\begin{equation} \label{Jspring}
\bs{J}_{\bs{F},p^*}(\x) = H \x p^*(\x),
\end{equation}
where the matrix $H$ is given explicitly as 
\begin{align} \label{Hspring}
    H &=S \begin{pmatrix}\kappa (\gamma + \kappa)(\epsilon_1^2 +\epsilon^2)&& - 2\gamma \epsilon_1^2 (\gamma + 2\kappa) - (\epsilon_1^2 + \epsilon_2^2)\kappa^2 \\ 2\gamma \epsilon_1^2 (\gamma + 2\kappa) + (\epsilon_1^2 + \epsilon_2^2)\kappa^2 && -\kappa (\gamma + \kappa)(\epsilon_1^2 +\epsilon_2^2)\end{pmatrix},
\end{align}
with the constant $S$ given by 
\begin{equation} \label{Sspring}
    S = \frac{(\epsilon_1^2 - \epsilon_2^2)\kappa}{4\gamma \epsilon_1^2\epsilon_2^2(\gamma + 2\kappa) + (\epsilon_1^2 + \epsilon_2^2)\kappa^2}.
\end{equation}
\begin{figure*}[t]
    \centering
            \begin{subfigure}[b]{.45\textwidth}
                \centering
                \includegraphics[width=6cm]{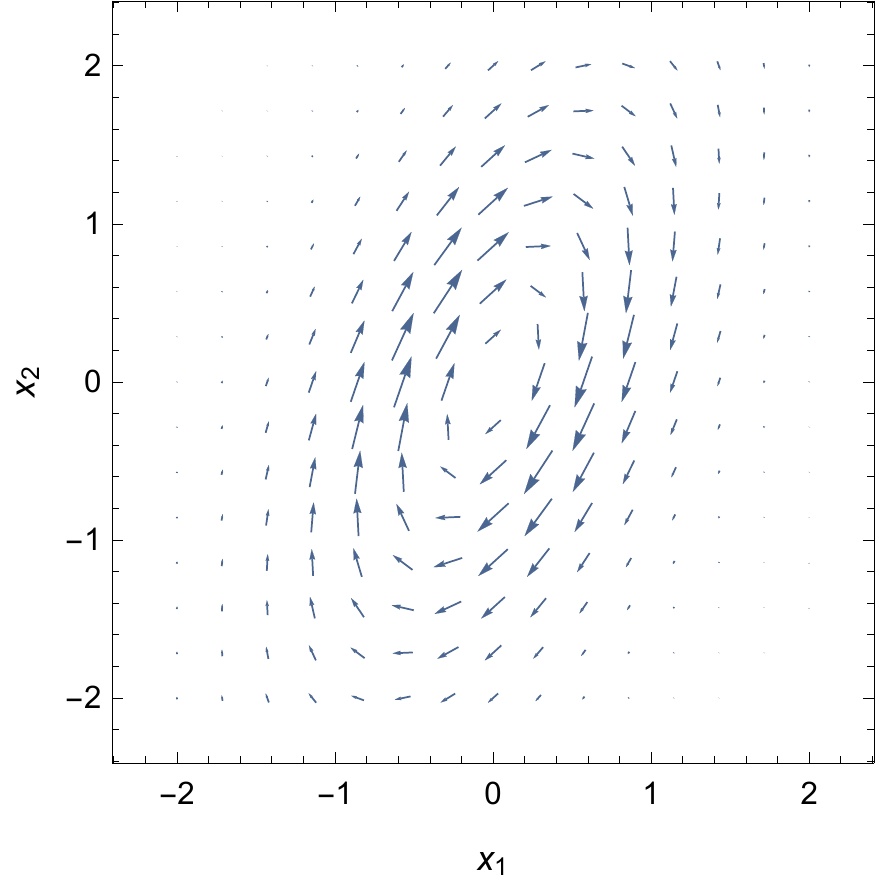}
                \caption{$\epsilon_1 = 1, \epsilon_2 = 2$}
                
            \end{subfigure}
            \begin{subfigure}[b]{.45\textwidth}
            \centering
                \includegraphics[width=6cm]{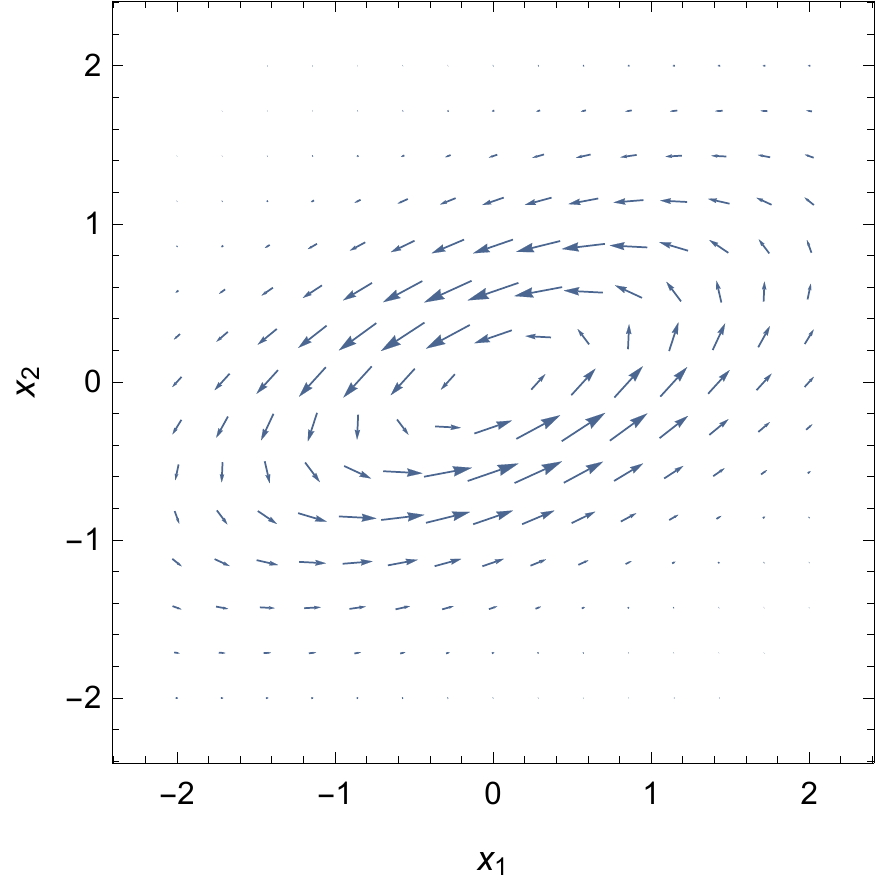}
                \caption{$\epsilon_1 = 2, \epsilon_2 = 1$}
            \end{subfigure}
        \caption{Vector plot of the stationary current for the spring system with different noise strengths for different values of $\epsilon_1$ and $\epsilon_2$. The friction parameter and coupling parameters are $\gamma =1$ and $ \kappa =1.$}
        \label{fig:Jspring}
\end{figure*} 
It can be seen from (\ref{Hspring}) and (\ref{Sspring}) that when $\kappa > 0$ and the noise strengths (temperatures) are different, that is $\epsilon_1 \neq \epsilon_2$ or $T_1 \neq T_2$, a stationary probability current exists and the process is nonequilibrium. If the noise strengths are identical then the system will have an equilibrium steady state for arbitrary $\kappa$. Likewise, for $\kappa = 0$ the system will be an equilibrium system even in the event that the noise strengths are non-identical; this is because for $\kappa =0$ the coordinates $X_1(t)$ and  $X_2(t)$ decouple and the system essentially reduces to two isolated systems in contact with separate heat baths. As such, it is the combination of the spring coupling controlled by $\kappa$ and the temperature difference related to the difference in noise strengths that is responsible for the nonequilibrium behavior. 
\par Finally, we observe a similar structure for the stationary current as for the stationary density: the flow is elliptical, with the angle of this ellipse and its eccentricity depending on the noise strengths in the same manner as for the density. The flow of the stationary probability current is clockwise for $\epsilon_1 < \epsilon_2$ and anti-clockwise for $\epsilon_1 > \epsilon_2$, as can be seen in Fig.~\ref{fig:Jspring}. 
\section{Large deviation principle}\label{sec:sec25}
We are interested in obtaining information regarding the probability density $P(A_T = a)$ associated with dynamical observables having the form (\ref{obs}). While explicitly calculating this density is usually difficult, it is known from the theory of large deviations~\cite{touchette2009large,touchette2018introduction,den2008large,dembo1998} that for this type of observable the density often satisfies the asymptotic expression 
\begin{equation} \label{asymp}
    P(A_T = a) = e^{-T I(a) + o(T)}
\end{equation}
as $T$ becomes large. The meaning of this asymptotic form is that the dominant contribution to the density $P(A_T = a)$ as $T$ becomes large is a decaying exponential in $T$, with the rate of this exponential decay controlled by the so-called \textit{rate function} $I(a)$. Corrections to the dominant exponential term are sub-exponential in the time $T$. If the density of $A_T$ has the asymptotic form (\ref{asymp}) we say that $A_T$ satisfies a \textit{large deviation principle} (LDP) with rate function $I$. Equivalently, $A_T$ satisfies an LDP if the limit 
\begin{equation}
    \lim_{T \rightarrow \infty} -\frac{1}{T} \ln P(A_T = a) = I(a)
\end{equation}
exists and is non-trivial, by which we mean that $I(a)$ is not everywhere equal to $0$ or $\infty$. 
\par In order to understand in more detail the asymptotic form (\ref{asymp}) we note that the rate function satisfies $I(a) \geq 0$. For values of $a$ for which $I(a) > 0$ we will have that $P(A_T = a)$ decays exponentially as $T$ grows, while values $\tilde{a}$ for which $I(\tilde{a}) = 0$ will not decay exponentially. Since the density $P(A_T = a)$ is normalized we will therefore find that $P(A_T = a)$ becomes increasingly concentrated at those points $\tilde{a}$ for which $I(\tilde{a}) = 0$ as $T$ grows. In this sense, zeroes of the rate function represent typical values of the observable $A_T$ in the long-time limit. 
\par In particular, if $I(a)$ has a unique zero $\tilde{a}$, then
\begin{equation}
    \lim_{T\rightarrow \infty} P(|A_T - \tilde{a}|< \epsilon) = 1 
\end{equation}
for all $\epsilon > 0$, so that $P(A_T = a)$ essentially converges to the Dirac delta function $\delta(a - \tilde{a})$. In the event that the process is ergodic, it is clear by comparison with (\ref{ergodicdensity}) and (\ref{ergodicexp}) that the typical value $\tilde{a}$ corresponds to the stationary expectation of the observable $A_T$, defined earlier as $a^*$ in (\ref{ergodicexp}). Given that we will consider only ergodic processes we will from now on simply use $a^*$ to indicate the typical value of the observable $A_T$. 
\par It should also be noted that the theory of large deviations serves as a generalization of the Gaussian theory of fluctuations with Gaussian fluctuations corresponding to the case where the rate function $I(a)$ is a parabola. Since this will not generally be the case it is therefore clear that the rate function can also characterize non-Gaussian fluctuations away from $a^*$. In this sense, the name `large deviation theory' derives from the fact that the rate function provides information regarding the probabilities of both small fluctuations close to the typical value $a^*$ and large fluctuations that deviate significantly from $a^*$.
\section{G\"{a}rtner-Ellis theorem}\label{sec:sec26}
Obtaining the rate function via a direct calculation of $P(A_T =a )$ is difficult even for simple systems and observables. As such we need an indirect method by which to find the rate function. To this end we introduce the scaled cumulant generating function (SCGF) associated with an observable $A_T$ and defined as 
\begin{equation} \label{SCGF}
    \lambda(k) = \lim_{T\rightarrow \infty} \frac{1}{T} \ln\mathbb{E}\left[e^{kTA_T} \right],
\end{equation}
where $\mathbb{E}\left[e^{kTA_T} \right]$ is the expectation of $e^{kTA_T}$. The importance of the SCGF in large deviation theory is due~\cite{ellis1984large,gartner1977large,touchette2009large} to the G\"{a}rtner-Ellis theorem, which states that if $\lambda(k)$ exists and is differentiable in $k$, then $A_T$ satisfies an LDP with rate function $I$ given by the Legendre-Fenchel transform of $\lambda(k)$:
\begin{equation}
    I(a) = \sup_{k \in \mathbb{R}} \{ka - \lambda(k)\}. 
\end{equation}
When $\lambda(k)$ is strictly convex in addition to being differentiable, the Legendre-Fenchel transform reduces to the well-known Legendre transform 
\begin{equation}
    I(a) = k(a)a - \lambda(k(a)),
\end{equation}
where $k(a)$ satisfies $\lambda'(k(a)) = a$. This will be the case for all systems considered in this dissertation. 
\par Of central importance in this dissertation is the \textit{moment generating function} associated with an observable $A_T$ and given by 
\begin{equation}
    G_k(\x,t) = \mathbb{E}_{\x}\left[e^{ktA_t} \right],
\end{equation}
where the subscript $\x$ in $\mathbb{E}_{\x}[\cdot]$ indicates that we are taking the expectation with respect to the process $\bs{X}(t)$ having the initial condition $\bs{X}(0) = \x$. We will often refer to $G_k(\x,t)$ simply as the generating function. \par Remarkably, the generating function has a semi-group structure and evolves in time according to the linear partial differential equation
\begin{equation} \label{FK}
    \partial_t G_k(\x,t) = \mathcal{L}_k G_k(\x,t),
\end{equation}
with the \textit{tilted generator} $\mathcal{L}_k$ given by 
\begin{equation} \label{tilted}
    \mathcal{L}_k = \bs{F}\cdot \left(\bs{\nabla} + k \bs{g}\right) + \frac{1}{2} \left(\bs{\nabla} + k \bs{g}\right)\cdot D \left(\bs{\nabla} +k \bs{g}\right) + kf,
\end{equation}
where the functions $f$ and $\g$ are those entering in the definition of the observable (\ref{obs}). The expression (\ref{FK}) is known as the Feynman-Kac formula~\cite{touchette2018introduction}. Exploiting the semi-group structure of $G_k(\x,t)$, and noting that $G_k(\x,0) = 1$, we can write the solution of (\ref{FK}) formally as
\begin{equation}
    G_k(\x,t) = \left(e^{t\mathcal{L}_k}1 \right)(\x),
\end{equation}
where $1$ denotes the constant function $h(\x) = 1$. 
\par Since the Feynman-Kac equation is linear, it is possible to expand $G_k(\x,t)$ in terms of the eigenfunctions $r_k^{(i)}$ and eigenvalues $\lambda_i(k)$ of $\mathcal{L}_k$ in the manner
\begin{equation} \label{specdec}
    G_k(\x,t) = \sum_{i} a_i e^{\lambda_i(k) t} r_k^{(i)}(\x).
\end{equation}
In the long-time limit the eigenvalue with largest real part will dominate the sum. As a result, it is clear by comparison with the definition (\ref{SCGF}) of the SCGF that $\lambda(k)$ is the dominant eigenvalue of the tilted generator. To make this clear, let $\lambda_0(k)$ be that eigenvalue of $\mathcal{L}_k$ with largest real part. Then 
\begin{equation} \label{Gklim}
    G_k(\x,t) \rightarrow  K e^{\lambda_0(k)t} r_k(\x)
\end{equation}
as $t$ becomes large, where $K$ is a constant and with $r_k(\x)$ the eigenfunction of $\mathcal{L}_k$ corresponding to $\lambda_0(k)$. We then have that 
\begin{equation}
    \lim_{t \rightarrow \infty} \frac{1}{t} \ln G_k(\x,t) = \lambda_0(k),
\end{equation}
which shows that the SCGF is in fact the eigenvalue of the tilted generator with largest real part. As a result, we will omit from now on the subscript $0$ to simply write the dominant eigenvalue as $\lambda(k)$. 
\par The SCGF can be calculated either by calculating the generating function $G_k(\x,t)$ explicitly and studying its behavior in the long-time limit or by solving the spectral problem for the dominant eigenvalue 
\begin{equation} \label{spec1}
    \mathcal{L}_k r_k(\x) = \lambda(k) r_k(\x).
\end{equation}
Given that the tilted generator $\mathcal{L}_k$ will not generally be Hermitian, this spectral equation has to be considered in conjunction with the adjoint equation 
\begin{equation} \label{spec2}
    \mathcal{L}_k^{\dagger} l_k(\x) = \lambda(k) l_k(\x),
\end{equation}
where $\mathcal{L}_k^{\dagger}$ is the adjoint of $\mathcal{L}_k$ with respect to the inner product (\ref{innerproduct}) and with $l_k$ the eigenfunction of $\mathcal{L}_k^{\dagger}$ associated with the eigenvalue $\lambda(k)$. The eigenfunctions $r_k$ and $l_k$ must satisfy the normalization conditions 
\begin{equation} \label{spec3}
    \int_{\om} r_k(\x) l_k(\x) d\x = 1
\end{equation}
and 
\begin{equation} \label{spec4}
    \int_{\om} l_k(\x) d\x = 1. 
\end{equation}
The problem of obtaining the rate function has therefore been reduced to the problem of solving a particular spectral problem or calculating explicitly the generating function associated with the observable of interest.
\par For future reference, we note that the generating function $G_k(\x,t)$ can also be written as an integral over all possible final values $\bs{X}(t) = \bs{y}$ in the manner 
\begin{equation} \label{Gkendint}
    G_k(\x,t) = \int d\bs{y} \, \mathbb{E}_{\x}\left[\delta(\bs{X}(t) - \bs{y})e^{ktA_t} \right],
\end{equation}
with the expectation value in the above integral now containing the delta function $\delta(\bs{X}(t) - \bs{y})$ so that for a particular value of $\bs{y}$ only those trajectories which satisfy $\bs{X}(t) = \bs{y}$ contributes. The integral taken over all final points then ensures that the contribution from each possible $\bs{y}$ is counted. From this expression we can define the \textit{end-point function} as
\begin{equation} \label{endpointfunc}
    G_k(\x,\bs{y},t)=\mathbb{E}_{\x}\left[\delta(\bs{X}(t) - \bs{y})e^{ktA_t} \right],
\end{equation}
so that 
\begin{equation}
    G_k(\x,t) = \int d\bs{y} \, G_k(\x,\bs{y},t).
\end{equation}
It is known~\cite{chetrite2015variational} that the end-point function also has a spectral decomposition of the form 
\begin{equation}
    G_k(\x,\bs{y},t) = \sum_{i}  a_i e^{\lambda_i(k)t} l_k^{(i)}(\bs{y}) r_k^{(i)}(\x),
\end{equation}
where $l_k^{(i)}$ are the eigenfunctions of the adjoint operator $\mathcal{L}_k$. As a result, in the long-time limit, we can write
\begin{equation} \label{fixedpointasymp}
    G_k(\x,\bs{y},t) \rightarrow  K \, l_k(\bs{y})r_k(\x) e^{\lambda(k)t},
\end{equation}
similarly to (\ref{Gklim}). 
\section{Effective process}\label{sec:sec27}
We discussed in the previous section an indirect method for obtaining the rate function, which describes the probabilities associated with fluctuations of $A_T$ occurring as $T$ becomes large. In dynamical large deviation theory we are interested not only in the likelihood of a fluctuation occurring, but also in the manner in which this fluctuation is created dynamically in time. We discuss here a construction known as the \textit{effective process}, introduced recently by Chetrite and Touchette~\cite{chetrite2015nonequilibrium,chetrite2015variational,chetrite2013nonequilibrium}, which provides this information. 
\par Consider a Markov diffusion $\bs{X}(t)$ satisfying an SDE (\ref{SDE})
and an associated dynamical observable $A_T$ (\ref{obs}) known to satisfy an LDP with rate function $I$. In order to understand the manner in which a particular fluctuation $A_T = a$ is produced we can consider the process $\bs{X}(t)$ conditioned on obtaining this fluctuation. The possible trajectories of this conditioned process correspond to exactly those trajectories of the original process for which $A_T = a$ holds. The effective process associated with the fluctuation $A_T = a$ is a conditioning-free process which corresponds asymptotically (as $T \rightarrow \infty$) to the original process conditioned on this fluctuation. As a consequence, the effective process has the value $a$ as a typical value, and therefore describes the manner in which fluctuations are manifested dynamically for large $T$. 
\par It was shown~\cite{chetrite2015nonequilibrium} that for a process evolving according to an SDE of the form (\ref{SDE}), the effective process $\bs{X}_k(t)$ associated with a particular fluctuation $A_T = a(k) = \lambda'(k)$ satisfies the SDE 
\begin{equation} \label{effproc}
    d\bs{X}_k(t) = \bs{F}_k(\bs{X}_k(t)) \, dt + \sigma d\bs{W}(t),
\end{equation}
which has the same noise matrix $\sigma$ as that of the original process, but with the \textit{effective drift} $\bs{F}_k$ given by 
\begin{equation} \label{effdrift}
    \bs{F}_k = \bs{F} + D(k\bs{g} + \bs{\nabla} \ln r_k), 
\end{equation}
with $r_k$ the eigenfunction of the tilted generator corresponding to the dominant eigenvalue and SCGF $\lambda(k)$. It is important to note that the stationary density $p^*_{k}$ associated with the effective process $\bs{X}_k(t)$ satisfies the relation
\begin{equation} \label{pklkrk}
    p^*_k(\x) = r_k(\x)l_k(\x),
\end{equation}
where $r_k$ and $l_k$ are the eigenfunctions of $\mathcal{L}_k$ and $\mathcal{L}_k^{\dagger}$ associated with the dominant eigenvalue (SCGF) $\lambda(k)$. Finally, we introduce the stationary current associated with the effective process. From (\ref{statcur}) it follows that this stationary current is given by 
\begin{equation} \label{effstatcur1}
    \bs{J}_{\bs{F}_k, p_k^*}(\x) = \bs{F}_k(\x) p_k^*(\x) - \frac{1}{2}D \bs{\nabla}p_k^*(\x),
\end{equation}
which can also be written in terms of the eigenfunctions $r_k$ and $l_k$ as 
\begin{equation} \label{effstatcur2}
    \bs{J}_{\bs{F}_k,l_k r_k}(\x) = \bs{F}_k(\x) \left(l_k(\x) r_k(\x) \right) - \frac{1}{2}D \bs{\nabla}\left(l_k(\x) r_k(\x) \right).
\end{equation}
This stationary current will be an object of central importance later in this dissertation. 
\par To summarize, the effective process corresponds asymptotically to the original process conditioned on obtaining a particular value of a fluctuation $A_T = a$ in the event that the observable under consideration satisfies an LDP. The spectral elements $\lambda(k)$ and $r_k$ are also seen to play distinct but complementary roles in the theory of dynamical large deviations: $\lambda(k)$ gives the rate function (via Legendre-Fenchel transform), which specifies the probabilities of fluctuations occurring, while $r_k$ determines the effective drift which describes the manner in which fluctuations are produced. 

\chapter[Large deviations of reflected diffusions]{Dynamical large deviations of reflected diffusions} \label{chap:chap3}
In Chap.~\ref{chap:chap2} we presented the theory of Markov diffusions evolving in some state space $\om$, with no mention of the way in which the boundaries of the state space are to be incorporated into the framework. We now clarify how the formalism is used to account for the presence of boundaries restricting or influencing the evolution of a Markov diffusion. We are particularly interested in the case where the system has reflecting boundaries and in the manner that these boundaries are incorporated into the large deviation spectral problem associated with obtaining the SCGF for a given dynamical observable. We have previously derived these boundary conditions as part of an MSc thesis~\cite{dubuissonmasters}, published in~\cite{Buisson2020}, for the case where the observable $A_T$ is purely additive, meaning that $f \neq 0$ and $\bs{g} = \bs{0}$ in (\ref{obs}). We summarize these results in this chapter and then present new results obtained for more general observables incorporating an additive term $f$ and a current term $\bs{g}$. These results were obtained in collaboration with Emil Mallmin and were published recently in~\cite{Mallmin2021}. Finally, we illustrate our results by discussing their application to a specific class of current-type observables for the heterogeneous single-file diffusion. This application was developed by Mallmin and is published in our joint paper~\cite{Mallmin2021}.

\section{Bounded Markov diffusions and reflecting boundaries}\label{sec:sec31}
We consider a diffusion $\bs{X}(t)$ evolving in some region $\om \subset \mathbb{R}^n$ with a smooth boundary $\po$. In the interior of $\om$ the diffusion $\bs{X}(t)$ evolves according to the SDE introduced in Chap.~\ref{chap:chap2} in (\ref{SDE}), which involves the drift $\bs{F}$ and the noise matrix $\sigma$. In order to complete the description of the system we must provide a prescription for the behavior of the system when it reaches the boundary, i.e., when $\bs{X}(t) \in \po$. This constitutes additional information that is not contained in the SDE. 
\par We know that the Fokker-Planck operator $\mathcal{L}^{\dagger}$ and the infinitesimal generator $\mathcal{L}$ act as differential operators according to (\ref{FPO}) and (\ref{IG}), respectively. However, the domains of these operators must be restricted to account for the boundary behavior present in the system. To understand this, we start from the duality relation between the Markov operators $\mathcal{L}$ and $\mathcal{L}^{\dagger}$ and discuss the boundary terms which arise in the integration by part procedure that relates these operators. We recall from Sec.~\ref{sec:sec23} that $\mathcal{L}$ and $\mathcal{L}^{\dagger}$ must be such that  
\begin{equation}
    \langle \mathcal{L}^{\dagger}p,h\rangle = \langle p, \mathcal{L}h\rangle
\end{equation}
for all densities $p$ in the domain of $\mathcal{L}^{\dagger}$ and all test functions $h$ in the domain of $\mathcal{L}$. The inner product $\langle p,h\rangle$ is defined as an integral in $\om$ as in (\ref{innerproduct}). 
\par As mentioned before, transforming the action of $\mathcal{L}$ on a test function $h$ to the action of $\mathcal{L}^{\dagger}$ on a density $p$ involves using integration by parts, which produces surface integrals over the boundary $\po$ in the process. This procedure is shown in App.~\ref{appendixA} and has the final result 
\begin{equation} \label{surface}
    \langle p, \mathcal{L}h\rangle = \langle \mathcal{L}^{\dagger}p, h\rangle - \int_{\po} h(\x) \bs{J}_{\bs{F}, p}(\x) \cdot \hat{\bs{n}}(\x)d\x  - \frac{1}{2} \int_{\po}  p(\x) D \bs{\nabla} h(\x) \cdot \hat{\bs{n}}(\x)d\x,
\end{equation}
where $\hat{\bs{n}}(\x)$ is the inward normal vector at $\x \in \po$ and $\bs{J}_{\bs{F},p}$ is the probability current as defined in (\ref{current}). 
\par In order for the Markov operators to be well-defined independently of a particular $p$ or $h$ we must have that the surface integral terms in the above expression vanish. For this to happen we must impose conditions on the densities and test functions, which amounts to a restriction of the domains of the Markov operators. Given a particular prescription for the behavior of the process on the boundary $\po$ there exists a corresponding restriction or \textit{boundary condition} on the domains of $\mathcal{L}$ and $\mathcal{L}^{\dagger}$. In this manner the boundary behavior of the process is encoded at the level of the Markov operators. 
\par For what follows in this chapter we will be interested in reflecting boundaries which are implemented at the level of the Fokker-Planck equation by requiring that, at the boundary, the probability current in the direction normal to the boundary is zero. In other words we require that the Fokker-Planck operator acts only on densities $p$ satisfying the condition 
\begin{equation} \label{bound1}
    \bs{J}_{\bs{F},p}\cdot \hat{\bs{n}}(\x)  = 0 \quad \forall \x \in \po. 
\end{equation}
Referring to the duality relation (\ref{surface}) and demanding that the surface terms vanish, we see that the above condition on the densities $p$ imply that test functions $h$ in the domain of the infinitesimal generator must satisfy 
\begin{equation}
    D\bs{\nabla}h(\x)\cdot\hat{\bs{n}}(\x) = 0 \quad \forall \x \in \po.
\end{equation}
Given that $D$ is a symmetric matrix, the above can be rewritten as
\begin{equation} \label{bound2}
    \bs{\nabla}h(\x)\cdot D\hat{\bs{n}}(\x) =0 \quad \forall \x \in \po
\end{equation}
so that, in contrast to the probability current which vanishes in the direction normal to the boundary, $\bs{\nabla}h$ vanishes in the direction $D\hat{\bs{n}}(\x)$, which is known as the co-normal direction~\cite{schuss2013}. \par Other types of boundary behavior, including absorption or partial reflection, can be treated similarly. For more information regarding other boundary types and the associated boundary conditions we refer the reader to~\cite{feller1954diffusion,schuss2013, karlin1981second, bou2020sticky,peskir2015boundary}. A comprehensive analysis of Brownian dynamics at boundaries is found in Schuss~\cite{schuss2013}.
\par Note that when the domain $\om$ in which the diffusion $\bs{X}(t)$ evolves is simply the entirety of $\mathbb{R}^n$ with no boundaries, so that $\om = \mathbb{R}^n$, the normalization condition 
\begin{equation} \label{normalization}
    \int_{\mathbb{R}^n}p(\x)d\x = 1
\end{equation}
on the probability densities $p$ and the condition 
\begin{equation} \label{finiteinner}
    \langle p,h\rangle < \infty
\end{equation}
are sufficient to ensure the vanishing of the boundary terms in (\ref{surface}). The normalization condition (\ref{normalization}) implies that 
\begin{equation} \label{decay1}
    \lim_{\norm{\x} \rightarrow \infty} p(\x) = 0.
\end{equation}
This, along with the fact that $p$ must be a positive function in order to be regarded as a probability density, then also implies that
\begin{equation}\label{decay2}
    \lim_{\norm{\x}\rightarrow \infty} \norm{\bs{J}_{\bs{F},p}(\x)} = 0
\end{equation}
for the probability current. Equations (\ref{decay1}) and (\ref{decay2}), along with (\ref{finiteinner}), ensure that the surface integrals in (\ref{surface}) vanish, since the `surface' $\po$ now lies at infinity. Thus for $\om = \mathbb{R}^n$ the domains of the Markov operators are defined simply by the requirements that densities and test functions are twice differentiable (given that $\mathcal{L}$ and its adjoint are second-order differential operators) and satisfy the relations (\ref{normalization}) and (\ref{finiteinner}). 
\section{Duality relation for large deviation operators}\label{sec:sec32}
In the previous section we discussed the surface terms arising via the duality relation between the Markov operators $\mathcal{L}$ and $\mathcal{L}^{\dagger}$ and the manner in which reflecting boundaries are incorporated at the level of these operators. We now turn to the issue of obtaining the proper boundary conditions on the large deviation operators $\mathcal{L}_k$ and $\mathcal{L}_k^{\dagger}$ associated with a dynamical observable $A_T$ of a reflected diffusion $\bs{X}(t)$. Understanding the boundary conditions for these operators is essential in properly defining the spectral problem described in (\ref{spec1}) and (\ref{spec2}) associated with the SCGF and thereby finding the rate function of $A_T$. We first present a summary of previous results~\cite{dubuissonmasters,Buisson2020} obtained for the case where $A_T$ is an additive observable and then describe the problems encountered in attempting to naively generalize these results to the case of current-type observables, indicating that a different argument is needed to account for these observables.
\subsection{Summary of previous results for additive observables}\label{sec:sec321}
We consider an additive observable of the form defined 
previously in (\ref{additive}), so  that $\bs{g} = \bs{0}$ for now. As described in the previous chapter, the SCGF $\lambda(k)$ of this observable is obtained by solving the spectral problem defined before in (\ref{spec1}) and (\ref{spec2}). This spectral problem involves the tilted generator $\mathcal{L}_k$ defined in (\ref{tilted}) and its adjoint $\mathcal{L}_k^{\dagger}$. For the case $\bs{g} = \bs{0}$ we have explicitly
\begin{equation}
    \mathcal{L}_k = \bs{F}\cdot \bs{\nabla} + \frac{1}{2} \bs{\nabla} \cdot D \bs{\nabla} + kf
\end{equation}
or equivalently 
\begin{equation}
    \mathcal{L}_k = \mathcal{L} + kf,
\end{equation}
where $\mathcal{L}$ is the infinitesimal generator (\ref{IG}) for the diffusion $\bs{X}(t)$. Using the definition of the adjoint as 
\begin{equation} 
    \langle \mathcal{L}_k^{\dagger}l,r\rangle = \langle l,\mathcal{L}_k r\rangle,
\end{equation}
where $r$ and $l$ lie in in the domains of $\mathcal{L}_k$ and its adjoint, respectively, it is clear that since the $kf$ term contains no derivatives (and is Hermitian under the inner product (\ref{innerproduct})) we have  
\begin{equation}
    \mathcal{L}_k^{\dagger} =\mathcal{L}^{\dagger} +kf,
\end{equation}
with $\mathcal{L}^{\dagger}$ the relevant Fokker-Planck operator, shown in (\ref{FPO}). 
\par From the normalization conditions (\ref{spec3}) and (\ref{spec4}) for the eigenfunctions of the large deviation operators $\mathcal{L}_k$ and $\mathcal{L}_k^{\dagger}$ we require 
\begin{equation}
    \int_{\om} l(\x) d\x = 1
\end{equation}
for all $l \in \mathcal{D}(\mathcal{L}_k^{\dagger})$ and 
\begin{equation}
    \int_{\om} l(\x)r(\x)d\x < \infty
\end{equation}
for $r \in \mathcal{D}(\mathcal{L}_k)$ and $l \in \mathcal{D}(\mathcal{L}_k^{\dagger})$. It should be clear that these conditions on the domains of the tilted generator and its adjoint are similar to those placed on the original Markov operators $\mathcal{L}$ and $\mathcal{L}^{\dagger}$, respectively. In fact, the calculation in the previous section leading to the duality relation (\ref{surface})
for the Markov operators can be repeated virtually exactly for the case of the tilted generators associated with the additive observable (\ref{additive}), leading to the relation 
\begin{equation} \label{dualitytilted}
    \langle l, \mathcal{L}_k r\rangle = \langle \mathcal{L}_k^{\dagger}l, r\rangle - \int_{\po} r(\x) \bs{J}_{\bs{F}, l}(\x) \cdot \hat{\bs{n}}(\x)d\x  - \frac{1}{2} \int_{\po}  l(\x) D \bs{\nabla} r(\x) \cdot \hat{\bs{n}}(\x)d\x.
\end{equation}
This follows because the tilted generator differs from the infinitesimal generator only by the Hermitian term $kf$, which contains no derivatives and therefore introduces no new boundary terms. 
\par From this result, we then obtain the boundary conditions on the operators $\mathcal{L}_k$ and $\mathcal{L}_k^{\dagger}$ in the presence of a reflecting boundary $\po$ by noting that the large deviation operators correspond to the Markov operators when $k = 0$~\cite{dubuissonmasters,Buisson2020}. Given that the boundary terms for this observable are independent of $k$ the condition that $\mathcal{L}_{k=0} = \mathcal{L}$ and $\mathcal{L}_{k=0}^{\dagger}=\mathcal{L}^{\dagger}$ is then sufficient to determine the boundary conditions on the large deviation operators for all $k$. \par Consequently, the operator $\mathcal{L}_k$ inherits the boundary conditions on $\mathcal{L}$ for all $k$, and similarly for $\mathcal{L}_k^{\dagger}$ and $\mathcal{L}^{\dagger}$. Explicitly then we have that in the presence of a reflecting boundary $\po$ the tilted generator $\mathcal{L}_k$ acts on functions $r$ satisfying a boundary condition of the form (\ref{bound2}), so that 
\begin{equation} \label{bound3}
    \bs{\nabla}r(\x)\cdot D\hat{\bs{n}}(\x) = 0 \quad \forall \x \in \po,
\end{equation}
while the adjoint operator $\mathcal{L}_k^{\dagger}$ acts on normalizeable densities $l$ satisfying a zero current boundary condition as in (\ref{bound1}) and given by
\begin{equation} \label{bound4}
    \bs{J}_{\bs{F},l}(\x)\cdot \hat{\bs{n}}(\x) = 0 \quad \forall \x\in\po. 
\end{equation}
These are the conditions that were derived in~\cite{dubuissonmasters} and later published in~\cite{Buisson2020}. 
\par As noted there, the boundary conditions (\ref{bound3}) and (\ref{bound4}) imply that the effective process, introduced before in Sec.~\ref{sec:sec27}, has a stationary current (\ref{effstatcur2}) that satisfies 
\begin{equation} \label{currenteffectivebound}
    \bs{J}_{\bs{F}_k,l_k r_k}(\x)\cdot\hat{\bs{n}}(\x) = 0 \quad \forall \x \in \po.
\end{equation}
In addition, the boundary condition (\ref{bound3}) implies that the effective drift $\bs{F}_k$ given in (\ref{effdrift}) satisfies on the boundary $\po$ the relation 
\begin{equation} \label{effdriftbound}
    \bs{F}_k(\x) \cdot \bs{\hat{n}}(\x) = \bs{F}(\x) \cdot \bs{\hat{n}}(\x) \quad \forall \x \in \po.
\end{equation}
The first result (\ref{currenteffectivebound}) shows that the effective process is also a reflected process. This makes sense intuitively, since the effective process corresponds asymptotically to the original process conditioned on manifesting a particular fluctuation in the long-time limit. The set of possible trajectories of the effective process therefore consists of a subset of the trajectories of the original (reflected) process and is therefore also a reflected process. 
\par The second result (\ref{effdriftbound}) shows that the component of the effective drift normal to a reflected boundary $\po$ is identical to the component of the original drift $\bs{F}$ normal to that boundary. In other words, the component of the drift normal to the boundary is not altered to manifest fluctuations.
\subsection{Failure of the argument for current-type observables}\label{sec:sec322}
We now turn our attention to the case of current-type observables defined in (\ref{currentobs}). We want to show here that attempts to obtain the appropriate boundary conditions on the large deviation operators solely on the grounds of arguments pertaining to the duality relation and the boundary terms arising therein fail for such observables. 
\par We start by noting that the tilted generator $\mathcal{L}_k$ associated with the observable $A_T$ is given from (\ref{tilted}) by 
\begin{equation} \label{tiltedcurrent}
    \mathcal{L}_k = \bs{F}\cdot \left( \bs{\nabla} + k\g\right) + \frac{1}{2}\left( \bs{\nabla} + k\g\right)\cdot D \left( \bs{\nabla} + k\g\right)
\end{equation}
and, as discussed previously, the adjoint operator $\mathcal{L}_k^{\dagger}$ acts on normalized densities $l$ while $\mathcal{L}_k$ acts on functions $r$ such that $\langle l,r \rangle < \infty$.
\par We are interested in obtaining the duality relation (including boundary terms) relating the operators $\mathcal{L}_k$ and its adjoint. This means carrying out an integration by parts procedure similar to that done for the Markov operators and the large deviation operators associated with an additive observable. This procedure is shown in App.~\ref{appendixB} and leads to
\begin{align} \label{tiltedcurrentduality}
    \langle l, \mathcal{L}_k r \rangle &= \langle \mathcal{L}_k^{\dagger} l,  r \rangle - \int_{\po}  \bigg\{l(\x) r(\x)\bigg(\bs{F}(\x) + k D \bs{g}(\x) \bigg) + \frac{1}{2} l(\x) D \bs{\nabla} r(\x) \nonumber \\ &\quad - \frac{1}{2} r(\x) D \bs{\nabla} l(\x) \bigg\}\cdot \hat{\bs{n}}(\x)d\x,
\end{align}
with the adjoint operator $\mathcal{L}_k^{\dagger}$ given as a differential operator by 
\begin{equation}
    \mathcal{L}_k^{\dagger} =(-\bs{\nabla} + k \bs{g})\cdot \bs{F} + \frac{1}{2}\left(-\bs{\nabla}+k\g\right)\cdot D\left(-\bs{\nabla}+k\g\right).
\end{equation} 
The boundary terms in (\ref{tiltedcurrentduality}) can then be written as
\begin{align}
    &\int_{\po}  \bigg\{l(\x) r(\x)\bigg(\bs{F}(\x) + k D \bs{g}(\x) \bigg) + \frac{1}{2} l(\x) D \bs{\nabla} r(\x)  - \frac{1}{2} r(\x) D \bs{\nabla} l(\x) \bigg\}\cdot \hat{\bs{n}}(\x)d\x \nonumber \\ 
    &= \int_{\po}r(\x) \bs{J}_{\bs{F},l}(\x)\cdot \hat{\bs{n}}(\x)d\x + \frac{1}{2}\int_{\po}l(\x) D\bs{\nabla}r(\x)\cdot\hat{\bs{n}}(\x) d\x  \nonumber \\ &\quad\quad +  k\int_{\po}l(\x)r(\x) D\bs{g}(\x)\cdot \hat{\bs{n}}(\x)d\x.
\end{align}
\par We observe that this expression contains the boundary terms that appeared in the duality relation (\ref{dualitytilted}) for the large deviation operators associated with an additive observable as well as an additional $k$-dependent boundary term appearing as a result of the fact that the tilted generator (\ref{tiltedcurrent}) now mixes $k\g$ and derivative terms. Because of the additional term the argument used for additive observables does not apply for current-type observables: the presence of the $k$-dependent boundary term indicates that the boundary conditions on the large deviation operators will necessarily depend on $k$, so the boundary conditions for $k=0$ are no longer sufficient to determine the boundary conditions for arbitrary $k$. For the case where $k=0$ the $k$-dependent boundary term vanishes, leading to the boundary conditions (\ref{bound3}) and (\ref{bound4}) that apply respectively to $\mathcal{L}_{k=0}$ and $\mathcal{L}_{k=0}^{\dagger}$. 
\par An alternative attempt to obtain the boundary conditions on the large deviation operators from the duality relation (\ref{tiltedcurrentduality}) involves exploring the consequences of this duality relation for the current of the effective process associated with manifesting fluctuations of the observable $A_T$. From the expression of the effective drift $\bs{F}_k$ in (\ref{effdrift}) and the current (\ref{effstatcur2}) it can be shown that the duality relation (\ref{tiltedcurrentduality}), as applied to the dominant eigenfunctions $r_k$ and $l_k$, can be expressed as  
\begin{equation}
    \langle l_k, \mathcal{L}_k r_k\rangle=\langle \mathcal{L}_k^{\dagger}l_k, r_k\rangle - \int_{\po} \bs{J}_{\bs{F}_k,l_k r_k}(\x)\cdot\hat{\bs{n}}(\x)d\x.
\end{equation}
The vanishing of the surface term in the above is equivalent to the statement that the net flow of probability across the boundary $\po$ is zero, so that probability is conserved for the effective process. This boundary term must vanish regardless of boundary type (in order for the large deviation operators to be properly defined), but the particular manner in which it is made to vanish corresponds to a specific boundary behavior of the process upon reaching $\po$. We observe that the boundary term can be made to vanish by requiring that the stationary current $\bs{J}_{\bs{F}_k,l_k r_k}$ satisfies
\begin{equation} \label{bound7}
    \bs{J}_{\bs{F}_k,l_k r_k}(\x)\cdot\hat{\bs{n}}(\x) = 0 \quad \forall \x \in \po.
\end{equation}
This relation was obtained previously (\ref{currenteffectivebound}) for the case of additive observables of a reflected process and expresses the fact that the effective process is also a reflected diffusion. 
\par We could assume that the above relation must also hold for the case of current type observables, and attempt to derive the appropriate boundary conditions on $r_k$ and $l_k$ using this assumption and the requirement that the boundary conditions must correspond for $k = 0$ to (\ref{bound3}) and (\ref{bound4}), respectively. Proceeding in this manner, and observing that 
\begin{align}
    \bs{J}_{\bs{F}_k,l_k r_k}(\x) &= \bigg(\left[\bs{F}(\x) + (1-c)kD\bs{g}(\x)\right]l_k(\x) - \frac{1}{2}D\bs{\nabla}l_k(\x)\bigg)r_k(\x)  \nonumber \\ &\quad + l_k(\x) \bigg(ck\, r(\x)D\g(\x) + \frac{1}{2}D\bs{\nabla}r_k(\x) \bigg)
\end{align}
for an arbitrary real number $c$, we find that the most general boundary condition on the functions $r_k$ and $l_k$ that is consistent with the requirement (\ref{bound7}) and the boundary conditions (\ref{bound3}) and (\ref{bound4}) for $k = 0$ is 	
\begin{equation} \label{bound8}
    \bigg(ck\,r(\x) D\g(\x) + \frac{1}{2}D\bs{\nabla}r_k(\x) \bigg)\cdot\hat{\bs{n}}(\x) = 0 \quad \forall\x \in\po
\end{equation}
and
\begin{equation} \label{bound9}
	\bigg(\left[\bs{F}(\x) + (1-c)kD\bs{g}(\x)\right]l_k(\x) - \frac{1}{2}D\bs{\nabla}l_k(\x)\bigg)\cdot \hat{\bs{n}}(\x) = 0\quad \forall\x \in\po,
\end{equation}
with $c$ arbitrary. This argument therefore fails in that it cannot uniquely determine the correct boundary conditions: it cannot single out a particular value of $c$ since all values of $c$ are consistent with the requirement that the effective process is a reflected diffusion with reflecting boundary $\po$ and with the boundary conditions (\ref{bound3}) and (\ref{bound4}) for $k = 0$. It is therefore necessary to find a new argument not based solely on duality arguments to determine the correct domains associated with the large deviation operators for a current-type observable. 
\par In our work~\cite{Mallmin2021} on the problem, we formulated two such arguments, leading to the boundary conditions
\begin{equation} \label{rboundint2}
    D\nabla r(\x) \cdot \hat{\bs{n}}(\x) = - k r(\x) D\bs{g}(\x)\cdot \hat{\bs{n}}(\x) \quad \textnormal{for all} \quad \x \in \po,
\end{equation}
for functions $r$ in the domain of $\mathcal{L}_k$ and 
\begin{equation} \label{lboundint2}
    \bigg(\left[\bs{F}(\x) + \frac{1}{2}kD\bs{g}(\x)\right]l(\x) - \frac{1}{2}D\bs{\nabla}l(\x)\bigg)\cdot \hat{\bs{n}}(\x) = 0\quad \forall\x \in\po
\end{equation}
for normalizable densities $l$ in the domain of $\mathcal{L}_k^{\dagger}$. These boundary conditions correspond to (\ref{bound8}) and (\ref{bound9}) for $c = 1/2$, respectively. 
\par The first argument is based on a discretization of the diffusion into a lattice model, which is then treated as a jump process. The second argument is based instead on a construction known as the local time and the generating function of $A_T$. We present next the second argument in detail, as it is the one I worked on during the PhD. The first argument is not presented here (see~\cite{Mallmin2021}) as it is mostly the work of Mallmin.

\section{Generating function approach to boundary conditions}\label{sec:sec33}
We provide here a derivation of the boundary condition on $r_k$ (and more generally all functions in the domain of the tilted generator) for a current-type observable $A_T$ of the form (\ref{currentobs}), proceeding directly from the definition of the generating function 
\begin{equation}
    G_k(\x,t) = \mathbb{E}_{\x}\left[\exp \left(k t A_t\right)\right].
\end{equation}
The generating function $G_k$ has the spectral decomposition (\ref{specdec}) in terms of the eigenfunctions of the tilted generator $\mathcal{L}_k$ and, as such, shares the boundary conditions placed on these eigenfunctions. By determining the behavior of the generating function $G_k$ at a reflecting boundary we can therefore also determine the appropriate boundary condition on functions in the domain of $\mathcal{L}_k$, with the corresponding boundary condition on the domain of $\mathcal{L}_k^{\dagger}$ then following directly from duality arguments. 
\par We employ a formulation of reflecting SDEs known as the local time formalism, introduced initially by Skorokhod~\cite{skorokhod1961stochastic} and Tanaka~\cite{tanaka1979}, which supplements an SDE with an additional term that accounts for the behavior of the process upon reaching the boundary and which compensates for the process' tendency to move across a boundary. In this formulation, we write a Markov diffusion undergoing reflection at a boundary $\po$ as \begin{equation} \label{eq:localtimesde}
   d\bs{X}(t) = \bs{F}(\bs{X}(t))dt + \sigma d \bs{W}(t) + \hat{\bs{\gamma}}(\bs{X}(t)) d L(t),
\end{equation}
where $\hat{\bs{\gamma}}(\x)$ is a smooth vector field (with unit norm for every $\x$) directed inwards into the domain $\om$ at every point on the boundary $\po$. The first two terms on the RHS describe the evolution of $\bs{X}(t)$ inside the domain $\om$, whereas the last term involves a non-decreasing random variable $d L(t)$, the so-called \textit{local time}, which increases only when the process reaches the boundary $\po$. Given that $\hat{\bs{\gamma}}(\x)$ is directed inwards into the domain on the surface $\po$ at point $\x$, the effect of this local time term is to push the process into the domain in the direction $\bs{\hat{\gamma}}(\x)$ in the event that the process reaches $\x \in \po$. This additional term in the SDE therefore explicitly accounts for the reflecting nature of the boundary, with the unit vector $\hat{\bs{\gamma}}(\x)$ determining the direction of reflection. 
\par Supposing that the process reaches $\x \in \po$ at time $t$, it is known~\cite{grebenkov2019probability} that the local time $d L(t)$ satisfies
\begin{equation}
    \mathbb{E}_{\bs{X}(t) = \x}\left[d L(t)\right]  = \mathcal{O}(\sqrt{dt}).
\end{equation}
Therefore the increment $d \bs{X}(t)$ satisfies
\begin{align} \label{eq:update-ref}
    \mathbb{E}_{\bs{X}(t) = \x}\left[d \bs{X}(t)\right] &= \bs{F}(\x) dt + \sigma \mathbb{E}_{\bs{X}(t) = \x}\left[d \bs{W}(t)\right] + \hat{\bs{\gamma}}(\x) \mathbb{E}_{\bs{X}(t)= \x}\left[d L(t)\right] \nonumber \\ &= \hat{\bs{\gamma}}(\x) \epsilon + \mathcal{O}(\epsilon^2) 
\end{align}
in the event that the process reaches $\x \in \po$ at time $t$, where we have used the fact that $\mathbb{E}_{\bs{X}(t) =\x}\left[d \bs{\bs{W}(t)}\right] = \bs{0}$ for all $t$, and where $\epsilon = \mathcal{O}(\sqrt{dt})$ such that $dt(\epsilon) = \mathcal{O}(\epsilon^2)$. From here onward we consider only the choice 
\begin{equation} \label{eq:co-normal}
    \hat{\bs{\gamma}}(\x) = \frac{D \hat{\bs{n}}(\x)}{\lvert D \hat{\bs{n}}(\x)\rvert},
\end{equation}
so that reflection occurs in the co-normal direction, given that this choice is necessary (see Theorem 2.6.1 of Schuss~\cite{schuss2013}) in order to preserve the zero current condition on $\po$ defined before in (\ref{bound1}). 
\par Our goal now is to understand the effect of the boundary dynamics on the generating function $G_k$ associated with the current-type observable $A_T$. To this end, consider  
\begin{equation}
    G_k(\x,t) = \mathbb{E}_{\x}\left[e^{ktA_t}\right] = \mathbb{E}_{\x}\left[\exp \left(k \int_0^t \bs{g}(\bs{X}(s))\circ d\bs{X}(s)\right)\right]
\end{equation}
for $\x \in \po$. We can write  
\begin{equation}
    G_k(\x,t) = \mathbb{E}_{\x}\left[\exp \left(k \int_0^{dt(\epsilon)} \bs{g}(\bs{X}(s))\circ d\bs{X}(s) +\int_{dt(\epsilon)}^t \bs{g}(\bs{X}(s))\circ d\bs{X}(s) \right)\right],
\end{equation}
isolating in the first integral the contribution from the reflection, which takes place over the infinitesimal time $dt(\epsilon)$. Using the Stratonovich discretization, as in (\ref{strato1}), we have
\begin{equation}
    \exp\left(k \int_0^{dt(\epsilon)} \bs{g}(\bs{X}(s))\circ d\bs{X}(s)\right) = \exp\left(k \bs{g}\left(\x + \frac{d \bs{X}(0)}{2} \right)\cdot d \bs{X}(0)\right),
\end{equation}
so that the generating function is given by
\begin{equation}
    G_k(\x,t) = \mathbb{E}_{\x}\left[\exp\left(k \bs{g}\left(\x + \frac{d \bs{X}(0)}{2} \right)\cdot d \bs{X}(0)\right) \exp\left(k\int_{dt(\epsilon)}^t \bs{g}(\bs{X}(s))\circ d\bs{X}(s) \right)\right].
\end{equation}
Integrating over all possible values of the increment $d \bs{X}(0) = \hat{\bs{\xi}}\delta$, the above expectation value can be written explicitly as
\begin{align} \label{eq:u-general1}
    G_k(\x,t) &= \int d (\hat{\bs{\xi}}\delta)\,  p\left(d\bs{X}(0) = \hat{\bs{\xi}}\delta | \bs{X}(0) = \x \right) \nonumber \\ &\quad \times e^{k\bs{g}\left(x + \frac{\hat{\bs{\xi}}\delta}{2} \right)\cdot\hat{\bs{\xi}}\delta} \mathbb{E}_{\bs{X}(dt) = \x + \hat{\bs{\xi}}\delta}\left[\exp\left(k\int_{dt(\epsilon)}^t \bs{g}(\bs{X}(s))\circ d\bs{X}(s) \right)\right],
\end{align}
using the conditional probability density of the first increment $d\mathbf{X}(0)$ from $\mathbf{X}(0)= \x\in\del\Omega$ to $\x + \hat{\bs{\xi}}(\x)$, which includes all the information about the reflections on the boundary. Using the definition of the generating function $G_k(\x,t)$ for the last factor in the above integral 
\begin{equation}
    \mathbb{E}_{\bs{X}(dt) = \x + \hat{\bs{\xi}}\delta}\left[\exp\left(k\int_{dt(\epsilon)}^t \bs{g}(\bs{X}(s))\circ d\bs{X}(s) \right)\right] = G_k(\x + \hat{\bs{\xi}}\delta, \, t - dt(\epsilon))
\end{equation}
we then obtain
\begin{equation} \label{eq:u-general2}
    G_k(\x,t) = \int d (\hat{\bs{\xi}}\delta)\,  p\left(d\bs{X}(0) = \hat{\bs{\xi}}\delta | \bs{X}(0) = \x \right) e^{k\bs{g}\left(x + \frac{\hat{\bs{\xi}}\delta}{2} \right)\cdot\hat{\bs{\xi}}\delta} G_k(\x + \hat{\bs{\xi}}\delta, \, t - dt(\epsilon)).
\end{equation}
\par We now perform Taylor expansions in both space and time. Taylor expanding up to first order in the magnitude $\delta$, we obtain
\begin{equation} \label{eq:taylor1}
    e^{k\bs{g}\left(x + \frac{\hat{\bs{\xi}}\delta}{2} \right)\cdot\hat{\bs{\xi}}\delta} = 1 + k\bs{g}(\x)\cdot \hat{\bs{\xi}} \delta + \mathcal{O}(\delta^2)
\end{equation}
and 
\begin{equation} \label{eq:taylor2}
    G_k(\x + \hat{\bs{\xi}} \delta, \, t - dt(\epsilon)) = G_k(\x, t - dt(\epsilon)) + \bs{\nabla} G_k(\x, t - dt(\epsilon)) \cdot \hat{\bs{\xi}} \delta + \mathcal{O}(\delta^2).
\end{equation}
Substituting \eqref{eq:taylor1} and \eqref{eq:taylor2} into \eqref{eq:u-general2} yields the result
\begin{align} \label{eq:u-taylor}
    G_k(\x,t) = \int d (\hat{\bs{\xi}}\delta)&\,  p\left(d\bs{X}(0) = \hat{\bs{\xi}}\delta | \bs{X}(0) = \x \right) \bigg[G_k(\x, t - dt(\epsilon)) + \bs{\nabla} G_k \cdot \hat{\bs{\xi}} \delta \nonumber \\ & \quad + k G_k(\x, t - dt(\epsilon))\bs{g}(\x)\cdot \hat{\bs{\xi}} \delta + \mathcal{O}(\delta^2)\bigg]
\end{align}
to first order in the spatial variable $\delta$. Given the normalization condition 
\begin{equation}
     \int d (\hat{\bs{\xi}}\delta)\,  p\left(d\bs{X}(0) = \hat{\bs{\xi}}\delta | \bs{X}(0) = \x \right) = 1
\end{equation}
and the fact that
\begin{equation}
    \int d (\hat{\bs{\xi}}\delta)\,  p\left(d\bs{X}(0) = \hat{\bs{\xi}}\delta | \bs{X}(0) = \x \right) \hat{\bs{\xi}}{\delta} = \mathbb{E}_{\bs{X}(0) = \x}\left[d \bs{X}(0)\right], 
\end{equation}
and noting that $\delta = \lvert d \bs{X}(0)\rvert$, we have from \eqref{eq:u-taylor},
\begin{align} \label{u-post-taylor}
    G_k(\x,t) =& G_k(\x, t - dt(\epsilon)) + \bs{\nabla} G_k(\x, t - dt(\epsilon)) \cdot \mathbb{E}_{\bs{X}(0) = \x}\left[d \bs{X}(0)\right]   \nonumber \\ &\quad + k G_k(\x, t - dt(\epsilon)) \bs{g}(\x)\cdot \mathbb{E}_{\bs{X}(0) = \x}\left[d \bs{X}(0)\right] + \mathbb{E}_{\bs{X}(0) = \x}\left[\lvert d \bs{X}(0)\rvert^2\right].
\end{align}
Since $\x \in \po$, we have that $\mathbb{E}_{\bs{X}(0) = \x}\left[d \bs{X}(0)\right]$ is given from \eqref{eq:update-ref} and \eqref{eq:co-normal} by 
\begin{equation}
    \mathbb{E}_{\bs{X}(0) = \x}\left[d \bs{X}(0)\right] = \frac{D \hat{\bs{n}}(\x)}{\lvert D \hat{\bs{n}}(\x)\rvert} \epsilon + \mathcal{O}(\epsilon^2).
\end{equation}
Substituting this into \eqref{u-post-taylor} then yields 
\begin{align}  \label{taylorexpand1}
    G_k(\x,t) =& G_k(\x, t - dt(\epsilon)) + \bs{\nabla}G_k(\x, t - dt(\epsilon)) \cdot \frac{D \hat{\bs{n}}(\x)}{\lvert D \hat{\bs{n}}(\x)\rvert} \epsilon  \nonumber \\ &\quad + k G_k(\x, t - dt(\epsilon)) \bs{g}(\x)\cdot \frac{D \hat{\bs{n}}(\x)}{\lvert D \hat{\bs{n}}(\x)\rvert} \epsilon  + \mathcal{O}(\epsilon^2).
\end{align}
\par Upon using the Feynman-Kac formula (\ref{FK}) for the time evolution of $G_k$ and Taylor expanding in the time $t$, we have that 
\begin{equation}
    G_k(\x,t - dt(\epsilon)) = G_k(\x,t) - \mathcal{L}_k dt(\epsilon) G_k(\x,t) + \mathcal{O}(dt(\epsilon)^2) = G_k(\x,t) + \mathcal{O}(\epsilon^2),
\end{equation}
where we have used the fact that $dt(\epsilon) = \mathcal{O}(\epsilon^2)$. Inserting the above expression into (\ref{taylorexpand1}), we obtain 
\begin{equation} \label{eq:u-intermediate}
    G_k(\x,t) = G_k(\x,t) + \bs{\nabla} G_k(\x, t) \cdot \frac{D \hat{\bs{n}}(\x)}{\lvert D \hat{\bs{n}}(\x)\rvert} \epsilon  + k G_k(\x, t ) \bs{g}(\x)\cdot \frac{D \hat{\bs{n}}(\x)}{\lvert D \hat{\bs{n}}(\x)\rvert} \epsilon  + \mathcal{O}(\epsilon^2).
\end{equation}
Subtracting $G_k(\x,t)$ on both sides of the above equation we find that  
\begin{equation}
    \bs{\nabla} G_k(\x, t) \cdot \frac{D \hat{\bs{n}}(\x)}{\lvert D \hat{\bs{n}}(\x)\rvert} \epsilon  = -  k G_k(\x, t ) \bs{g}(\x)\cdot \frac{D \hat{\bs{n}}(\x)}{\lvert D \hat{\bs{n}}(\x)\rvert} \epsilon
\end{equation}
up to first order in $\epsilon$. Using the symmetry of $D$ and multiplying on both sides by $\lvert D \hat{\bs{n}}(\x)\rvert/\epsilon$, we finally obtain 
\begin{equation} \label{eq:u-main-result}
    D\bs{\nabla} G_k(\x, t) \cdot \hat{\bs{n}}(\x) = - k G_k(\x,t) D\g(\x) \cdot \hat{\bs{n}}(\x) \quad \forall \x \in \po. 
\end{equation}
This is the main result of this section, showing explicitly the boundary behavior of the generating function, which depends in a $k$-dependent manner on the function $\bs{g}$. From this boundary condition on the generating function $G_k$ the appropriate boundary conditions on the eigenfunctions $r_k$ and $l_k$ can be obtained.
\par Given that the generating function $G_k(\x,t)$ and the functions in the domain of the tilted generator $\mathcal{L}_k$ share the same boundary conditions we have 
\begin{equation} \label{rboundint}
    D\nabla r(\x) \cdot \hat{\bs{n}}(\x) = - k r(\x) D\bs{g}(\x)\cdot \hat{\bs{n}}(\x) \quad \textnormal{for all} \quad \x \in \po,
\end{equation}
for arbitrary $r \in \mathcal{D}(\mathcal{L}_k)$. In particular, for the dominant eigenfunction $r_k$ associated with the eigenvalue $\lambda(k)$ we have 
\begin{equation}
    D\nabla r_k(\x) \cdot \hat{\bs{n}}(\x) = - k r_k(\x) D\bs{g}(\x)\cdot \hat{\bs{n}}(\x) \quad \textnormal{for all} \quad \x \in \po.
\end{equation}
Note that this corresponds to the case $c = 1/2$ of the boundary condition (\ref{bound8}) obtained before. Arguing that the surface terms in the duality relation (\ref{tiltedcurrentduality}) must vanish for arbitrary $r$ and $l$ it is clear that the corresponding boundary condition on a normalized density $l$ in the domain of $\mathcal{L}_k^{\dagger}$ is the condition (\ref{bound9}) for $c = 1/2$, given explicitly as  
\begin{equation}
    \bigg(\left[\bs{F}(\x) + \frac{1}{2}kD\bs{g}(\x)\right]l(\x) - \frac{1}{2}D\bs{\nabla}l(\x)\bigg)\cdot \hat{\bs{n}}(\x) = 0\quad \forall\x \in\po.
\end{equation}
\par We have therefore succeeded in deriving the appropriate boundary conditions for the spectral problem described in (\ref{spec1}) and (\ref{spec2}) associated with calculating the SCGF $\lambda(k)$ of current-type observable $A_T$. In particular, we obtain again the physically reasonable result (\ref{bound7}) which shows that the effective process is again a reflected process. Furthermore, the relation (\ref{effdriftbound}) holds here as well, again showing that the component of the effective drift normal to the reflecting boundary is identical to that of the original drift. These results are therefore found to be true regardless of whether the observable under consideration is an additive or current-type observable. 
\par It is interesting to note that the calculation presented here can be repeated for an additive observable of the form (\ref{additive}) and reproduces the boundary condition (\ref{bound3}) obtained via purely duality arguments. Our new results therefore serve as a substantial generalization of our previous results~\cite{dubuissonmasters,Buisson2020}, and provides a further verification of their validity.\par In principle, the calculation in this section can also be done for an arbitrary reflection direction $\hat{\bs{\gamma}}(\x)$, in which case the boundary condition on the generating function $G_k$, and hence any function in the domain of $\mathcal{L}_k$, becomes 
\begin{equation}
    \nabla G_k(\x, t) \cdot \hat{\bs{\gamma}}(\x)   = - k G_k(\x, t ) \bs{g}(\x)\cdot \hat{\bs{\gamma}}(\x).
\end{equation}
Comparing with the duality relation (\ref{surface}) (for the case where $k = 0$) then shows that we only obtain a sensible zero-current condition $\bs{J}_{\bs{F},p}\cdot \hat{\bs{n}}(\x) = 0$ at the boundary in the event that the chosen direction for reflection is the co-normal direction $D\hat{\bs{n}}(\x)$, as mentioned before. 
\section{Application to heterogeneous single-file diffusion on a ring}
We now illustrate the results obtained in this chapter for the large deviations of current-type observables for an exactly solvable system having reflecting boundaries, known as the heterogeneous single-file diffusion on a ring~\cite{Mallmin2021}.
\par
\begin{figure}[t]
	\centering
	\includegraphics[]{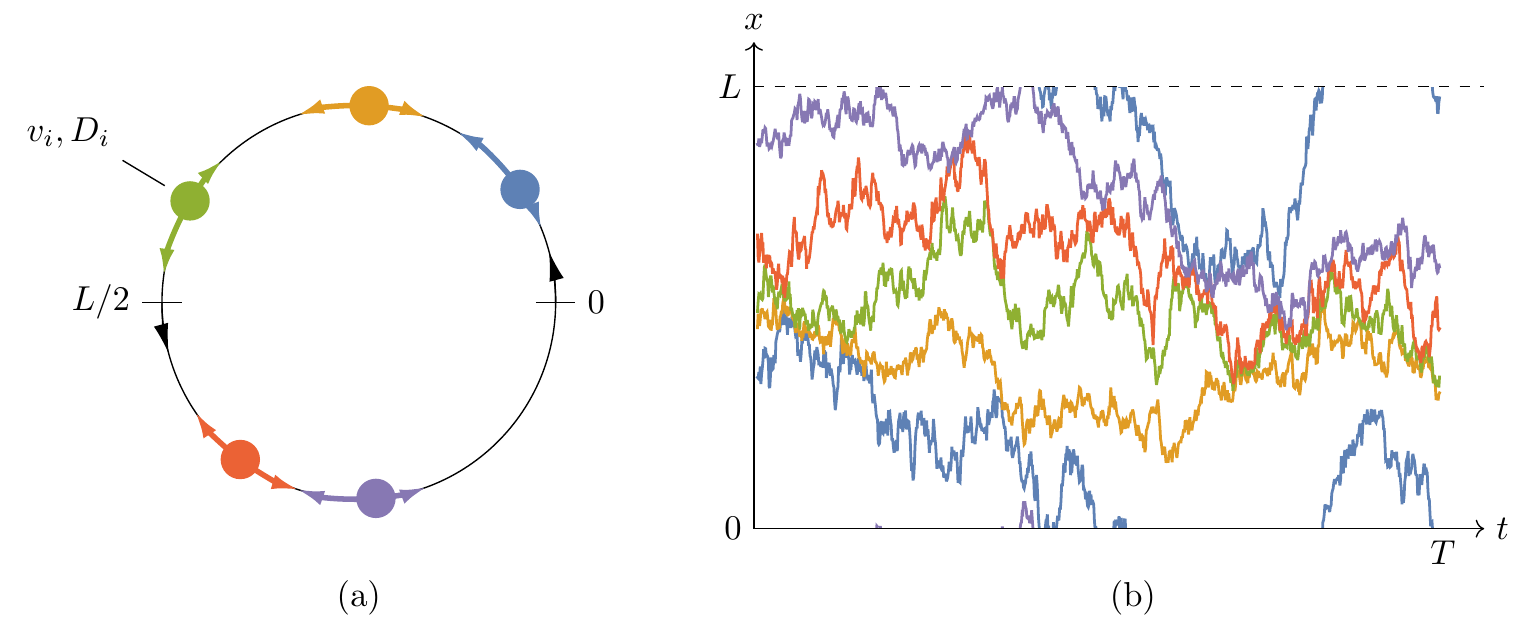}
	\caption{(a) Illustration of heterogeneous single-file diffusion on a ring. (b) Space-time plot of a typical realization of the particle positions. Due to hardcore exclusion, i.e.\ reflection, the particles' paths do not cross.}\label{fig:SFD}
\end{figure}

 The heterogeneous single file diffusion consists of $N$ particles undergoing diffusion on a ring with circumference $L$, with the position $X_i(t) \in [0,L)$ of the $i$th particle evolving according to the SDE
\begin{equation}
    d X_i(t) = v_i \, dt + \sigma_i \, dW_i(t),
\end{equation}
with $v_i$ representing the intrinsic velocity of the $i$th particle and $\sigma_i > 0$ representing the diffusivity of that same particle.  The system is illustrated in Fig.~\ref{fig:SFD}. The diffusion with state $\bs{X}(t) \in [0,L)^N$ therefore has the drift $\bs{F} = \bs{v} = (v_1, v_2, \ldots, v_N)^{\mathsf{T}}$, while the diffusion matrix is given as $D = \tn{diag}(\sigma_1^2, \sigma_2^2, \ldots, \sigma_N^2)$. Moreover, the particles are assumed to interact via hardcore exclusion: each particle acts as a reflecting barrier for the adjacent particles and as such particles cannot overtake one another. We can assume without loss of generality that the particles are ordered according to $X_1(t) \leq X_2(t) \leq \ldots \leq X_N(t)$ (modulo $L$).
\par The reflecting boundary $\po$ present in the system consists of all those configurations for which two or more particles have equal positions. To make this more concrete, consider the case where $N = 2$. We then have that the boundary $\po$ is given by all configurations with $X_1(t) = X_2(t)$, which consists of a diagonal in $[0,L)^2$, so that the normal vector of this boundary $\hat{\bs{n}}$ is given by 
\begin{equation}
    \hat{\bs{n}} = \frac{\hat{\bs{e_2}}- \hat{\bs{e_1}}}{\sqrt{2}}.
\end{equation}
This can be generalized to $N$ particles, for which the appropriate reflecting boundary condition (\ref{bound1}) on probability densities $p$ in the domain of the Fokker-Plank operator $\mathcal{L}^{\dagger}$ therefore becomes, from the above discussion, the condition 
\begin{equation} \label{sfdref}
    \hat{\bs{e_i}}\cdot \bs{J}_{\bs{F},p}(\x,t)|_{x_i = x_j} = \hat{\bs{e_j}}\cdot \bs{J}_{\bs{F},p}(\x,t)|_{x_i = x_j}.
\end{equation}
Furthermore, appropriate densities $p$ must satisfy the periodic boundary condition 
\begin{equation} \label{sfdper}
    p(\x,t) = p(\x + L\bs{1},t)
\end{equation}
for all $\x \in [0,L)^N$, where $\bs{1}$ represents the vector of $1$s, owing to the periodic nature of the ring. 
\par It was found in~\cite{Mallmin2021a} by Mallmin et al.~that the heterogeneous single-file diffusion has a stationary density $p^*$ satisfying 
\begin{equation} \label{sfdstat}
    p^*(\x) \propto \exp\left(\bs{d} \cdot \x \right),
\end{equation}
with the vector $\bs{d}$ having the entries 
\begin{equation}
    d_i = \frac{v_i - \tilde{v}}{D_i},
\end{equation}
where $D_i = \sigma_i^2$ and with $\tilde{v}$ given by 
\begin{equation}
    \frac{\tilde{v}}{\tilde{D}} = \sum_{i = 1}^N \frac{v_i}{D_i},
\end{equation}
and 
\begin{equation}
    \frac{1}{\tilde{D}} = \sum_{i = 1}^N \frac{1}{D_i}.
\end{equation}
\par It is important to note that, from (\ref{statcur}), the stationary current associated with $p^*$ satisfies 
\begin{equation} \label{sfdcurbound}
    \hat{\bs{e_i}}\cdot \bs{J}_{\bs{F},p^*}(\x) = \tilde{v} p^*(\x)
\end{equation}
for all $i \in \{1,2,\ldots,N\}$. As such, the quantity $\tilde{v}$ represents a net velocity common to all the particles in the long-time limit. The fact that the particles all have a single common velocity in this limit is an intuitive result given the geometric constraints present in the system. Furthermore, this shows that the stationary density does indeed satisfy the required reflecting condition (\ref{sfdref}). We observe also that 
\begin{equation}
    \bs{d}\cdot \bs{1} = 0
\end{equation}
so that $p^*$ satisfies the periodicity condition (\ref{sfdper}). 
\par We are interested in obtaining, for the system under consideration, the large deviations associated with current-type observables $V_T$ having the form 
\begin{equation} \label{sfdobs}
    V_T = \frac{1}{T} \int_0^T \bs{g} \circ d\bs{X}(t),
\end{equation}
where $\bs{g}$ is any constant vector satisfying $\bs{g}\cdot \bs{1} = 1$. This class of observables includes the so-called empirical velocity, which is obtained through the choice $g_j = \delta_{ij}$. We expect the large deviations to be the same for any observable satisfying the condition $\bs{g}\cdot\bs{1}=1$, given that the particles have a common velocity in the long-time limit. 
\par The tilted generator $\mathcal{L}_k$ associated with the observable (\ref{sfdobs}) is given from (\ref{tilted}) by 
\begin{equation}
    \mathcal{L}_k = \bs{v}\cdot \bs{\nabla} + k \bs{v}\cdot \g + \frac{1}{2}\bs{\nabla}\cdot D\bs{\nabla} + 2k\g\cdot D \bs{\nabla} + k^2 \g\cdot D \g
\end{equation}
and acts on functions $r$ satisfying the reflecting boundary condition (\ref{rboundint2})
\begin{equation} \label{sfdrkbound1}
    D\left(\bs{\nabla} r(\x) + k \g r(\x)\right)\cdot \bs{\hat{n}}(\x) = 0,
\end{equation}
along with the periodic boundary condition 
\begin{equation} \label{sfdrkbound2}
    r(\x) = r(\x + L\bs{1}). 
\end{equation}
To obtain the dominant eigenfunction $\lambda(k)$ and the corresponding eigenfunction $r_k$, we use the ansatz 
\begin{equation} \label{sfdrk}
    r_k(\x) = \exp(\bs{a}\cdot \x).
\end{equation}
This is a reasonable choice given that for $k=0$ the dominant eigenfunction of the tilted generator (corresponding to the dominant eigenfunction of the infinitesimal generator $\mathcal{L}$) is $e^0 = 1$. We find for the ansatz (\ref{sfdrk}) that  
\begin{equation} \label{applic}
    \mathcal{L}_k r_k(\x) = \left(\bs{v}\cdot \bs{a} + k \bs{v}\cdot \bs{g} + \frac{1}{2} \bs{a}\cdot D \bs{a} + 2k\bs{g} \cdot D\bs{a} + \frac{k^2}{2} \g \cdot D \g\right) r_k(\x)
\end{equation}
and, since the term in brackets contains only $\x$-independent quantities, the ansatz is clearly a valid candidate for an eigenfunction, if $\bs{a}$ can be chosen in such a manner that the boundary conditions (\ref{sfdrkbound1}) and (\ref{sfdrkbound2}) are satisfied. 
\par Noting that $\bs{1}\cdot \hat{\bs{n}}(\x) = 0$, and that $\bs{1}$ is the only vector having this property, we have that the boundary condition (\ref{sfdrkbound1}) reduces to 
\begin{equation}
    D\left( \bs{a} + k \g\right) = \alpha \bs{1},
\end{equation}
for some $\alpha$. We therefore have 
\begin{equation}
    \bs{a} = \alpha D^{-1}\bs{1} - k \bs{g},
\end{equation}
with the periodic boundary condition (\ref{sfdrkbound2}) requiring in addition that $\bs{a}\cdot \bs{1} = 0$, so that we must have 
\begin{equation}
    \alpha D^{-1}\bs{1}\cdot \bs{1} - k \g \cdot \bs{1}.
\end{equation}
Using now the fact that $\g \cdot \bs{1} = 1$ by assumption, we obtain 
\begin{equation}
    \alpha = k \tilde{D}
\end{equation}
so that 
\begin{equation}
    \bs{a} = k \tilde{D} D^{-1}\bs{1} - k\g.
\end{equation}
It can now be verified that $r_k$ with $\bs{a}$ thus defined is an eigenfunction of $\mathcal{L}_k$ with eigenvalue $\lambda(k)$ given from (\ref{applic}) by
\begin{equation} \label{sfdscgf}
    \lambda(k) = k \tilde{v} + \frac{1}{2} k^2 \tilde{D}.
\end{equation}
This is seen not to depend explicitly on $g$, so long as $g$ satisfies the condition $\g\cdot \bs{1} = 1$. The rate function can be obtained as the Legendre transform of $\lambda(k)$ and is given by 
\begin{equation}
    I(v) = \frac{(v - \tilde{v})^2}{2\tilde{D}},
\end{equation}
which shows that the fluctuations of the current are Gaussian and centered on the long-time velocity common to all particles, $\tilde{v}$. In fact, the rate function obtained thus is the same as in the event where we consider a system with a single particle having intrinsic velocity $\tilde{v}$ and diffusivity $\tilde{D}$. 
\par To understand how fluctuations are created, we consider the effective drift (\ref{effdrift}) associated with the fluctuation $V_T = v$. We have that 
\begin{equation}
    \bs{F}_k = \bs{v} + (v(k) - \tilde{v})\bs{1} = \bs{v}_k,
\end{equation}
so that the current fluctuation $V_T = v(k)$ is manifested by changing the velocity of all the particles by the same amount $v(k) - \tilde{v}$. This system is again a heterogeneous single-file diffusion, and thus has stationary density given from (\ref{sfdstat}) for $\bs{F}_k = \bs{v}_k$. Remarkably, it is found that 
\begin{equation}
    p_k^*(\x) = p^*(\x)
\end{equation}
so that the effective process has the same stationary density as that of the original process. For the stationary current associated with the effective process we thus obtain
\begin{equation}
    \bs{J}_{\bs{F}_k, p_k^*}(\x) = \bs{J}_{\bs{F}_k, p^*}(\x) = \frac{v(k)}{\tilde{v}} \bs{J}_{\bs{F},p^*}(\x), 
\end{equation}
showing how the current is altered to account for a given current fluctuation: by a simple rescaling. It is important to note that the stationary current for the effective process satisfies the reflecting boundary condition (\ref{sfdcurbound}), given that the effective current is proportional to the original current. This shows explicitly that the effective process is again a reflected process, in accordance with the results obtained in Sec.~\ref{sec:sec33}. Furthermore, given that $\bs{n}(\x)\cdot \bs{1} = 0$, we have that 
\begin{equation}
    \bs{F}_k\cdot \hat{\bs{n}}(\x) = \bs{F}\cdot \hat{\bs{n}}(\x),
\end{equation}
which serves as an explicit verification of the relation (\ref{effdriftbound}) for current-type observables. 
\par Finally, we note that since $p_k^* = p^*$ and also $p_k^* = l_k r_k$, we can obtain the eigenfunction $l_k$ as 
\begin{equation}
    l_k(\x) \propto \exp((\bs{d}-\bs{a})\cdot \x).
\end{equation}
This can be explicitly verified to be an eigenfunction of the adjoint operator $\mathcal{L}_k^{\dagger}$ given by 
\begin{equation}
    \mathcal{L}_k^{\dagger} = \bs{v}\cdot (-\bs{\nabla} + k\bs{g}) + \frac{1}{2}(-\bs{\nabla} + k\g) \cdot D (-\bs{\nabla} + k\g) 
\end{equation}
with eigenvalue $\lambda(k)$ as given in (\ref{sfdscgf}) and satisfying the appropriate reflecting boundary condition (\ref{lboundint2}) on the domain of the adjoint operator, in addition to the periodicity condition
\begin{equation}
    l_k(\x) = l_k(\x + L \bs{1}). 
\end{equation}

\chapter[Large deviations of linear diffusions]{Dynamical large deviations of linear diffusions}\label{chap:chap4}
In this chapter, we move to the second problem of the dissertation, which is to study the large deviations of linear diffusions. We consider two classes of additive observables, being respectively linear and quadratic in the state of the process, and current-type observables which are linear in the state. Exact expressions for the generating function of each of these types of observables is obtained via the Feynman-Kac formula, after which the SCGF and its corresponding eigenfunction can be found by studying the long-time limit of the generating function. This allows us to determine both the rate function which gives the probability of fluctuations as well as the effective process which is responsible for manifesting fluctuations. We are particularly interested in studying the stationary density and current associated with the effective process and in understanding the manner in which these quantities differ from those of the original process. Finally, we obtain exact expressions for the asymptotic mean and variance associated with all the observables studied here. In the next chapter we will illustrate the general formalism developed here by applying these results to a variety of specific linear systems and observables. 

\section{Exact Results for the moment generating function}\label{sec:sec41}
We derive in this section an exact expression for the moment generating function $G_k$ associated with a variety of physically relevant observables for linear diffusions, defined before by the SDE (\ref{eqSDE}). We consider three classes of observables:
\begin{itemize}
    \item The linear additive observable 
    \begin{equation} \label{linadd}
    A_T = \frac{1}{T} \int_0^T \langle \bs{\eta},\bs{X}(t)\rangle dt,
\end{equation}
where $\langle \cdot,\cdot\rangle$ indicates the standard vector inner product on $\mathbb{R}^n$ and $\bs{\eta}$ is an arbitrary vector in $\mathbb{R}^n$, 
\item The quadratic additive observable
\begin{equation} \label{quadadd}
    A_T = \frac{1}{T}\int_0^T \langle \bs{X}(t),Q\bs{X}(t)\rangle dt,
\end{equation}
with $Q$ being a symmetric $n \times n$ matrix.
\item The linear current-type observable
\begin{equation} \label{lincurobs}
    A_T = \frac{1}{T}\int_0^T \Gamma\bs{X}(t) \circ d\bs{X}(t),
\end{equation} 
where $\Gamma$ is an arbitrary $n \times n$ matrix. 
\end{itemize}
\par These observables are important in physics as they include many quantities that can be measured in practice, such as the mechanical work done on a nonequilibrium system, the heat transferred between a system and its environment, and the entropy production, which is an important quantity in nonequilibrium statistical mechanics related to the irreversibility of stochastic processes~\cite{seifert2012stochastic, sekimoto2010stochastic}. Mathematically, these observables are also interesting as we shall see that the effective process associated with their large deviations is also a linear diffusion. In order to arrive at this result,  we shall consider next each class of observable separately, starting with the linear additive observable. 

\subsection{Linear additive observable}\label{sec:sec411}
To calculate the generating function $G_k$ we will exploit the fact that $G_k$ evolves according to the Feynman-Kac equation given in (\ref{FK}) in Sec.~\ref{sec:sec26} with initial condition $G_k(\x,0) = 1$. Here $\mathcal{L}_k$ is the tilted generator, which for the linear additive observable (\ref{linadd}) and linear diffusion (\ref{eqSDE}), is given by 
\begin{equation}
    \mathcal{L}_k = -M \x \cdot \bs{\nabla} + \frac{1}{2} \bs{\nabla} \cdot D \bs{\nabla} + k \langle \bs{\eta}, \x \rangle.  
\end{equation}
We will obtain an explicit solution $G_k(\x,t)$ by solving the Feynman-Kac equation iteratively in time steps $\Delta t = t/n$ starting from the initial condition $G_k(\x,0) = 1$. At the end of the procedure we take the $\Delta t \rightarrow 0$ limit to obtain a continuous time expression for $G_k(\x,t)$ for all times $t$. 
\par For the first time step, we use the fact that 
\begin{equation}
    G_k(\x,t) = \left(e^{t\mathcal{L}_k}1\right)(\x)
\end{equation}
and
\begin{equation}
    e^{\Delta t \mathcal{L}_k} = 1 + \Delta t \mathcal{L}_k + \mathcal{O}(\Delta t^2),
\end{equation}
to write 
\begin{align}
    G_k(\x,\Delta t) &= e^{\Delta t\mathcal{L}_k} 1 \nonumber \\ &= (1 + \Delta t \mathcal{L}_k) G_k(\x,0) \nonumber \\ &= (1 + \Delta t \mathcal{L}_k) 1 \nonumber \\ &= (1 + \Delta t \mathcal{L}_k)e^{\langle \bs{v}^{(0)}_k, \x\rangle}
\end{align}
up to first order in $\Delta t$, and with the vector $\bs{v}^{(0)}_k = \bs{0}$. The reason for introducing this vector will become clear. Proceeding, and using the fact that $\mathcal{L}_k 1 = k\langle \bs{\eta}, \x\rangle $, we have  
\begin{equation}
    G_k(\x,\Delta t) = (1 + \Delta t k\langle \bs{v}, x\rangle )\exp(\langle \bs{v}^{(0)}_k, \x\rangle) = \exp(\langle \bs{v}^{(1)}_k, \x \rangle),
\end{equation}
where 
\begin{equation}
    \bs{v}^{(1)}_k = \bs{v}^{(0)}_k + \Delta t \, k \bs{\eta} = \Delta t \, k \bs{\eta}.
\end{equation}
\par Continuing, we now time evolve $G_k(\x, \Delta t)$, obtaining
\begin{equation}
    G_k(\x, 2\Delta t) = (1 +\Delta t \mathcal{L}_k)G_k(\x,\Delta t) =(1 +\Delta t \mathcal{L}_k) \exp(\langle \bs{v}^{(1)}_k, \x \rangle).
\end{equation}
Observing that 
\begin{equation}\label{der1}
    \bs{\nabla} \exp(\langle \bs{w}, \x \rangle) = \bs{w}\exp(\langle \bs{w}, \x \rangle)
\end{equation}
and 
\begin{equation}\label{der2}
    \bs{\nabla}\cdot D\bs{\nabla} \exp(\langle \bs{w}, \x \rangle) = \langle \bs{w}, D\bs{w}\rangle\exp(\langle \bs{w},\x\rangle)
\end{equation}
for arbitrary $\bs{w} \in \mathbb{R}^n$, we then obtain  
\begin{align} \label{intermediatenew1}
    \mathcal{L}_k \exp(\langle \bs{v}^{(1)}_k, \x \rangle) &= \left(-M\x \cdot \bs{\nabla} + \frac{1}{2} \bs{\nabla}\cdot D \bs{\nabla} + k\langle \bs{\eta}, \x\rangle \right)\exp(\langle \bs{v}^{(1)}_k, \x \rangle) \nonumber \\
    &= \left(-\langle \bs{v}^{(1)}_k, M\x\rangle + \frac{1}{2} \langle \bs{v}^{(1)}_k, D \bs{v}^{(1)}_k \rangle + k \langle \bs{\eta}, \x\rangle \right)\exp(\langle \bs{v}^{(1)}_k, \x \rangle) \nonumber \\
    &= \left(k \langle \bs{\eta}, \x\rangle - \langle M^{\mathsf{T}}\bs{v}^{(1)}_k, \x\rangle + \frac{1}{2} \langle \bs{v}^{(1)}_k, D \bs{v}^{(1)}_k \rangle\right)\exp(\langle \bs{v}^{(1)}_k, \x \rangle)
\end{align}
and therefore
\begin{align} \label{intermediatenew2}
    G_k(\x,2\Delta t) &= \exp(\langle \bs{v}^{(1)}_k, \x \rangle)\exp\left(\Delta t\left(k \langle \bs{\eta}, \x\rangle - \langle M^{\mathsf{T}}\bs{v}^{(1)}_k, \x\rangle\right)\right)\exp\left(\frac{1}{2} \Delta t\langle \bs{v}^{(1)}_k, D \bs{v}^{(1)}_k \rangle\right)\nonumber \\
    &= \exp(\langle \bs{v}^{(2)}_k, \x \rangle) \exp\left(\frac{1}{2} \Delta t\langle \bs{v}^{(1)}_k, D \bs{v}^{(1)}_k \rangle\right),
\end{align}
where 
\begin{equation}\label{equationnnn}
    \bs{v}^{(2)}_k = \bs{v}^{(1)}_k + \Delta t \left(k\bs{\eta} - M^{\mathsf{T}}\bs{v}^{(1)}_k\right). 
\end{equation}
Generalizing this procedure to $m$ steps, it is easy to see that we have
\begin{equation} \label{genlingen}
    G_k(\x,m \Delta t) = \exp(\langle \bs{v}^{(m)}_k, \x \rangle)\exp\left(\frac{1}{2} \sum_{i = 0}^{m - 1}\Delta t\langle \bs{v}^{(i)}_k, D \bs{v}^{(i)}_k \rangle  \right),
\end{equation}
where $\bs{v}^{(m)}_k$ satisfies 
\begin{equation} \label{eq1}
    \bs{v}^{(m)}_k = \bs{v}^{(m-1)}_k + \Delta t \left(k\bs{\eta} - M^{\mathsf{T}}\bs{v}^{(m-1)}_k\right).
\end{equation}
This can also be shown explicitly using an induction step, which we omit given that this step is essentially the same as the calculation done already in obtaining (\ref{intermediatenew2}) with (\ref{equationnnn}). We note that the result (\ref{eq1}) holds also for $\bs{v}^{(1)}_k$ since $\bs{v}^{(0)}_k = \bs{0}$. \par At this point, we can rewrite (\ref{eq1}) as 
\begin{equation} \label{difference1}
    \frac{\bs{v}^{(m)}_k - \bs{v}^{(m-1)}_k}{\Delta t} = k\bs{\eta} - M^{\mathsf{T}}\bs{v}^{(m-1)}_k.
\end{equation}
Writing now $\bs{v}_k(i\Delta t) = \bs{v}_k^{(i)}$ and recalling that $\Delta t = t/n$ we have  
\begin{equation}
    G_k(\x,n\Delta t) = \exp(\langle\bs{v}_k(t), \x\rangle) \exp\left(\frac{1}{2} \sum_{i = 0}^{n - 1}\Delta t\langle \bs{v}_k(i\Delta t), D \bs{v}_k(i\Delta t) \rangle  \right).
\end{equation}
Taking the continuum limit, $n \rightarrow \infty$ and $\Delta t\rightarrow 0$, we then obtain the final result
\begin{equation} \label{GKlinear}
    G_k(\x,t) = \exp(\langle\bs{v}_k(t), \x\rangle)\exp\left(\frac{1}{2} \int_0^t \langle \bs{v}_k(s), D \bs{v}_k(s) \rangle ds \right),
\end{equation}
with $\bs{v}_k(t)$ satisfying the differential equation 
\begin{equation} \label{differential1}
    \frac{d \bs{v}_k(t)}{dt} = k\bs{\eta} - M^{\mathsf{T}}\bs{v}_k(t),
\end{equation}
with initial condition $\bs{v}_k(0) = \bs{0}$, which represents the continuous time limit of (\ref{difference1}). 
\par The result (\ref{GKlinear}) gives the generating function of $A_t$ for any time $t$ via the differential equation (\ref{differential1}) and the initial condition on $\bs{v}_k$. To find the SCGF $\lambda(k)$, we note that the differential equation (\ref{differential1}) has a stationary solution $\bs{v}_k^*$ satisfying 
\begin{equation}
    \frac{d \bs{v}_k^*}{dt} = \bs{0}
\end{equation}
and given explicitly by
\begin{equation} \label{wkstat}
    k\bs{\eta} - M^{\mathsf{T}}\bs{v}_k^* = \bs{0} \implies \bs{v}_k^* = k (M^{\mathsf{T}})^{-1} \bs{\eta}. 
\end{equation}
This long time stationary solution is attained if the differential equation (\ref{differential1}) is stable, which requires that the matrix $M$ be positive definite. Given that this is the case by assumption, we then have $\bs{v}_k(t) \rightarrow \bs{v}_k^*$ as $t \rightarrow \infty$. As a result, we have from the definition (\ref{SCGF}) of $\lambda(k)$ that 
\begin{align} \label{SCGFlin}
   \lambda(k) &= \lim_{t \rightarrow \infty}\frac{1}{t} \ln G_k(\x,t) 
   \nonumber \\ &= \lim_{t\rightarrow \infty} \frac{1}{t} \left[\langle \bs{v}_k(t), \x \rangle + \frac{1}{2} \int_0^t \langle \bs{v}_k(s), D \bs{v}_k(s) \rangle ds \right] \nonumber \\
   &= \lim_{t \rightarrow \infty} \frac{1}{2t} \int_0^t \langle \bs{v}_k(s), D \bs{v}_k(s) \rangle ds \nonumber \\ &= \frac{1}{2} \langle \bs{v}_k^*, D \bs{v}_k^* \rangle. 
\end{align}
The last step follows because the transient part of the evolution of $\bs{v}_k(s)$ does not contribute to $\lambda(k)$, since this part of the integral vanishes when normalized by $1/t$ in the limit where $t \rightarrow \infty$. Only the stationary part of the evolution contributes, since once $\bs{v}_k(t) \approx \bs{v}_k^*$ the above integral scales as $\mathcal{O}(t)$ and so attains a finite value when normalized with $1/t$ in the limit $t \rightarrow \infty$. Using the form of $\bs{v}_k^*$ shown in (\ref{wkstat}) we then find 
\begin{equation} \label{SCGFlin2}
    \lambda(k) = \frac{k^2}{2} \left\langle\left(M^{\mathsf{T}}\right)^{-1}\bs{\eta}, D\left(M^{\mathsf{T}}\right)^{-1} \bs{\eta}  \right\rangle.
\end{equation}
Since the expression of $\lambda(k)$ is quadratic in $k$, we immediately conclude that the rate function $I(a)$, found as the Legendre transform of the SCGF, is quadratic in $a$ and is in fact equal to 
\begin{equation} \label{ratelin}
    I(a) = \frac{a^2}{2 \left\langle\left(M^{\mathsf{T}}\right)^{-1}\bs{\eta}, D\left(M^{\mathsf{T}}\right)^{-1} \bs{\eta}  \right\rangle}.
\end{equation}
Moreover, it is clear from (\ref{Gklim}) that the eigenfunction $r_k$ associated with $\lambda(k)$ has the form
\begin{equation}
    r_k(\x) = \exp(\langle \bs{v}_k^*,\x\rangle).
\end{equation}
It can be verified explicitly that $r_k$ satisfies the spectral equation given in (\ref{spec1}) for $\mathcal{L}_k$ with $\lambda(k)$ as given in (\ref{SCGFlin}).
\par From the expression of $r_k$, we can study the form of the effective process, as defined in Sec.~\ref{sec:sec27}. From the expression (\ref{effdrift}) for the drift of this process, we find that 
\begin{equation}
    \bs{F}_k(\x) = -M \left(\x -  M^{-1} D \bs{v}_k^*\right). 
\end{equation}
As a result, we see that the effective process is also a linear process with the same drift matrix $M$ as the original process, but with a fixed point in the drift which is now located at 
\begin{equation}
    \x_k = M^{-1} D \bs{v}_k^*
\end{equation}
instead of being located at $\x = \bs{0}$. Its stationary density $p_k^*$ is therefore the same as the original process, except for the translation $\x \rightarrow \x - \x_k$ and is obtained from (\ref{statden}) under this substitution. The stationary density $p_k^*$ of this process is therefore 
\begin{align}
    &p_k^*(\x) = \sqrt{\frac{1}{(2\pi)^n \tn{det}C}} \exp\left(\left\langle \x - \x_k, C^{-1}\left( \x - \x_k \right)\right\rangle \right),
\end{align}
with $C$ being the stationary covariance matrix of the original process. Similarly the stationary current associated with the effective process is simply a translated version of the stationary current of the original process and is obtained from (\ref{statcurlin}) via the substitution $\x \rightarrow \x - \x_k$, giving 
\begin{equation}
    \bs{J}_{\bs{F}_k,p^*_k}(\x) = \left(\frac{D}{2}C^{-1} - M \right)(\x - \x_k) p_k^*(\x).
\end{equation}
In other words, for this type of observable, the stationary density and current for the effective process can be obtained simply by the substitution $\x \rightarrow \x - \x_k$ in the stationary density and current of the original process. 
\par The calculation performed here can also be extended easily to include the case where the drift matrix $M$, diffusion matrix $D$ and $\bs{\eta}$ are explicitly time-dependent. The result in this case is 
\begin{equation}
    G_k(\x,t) = e^{\langle \bs{v}_k(t),\x \rangle} \exp\left(\frac{1}{2} \int_0^t\langle\bs{v}_k(s), D(s) \bs{v}_k(s) \rangle  \, ds \right) 
\end{equation}
where the vector $\bs{v}_k(t)$ now satisfies 
\begin{equation}
    \frac{d \bs{v}_k(t)}{dt} = k\bs{\eta}(t) - M^{\mathsf{T}}(t) \bs{v}_k(t), \quad \bs{v}_k(0) = 0. 
\end{equation}

\subsection{Quadratic additive observable}\label{sec:sec412}
We now consider the class of quadratic observables defined as in (\ref{quadadd}) and assume that $Q$ is a symmetric $n \times n$ matrix. This can be done without loss of generality since only the symmetric part of $Q$ given by 
\begin{equation}
    Q^{+}=\frac{Q + Q^{\mathsf{T}}}{2}
\end{equation}
contributes to quadratic forms, so that 
\begin{equation}
    \langle \x ,Q \x \rangle = \langle \x , Q^{+} \x \rangle. 
\end{equation}
This follows because the antisymmetric part of $Q$ given by 
\begin{equation}
    Q^{-} = \frac{Q - Q^{\mathsf{T}}}{2}
\end{equation}
is such that
\begin{equation}
    \langle \x, Q^{-}\x\rangle = x_i \left(Q^{-}\right)_{ij} x_j = - x_i \left(Q^{-}\right)_{ji} x_j = -\langle \x,Q^{-}\x\rangle ,
\end{equation}
so that $ \langle \x, Q^{-}\x\rangle =0$. Note our use of the Einstein summation convention above.
\par To obtain the generating function of quadratic observables for linear diffusions, we repeat the calculation for the previous observable using now the tilted generator 
\begin{equation}
    \mathcal{L}_k = -M\x \cdot \bs{\nabla} + \frac{1}{2} \bs{\nabla}\cdot D \bs{\nabla} + k\langle \x,Q \x\rangle.
\end{equation}
associated with the quadratic observable (\ref{quadadd}). Rather than explicitly going through all the steps that we presented previously, we here simply prove the desired result via induction. Our induction hypothesis is that 
\begin{equation} \label{inductionmain1}
    G_k(\x, m \Delta t) = \exp\left(\left\langle \x, B^{(m)}_k\x \right\rangle\right) \exp \left(\sum_{i=0}^{m-1}\Delta t \tn{Tr}\left(DB^{(i)}_k\right)\right)
\end{equation}
where 
\begin{equation} \label{induction1}
    B^{(m)}_k = B^{(m-1)}_k + \Delta t\bigg(2 B^{(m-1)}_k D B^{(m-1)}_k - (M^{\mathsf{T}}B^{(m-1)}_k + B^{(m-1)}_k M) + kQ\bigg),
\end{equation}
with $B^{(i)}_k$ being symmetric for all $i$ and $B^{(0)}_k = 0$. The induction hypothesis stated by (\ref{inductionmain1}) and (\ref{induction1}) is only claimed to be valid up to first order in $\Delta t$: we neglect all higher order contributions given that we eventually take the continuum limit $\Delta t \rightarrow 0$. 
\par It can be seen that $B^{(i)}_k$ satisfying (\ref{induction1}) must be symmetric for all $i$ in the following manner: from the above we have that 
\begin{equation}
    B^{(1)}_k = kQ,
\end{equation}
which is symmetric. But then $B^{(2)}_k$ must also be symmetric, since all terms on the RHS of (\ref{induction1}) are symmetric, including the term
\begin{equation}
    (M^{\mathsf{T}}B^{(1)}_k + B^{(1)}_k M) = \left(M^{\mathsf{T}}B^{(1)}_k\right) + \left(M^{\mathsf{T}}B^{(1)}_k\right)^{\mathsf{T}}.
\end{equation}
The same logic can then be applied to show that $B^{(i)}_k$ is symmetric for all $i \geq 3$. 
\par Proceeding now with the induction proof, we first we show that (\ref{inductionmain1}) and (\ref{induction1}) holds for $m = 1$. We have 
\begin{align}
    G_k(\x,\Delta t) &= (1 + \Delta t \mathcal{L}_k) 1 \nonumber \\&= (1 + \Delta t \,k\langle \x, Q \x\rangle )1 \nonumber \\ &= e^{\langle \x, B^{(1)}_k \x\rangle},
\end{align}
where $B^{(1)}_k = kQ.$ Clearly, since $B^{(0)}_k = 0$, (\ref{induction1}) is satisfied and (\ref{inductionmain1}) holds for $m =1$. 
\par Before proceeding further with the induction argument we first note that 
\begin{equation}\label{eq3}
    \bs{\nabla}\exp(\langle \x, A\x\rangle) = 2 A\x \exp(\langle \x, A\x\rangle)
\end{equation}
for any symmetric $n\times n$ matrix $A$. Further, for such $A$ we also have 
\begin{align}\label{eq4}
    \bs{\nabla}\cdot D\bs{\nabla}\exp(\langle \x, A\x\rangle) &= 4\langle A\x, DA\x\rangle \exp(\langle \x, A\x\rangle) \nonumber \\ &= 4 \langle \x, ADA\x\rangle \exp(\langle \x, A\x\rangle),
\end{align}
where we have used the symmetry of $A$ in the last step. 
\par We now proceed with the induction argument. Assuming that (\ref{inductionmain1}) holds for $m = j$, we will then show that it holds for $m = j + 1$. We have 
\begin{align}
    G_k(\x,(j+1) \Delta t) &= (1 + \Delta t \mathcal{L}_k)G_k(\x,j\Delta t) \nonumber \\ &=(1 + \Delta t \mathcal{L}_k)\exp(\langle \x, B^{(j)}_k\x \rangle) \exp \left(\sum_{i=0}^{j-1}\Delta t \tn{Tr}\left(DB^{(i)}_k\right)\right)
\end{align}
by assumption. Now,
\begin{align}\label{intermediate}
    \mathcal{L}_k \exp(\langle \x, B^{(j)}_k\x \rangle) &= \bigg(-M\x \cdot \bs{\nabla} + \frac{1}{2} \bs{\nabla}\cdot D \bs{\nabla} + k\langle \x,Q \x\rangle\bigg)\exp(\langle \x, B^{(j)}_k\x \rangle)\nonumber \\&=\bigg(-2\langle \x, M^{\mathsf{T}}B^{(j)}_k\x \rangle + 2\langle \x, B^{(j)}_k D B^{(j)}_k\x \rangle +k\langle \x, Q \x \rangle + \tn{Tr}\left(DB^{(j)}_k\right)\bigg) \nonumber\\ &\quad \times\exp(\langle \x, B^{(j)}_k\x \rangle)
\end{align}
where we have used (\ref{eq3}) and (\ref{eq4}). Furthermore, since only the symmetric part of a matrix contributes inside an inner product as discussed earlier, it is possible to write
\begin{equation}
    \langle \x, M^{\mathsf{T}}B^{(j)}_k\x \rangle = \frac{1}{2}\left\langle\x, \left(M^{\mathsf{T}}B^{(j)}_k + B^{(j)}_k M\right)\x\right\rangle
\end{equation}
so that (\ref{intermediate}) becomes 
\begin{align}
     \mathcal{L}_k \exp(\langle \x, B^{(j)}_k\x \rangle) &= \bigg(-\left\langle\x, \left(M^{\mathsf{T}}B^{(j)}_k + B^{(j)}_k M\right)\x\right\rangle  + 2\langle \x, B^{(j)}_k D B^{(j)}_k\x \rangle \nonumber \\ &\quad  +k\langle \x, Q \x \rangle  + \tn{Tr}\left(DB^{(j)}_k\right)\bigg) \exp(\langle \x, B^{(j)}_k\x \rangle).
\end{align}
We therefore have 
\begin{align} \label{Gkquaddisc}
    G_k(\x,(j+1)\Delta t) &= \left(1 + \Delta t \mathcal{L}_k\right)G_k(\x,j\Delta t)\nonumber \\&=\bigg[1 +\Delta t\bigg(-\left\langle\x, \left(M^{\mathsf{T}}B^{(j)}_k + B^{(j)}_k M\right)\x\right\rangle  + 2\langle \x, B^{(j)}_k D B^{(j)}_k\x \rangle +\nonumber \\ &\quad k\langle \x, Q \x \rangle   + \tn{Tr}\left(DB^{(j)}_k\right)\bigg) \bigg]\exp(\langle \x, B^{(j)}_k\x \rangle) \exp\left(\sum_{i=0}^{j-1} \tn{Tr}\left(DB^{(i)}_k\right)\right)\nonumber \\ &= \exp(\langle \x, B^{(j+1)}_k\x \rangle)\exp\left(\sum_{i=0}^{j} \tn{Tr}\left(DB^{(i)}_k\right)\right),
\end{align}
with $B^{(j+1)}_k$ satisfying 
\begin{equation} \label{Akquaddisc}
    B^{(j+1)}_k = B^{(j)}_k + \Delta t\bigg(2 B^{(j)}_k D B^{(j)}_k - (M^{\mathsf{T}}B^{(j)}_k + B^{(j)}_k M^{\mathsf{T}}) + kQ\bigg),
\end{equation}
so that (\ref{Gkquaddisc}) and (\ref{Akquaddisc}) corresponds to (\ref{inductionmain1}) and (\ref{induction1}), respectively, for $m = j +1$. The induction claim is therefore proven. 
\par Taking now the continuum limit in $G_k(\x,n\Delta t)$, we obtain 
\begin{equation} \label{Gkquad}
    G_k(\x,t) = \exp(\langle \x, B_k(t) \x\rangle) \exp\left(\int_0^t \tn{Tr}(DB_k(s)) ds\right),
\end{equation}
with $B_k(t)$ satisfying the differential Ricatti equation 
\begin{equation} \label{Akquaddiff}
    \frac{dB_k(t)}{dt} = 2 B_k(t) D B_k(t) - (M^{\mathsf{T}}B_k(t) + B_k(t)M) + kQ, \quad B_k(0) = 0,
\end{equation}
which represents the continuous time limit of (\ref{Akquaddisc}). The stationary solution $B_k^*$ for this differential equation satisfies an algebraic Riccati equation 
\begin{equation} \label{Akquadricc}
    2B_k^* D B_k^* - (M^{\mathsf{T}}B_k^* + B_k^*M) + kQ = 0.
\end{equation}
In general this equation has multiple possible solutions, with the correct solution found as that $B_k^*$ which also satisfies $B_0^* = 0$, since for $k = 0$ we have $G_{k=0}(\x,t) = 1$ for all $\x$ and $t$. This can also be seen explicitly in the differential equation (\ref{Akquaddiff}): if $k = 0$ then, given the initial condition $B_k(0) = 0$, we have that $B_k(t)$ remains zero for all times $t$. 
\par In the event that $B_k(t)$ converges to this appropriate solution $B_k^*$ we find, using similar reasoning as that employed in the long-time limit for the linear additive observable, that 
\begin{equation} \label{quadgenlim}
    G_k(\x,t) \rightarrow e^{\langle\x, B_k^* \x\rangle} e^{t \tn{Tr}(DB_k^*)}
\end{equation}
so that the SCGF is found from (\ref{SCGF}) to be 
\begin{equation}\label{quadscgf}
    \lambda(k) = \tn{Tr}(DB_k^*).
\end{equation}
From this expression, we can obtain the rate function $I(a)$ by Legendre transform. The result is not as explicit as for linear additive observables because $B_k^*$ must now be found first by solving the algebraic Riccati equation (\ref{Akquadricc}). 
\par To find the effective process associated with the fluctuations of the quadratic observable, we note from (\ref{quadgenlim}) that the eigenfunction $r_k$ of $\mathcal{L}_k$ is given (up to multiplication by a constant) by
\begin{equation} \label{quadrk}
    r_k(\x) = \exp(\langle \x, B_k^*\x\rangle).
\end{equation}
It can be checked that this solves the spectral equation (\ref{spec1}) with the eigenvalue given in (\ref{quadscgf}). Moreover from (\ref{effdrift}) we find 
\begin{equation} \label{effdriftquad}
    \bs{F}_k(\x) = -M_k \x,
\end{equation}
where
\begin{equation}\label{mkquad}
    M_k = M - 2DB_k^*
\end{equation}
is the modified drift matrix. Consequently we see again that the effective process is a linear diffusion. For this process to be ergodic, $M_k$ must be positive definite. In this case, and noting that $D$ is positive definite by assumption, the stationary density for the effective process 
with drift (\ref{effdriftquad}) and appropriate $B_k^*$ is given in accordance with (\ref{statden}) by 
\begin{equation} \label{rhoquad}
p_k^{*}(\x) = \sqrt{\frac{1}{(2\pi)^n \tn{det}C_k}}\exp\left(-\frac{1}{2} \langle \x, C_k^{-1} \x\rangle \right),
\end{equation}
with $C_k$ the stationary covariance matrix satisfying now the Lyapunov equation
\begin{align} \label{ricccovquad}
    D &= M_k C_k + C_k M_k^{\mathsf{T}} \nonumber \\
    &= (M-2DB_k^*) C_k + C_k(M-2DB_k^*)^{\mathsf{T}},
\end{align}
similar to (\ref{ricccov}). We note also that since $p_k^* = l_k r_k$, we have that the eigenfunction $l_k$ corresponding to the eigenvalue $\lambda(k)$ is given (up to a normalization constant) by 
\begin{equation} \label{lkquad}
    l_k(\x) = \exp\left(-\frac{1}{2} \langle \x, (C_k^{-1} + 2B_k^*) \x \rangle \right).
\end{equation}
It can be verified explicitly that this is indeed an eigenfunction of $\mathcal{L}_k^{\dagger}$ with eigenvalue $\lambda(k)$. Finally, the stationary current $\bs{J}_{\bs{F}_k, p^*_k}$ for the effective process is given by 
\begin{equation} \label{Jquad}
    \bs{J}_{\bs{F}_k, p^*_k}(\x) = \left(\frac{D}{2} C_k^{-1} - M +2DB_k^*\right)\x p_k^*(\x).
\end{equation}
We observe that here both the stationary density and current are modified in a non-trivial way, in contrast with the case where the linear additive observable was considered. 
\par In the event that the differential equation (\ref{Akquaddiff}) cannot be solved exactly and the explicit convergence of $B_k(t)$ to the solution $B_k^*$ satisfying $B_0^* = 0$ and such that $M_k$ is positive definite is difficult to confirm, we can reason as follows. Assume that $B_k^*$ is a solution to the algebraic Riccati equation (\ref{Akquadricc}) that is continuous in $k$ and which satisfies $B_0^* = 0$. Furthermore, assume that this $B_k^*$ is such that the matrix $M_k$ as defined in (\ref{mkquad}) is positive definite. It is known~\cite{abou2012matrix} that the algebraic Riccati equation (\ref{Akquadricc}) has (at most) one such solution, so that the $B_k^*$ satisfying these conditions is unique, if it exists. The condition that $M_k$ is positive definite ensures that the Lyapunov equation (\ref{ricccovquad}) has a unique positive definite solution $C_k$. If, in addition, this $C_k$ is such that $C_k^{-1} + 2B_k^*$ is positive definite, then it can be shown that $r_k$ as in (\ref{quadrk}) and $l_k$ as in (\ref{lkquad}) are eigenfunctions of $\mathcal{L}_k$ and $\mathcal{L}_k^{\dagger}$, respectively, with eigenvalue $\tn{Tr}(DB_k^*)$ and that these eigenfunctions are normalizable in the sense of (\ref{spec3}) and (\ref{spec4}), therefore constituting valid solutions to the spectral problem (\ref{spec1}) and (\ref{spec2}). It then follows that $\tn{Tr}(DB_k^*)$ is an eigenvalue of $\mathcal{L}_k$ which is continuous and corresponds to $0$ for $k =0$. As such, this eigenvalue must be the SCGF and we have that the generating function has the long-time form (\ref{quadgenlim}). The effective drift, density and current can be obtained as before. For those $B_k^*$ such that $M_k$ is not positive definite the Lyapunov equation (\ref{ricccovquad}) will not have a positive definite solution $C_k$ and as such the corresponding eigenfunctions $l_k$ and $l_k r_k$, formally defined as in (\ref{quadrk}) and (\ref{lkquad}), will not constitute valid eigenfunctions of the spectral problem associated with the SCGF, given that $l_k r_k$ will not be normalizable. 
\par As for the linear additive observable, it is again possible to obtain an explicit form for the generating function $G_k(\x,t)$ for a time dependent $Q= Q(t)$ and an explicitly time-dependent SDE. The result for the moment generating function is then
\begin{equation}
    G_k(\x,t) = \exp(\langle \x, B_k(t) \x\rangle) \exp\left(\int_0^t \tn{Tr}\left(D(s) B_k(s)\right) ds\right)
\end{equation}
with 
\begin{align}
    \frac{dB_k(t)}{dt} &= 2 B_k(t) D(t) B_k(t) - (M^{\mathsf{T}}(t)B_k(t) + B_k(t)M(t)) + kQ(t) 
\end{align}
and initial condition $B_k(0) = 0$.
\subsection{Linear current-type observable}\label{sec:sec413}
We conclude our study by considering linear current-type observables $A_T$, as defined in (\ref{lincurobs}), which involve an $n \times n $ matrix $\Gamma$. We first address the case where $\Gamma$ is purely antisymmetric so that $\Gamma = -\Gamma^{\mathsf{T}}$. The case where $\Gamma$ also has a non-zero symmetric part is more involved and is therefore treated separately after. 
\subsubsection{Antisymmetric $\Gamma$}
For the observable $A_T$ given in (\ref{lincurobs}) and with $\Gamma$ assumed to be purely antisymmetric, the associated tilted generator $\mathcal{L}_k$ is given by 
\begin{equation}
    \mathcal{L}_k = -k\langle M\x,\Gamma\x\rangle + (-M + kD\Gamma)\x\cdot\bs{\nabla} + \frac{1}{2} \bs{\nabla}\cdot D \bs{\nabla} + \frac{k^2}{2} \langle \Gamma\x,D\Gamma\x\rangle.
\end{equation}
This can be written in a slightly more convenient form as 
\begin{equation} \label{tiltedcurrentoperator}
    \mathcal{L}_k = -\frac{k}{2}\langle\x, (M^{\mathsf{T}}\Gamma - \Gamma M)\x\rangle +  (-M + kD\Gamma)\x\cdot\bs{\nabla} + \frac{1}{2} \bs{\nabla}\cdot D \bs{\nabla} + \frac{k^2}{2} \langle \x, \Gamma^{\mathsf{T}}D\Gamma\x\rangle,
\end{equation}
given that 
\begin{equation}
    \langle M\x,\Gamma\x\rangle = \langle \x,M^{\mathsf{T}}\Gamma\x\rangle = \frac{1}{2}\langle\x, (M^{\mathsf{T}}\Gamma +  \Gamma^{\mathsf{T}}M)\x\rangle = \langle\x, (M^{\mathsf{T}}\Gamma - \Gamma M)\x\rangle,
\end{equation}
where we have used the antisymmetry of $\Gamma$ in the last line. 
\par The solution of the Feynman-Kac equation with the tilted generator (\ref{tiltedcurrentoperator}) can be found, similarly to the case of the quadratic additive observable, by induction. The calculation in this case is similar and is therefore presented in App.~\ref{appendixC}. The final result for the generating function is 
\begin{equation}\label{GKcur}
    G_k(\x,t) = \exp(\langle \x, B_k(t) \x\rangle) \exp\left(\int_0^t \tn{Tr}(DB_k(s)) ds\right),
\end{equation}
with $B_k(t)$ satisfying the differential Riccati equation
\begin{align} \label{riccatidiffcur}
    \frac{dB_k(t)}{dt} &= \frac{k^2}{2} \Gamma^{\mathsf{T}}D\Gamma \,-\, \frac{k}{2}(M^{\mathsf{T}}\Gamma-\Gamma M) \,+ \,(-M + kD\Gamma)^{\mathsf{T}}B_k(t) \nonumber \\& \quad + B_k(t)(-M + kD\Gamma) + 2B_k(t) DB_k(t),
\end{align}
with initial condition $B_k(0) = 0$. A similar result was obtained by path-integral methods for a particular type of linear current-type observable, namely the nonequilibrium work, by Kwon, Noh and Park~\cite{kwon2011nonequilibrium}. These authors did not explicitly consider the stationary solution of (\ref{riccatidiffcur}), instead obtaining the long-time behavior for a specific system via numerical integration. We now study the general properties of the stationary solution analytically and obtain, as a consequence, the SCGF and effective process.
\par The stationary solution $B_k^*$ corresponding to (\ref{riccatidiffcur}) satisfies in this case 
\begin{align} \label{ricccur}
    0 &=\frac{k^2}{2} \Gamma^{\mathsf{T}}D\Gamma \,-\, \frac{k}{2}(M^{\mathsf{T}}\Gamma-\Gamma M) \,+ \,(-M + kD\Gamma)^{\mathsf{T}}B_k^* + B_k^*(-M + kD\Gamma) \nonumber \\ &\quad + 2B_k^* DB_k^*.
\end{align}
with $B_0^* = 0$. It is interesting to note that the form of the generating function obtained here is the same as that obtained in (\ref{Gkquad}) for the quadratic additive observable, with only the differential Riccati equation satisfied by $B_k(t)$ being different. 
\par From this result, we follow the same logic as in the previous sections and obtain 
\begin{equation} \label{curscgf}
    \lambda(k) = \tn{Tr}(DB_k^*).
\end{equation}
for the SCGF and 
\begin{equation} \label{currk}
    r_k(\x) = \exp(\langle \x, B_k^* \x\rangle)
\end{equation}
for the associated eigenfunction (up to a multiplicative constant). As a result, we find that the effective drift $\bs{F}_k$ given from (\ref{effdrift}) is a linear vector field
\begin{equation} \label{effdriftcur}
    \bs{F}_k(\x) = -M_k\x,
\end{equation}
where now the drift matrix $M_k$ is given by 
\begin{equation}\label{mkcur}
    M_k = M - 2DB_k^* - k D \Gamma.
\end{equation}
\par For those $k$ for which $M_k$ is positive definite we have that the effective process is ergodic and large deviations exist. The stationary density $p^*_k$ has then the same form as (\ref{rhoquad}), and is given as \begin{equation} \label{statdencur}
    p^*_k(\x) = \sqrt{\frac{1}{(2\pi)^n\tn{det}C_k}} \exp\left(-\frac{1}{2}\langle\x,C_k^{-1}\x\rangle \right),
\end{equation}
with the stationary covariance matrix $C_k$ satisfying here the Lyapunov equation 
\begin{align} \label{ricccovcur}
    D &= M_k C_k + C_k M_k^{\mathsf{T}} \nonumber \\ 
    &= (M - 2DB_k^* - kD\Gamma)C_k + C_k(M - 2DB_k^* - kD\Gamma)^{\mathsf{T}}.
\end{align}
For the stationary current $\bs{J}_{\bs{F}_k,p_k^*}$ associated with the effective process we obtain
\begin{equation}
    \bs{J}_{\bs{F}_k,p^*_k}(\x) = \left(\frac{D}{2} C_k^{-1} - M +2DB_k^* + kD\Gamma\right)\x p_k^*(\x).
\end{equation}
\par It will prove useful later to have an expression also for the eigenfunction $l_k$ associated with the adjoint operator $\mathcal{L}_k^{\dagger}$ and corresponding to the eigenvalue $\lambda(k)$. To obtain this eigenfunction we recall, as in the previous section, the relation $p_k^*(\x) =l_k(\x) r_k(\x)$ from which it follows that $l_k$ has the same form as in (\ref{lkquad}). Noting that we must apply the normalization conditions (\ref{spec3}) and (\ref{spec4}) and given that the density (\ref{statdencur}) is already properly normalized, we can therefore write  
\begin{equation} \label{normlk}
    l_k(\x) = \mathcal{N} \sqrt{\frac{1}{(2\pi)^n\tn{det}C_k}} \exp\left(-\frac{1}{2}\langle\x,\left(C_k^{-1}+2B_k^*\right)\x\rangle \right) ,
\end{equation}
with $\mathcal{N}$ a normalization constant such that (\ref{spec4}) holds. The corresponding (properly normalized) $r_k$ can then be written as 
\begin{equation} \label{normrk}
    r_k(\x) = \mathcal{N}^{-1} \exp\left(\langle \x, B_k^* \x\rangle \right).
\end{equation}
\par The exact expression for the normalization constant $\mathcal{N}$ is not necessary for us since all $r_k$ proportional to $\exp(\langle\x,B_k^*\x\rangle)$ produce the same effective drift. This can be seen simply by noting that the contribution to the effective drift owing to the eigenfunction $r_k$ is given by 
\begin{equation}
    D\bs{\nabla}\ln r_k(\x) = D\bs{\nabla}\ln \left(\mathcal{M}\exp(\langle \x, B_k^*\x\rangle \right) = 2DB_k^*\x
\end{equation}
for all constants $\mathcal{M}$. As such all of the results for the effective drift and the associated stationary current and density hold true even though we did not use a properly normalized $r_k$. 
\par Finally, to obtain expressions for (\ref{GKcur}) and (\ref{riccatidiffcur}) in the event of time-dependent $\Gamma(t), M(t)$, and $\sigma(t)$, we can simply substitute the time dependent quantities in the place of their formerly time-independent versions in the expressions (\ref{GKcur}) and (\ref{riccatidiffcur}), as seen in the previous sections. Note that in order for the expression (\ref{GKcur}) to generalize in this straightforward manner to the time-dependent case, we must demand that $\Gamma(t)$ featuring then in the definition of the current-type observable must remain antisymmetric for all times $t$. 

\subsubsection{General $\Gamma$}
We now address the case where the matrix $\Gamma$ featuring in the definition (\ref{lincurobs}) of the observable $A_T$ has a non-zero symmetric component. In this case we write 
\begin{equation}
    \Gamma = \Gamma^{-} + \Gamma^{+},
\end{equation}
where $\Gamma^{-} = (\Gamma - \Gamma^{\mathsf{T}})/2$ is the antisymmetric component and $\Gamma^{+} = (\Gamma + \Gamma^{\mathsf{T}})/2$ the symmetric component of $\Gamma$. With this decomposition, we write $A_T$ as 
\begin{align} \label{genlinobs}
    A_T  = A_T^{-} + A_T^{+},
\end{align}
where  
\begin{equation}
    A_T^{\pm} = \frac{1}{T} \int_0^T \Gamma^{\pm} \bs{X}(t)\circ d\bs{X}(t),
\end{equation}
so that $A_T^{-}$ represents the part of the observable associated with the antisymmetric $\Gamma^{-}$, while $A_T^{+}$ represents the symmetric part of the observable.
\par Using the fact that 
\begin{equation}
    \frac{1}{2}\bs{\nabla}\langle \x, \Gamma^{+} \x \rangle = \Gamma^{+}\x
\end{equation}
and given that the Stratonovich convention used for the definition of the observable preserves the standard rules of calculus as in (\ref{strato2}), we have that the symmetric part $A_T^{+}$ can be written as 
\begin{align}
    A_T^{+} &= \frac{1}{T}\int_0^T \frac{1}{2}\bs{\nabla}\left\langle \bs{X}(t), \Gamma^{+} \bs{X}(t) \right\rangle \circ d\bs{X}(t) \nonumber \\ &= \frac{1}{2T} \left\langle \bs{X}(T),\Gamma^{+} \bs{X}(T)\right\rangle -\frac{1}{2T} \left\langle\bs{X}(0),\Gamma^{+}\bs{X}(0)\right\rangle,
\end{align}
so that this observable reduces to a difference of boundary terms. Based on this we might naively think that this part of the observable will not contribute in a meaningful way to the large deviations given that the factor $1/T$ will make the boundary terms vanish as $T \rightarrow \infty$. However we will see that this is not exactly the case and that these boundary terms can in fact affect the large deviations.
\par To see this, we go back to the full observable $A_T$ and its decomposition (\ref{genlinobs}) to write the generating function as 
\begin{align}
    G_k(\x,t) = \mathbb{E}_{\x}\left[\exp\left(k t A_T^{-} + \frac{k}{2}\langle \bs{X}(t),\Gamma^{+} \bs{X}(t) \rangle - \frac{k}{2}\langle \bs{X}(0), \Gamma^{+}\bs{X}(0)\rangle \right)\right].
\end{align}
We can write the above expectation value in the form of an integral over all possible values of the final point $\bs{X}(t) = \bs{y}$, as done in (\ref{Gkendint}), and since we also have $\bs{X}(0) = \x$, we can write
\begin{equation} \label{genint}
    G_k(\x,t) = \int d\bs{y} \exp\left(\frac{k}{2}\langle\bs{y},\Gamma^{+}\bs{y}\rangle - \frac{k}{2}\langle \bs{x},\Gamma^{+}\bs{x}\rangle \right) G_k^{-}(\x,\bs{y},t),
\end{equation}
where 
\begin{equation}
    G_k^{-}(\x,t) = \mathbb{E}_{\x}\left[e^{ktA_T^{-}} \right],
\end{equation}
denotes the generating function associated with $A_T^{-}$, and
\begin{equation}
    G_k^{-}(\x,\bs{y},t) = \mathbb{E}_{\x}\left[\delta(\bs{X}(t) - \bs{y})e^{ktA_T^{-}} \right] 
\end{equation}
is the corresponding end-point function (\ref{endpointfunc}).
\par In order to determine the behavior of the generating function (\ref{genint}) as $t$ becomes large, we must determine the asymptotic behavior of $G_k^{-}(\x,\bs{y},t)$. From the discussion leading up to (\ref{fixedpointasymp}) we know that the end-point function satisfies the asymptotic form 
\begin{equation}
    G_k^{-}(\x,\bs{y},t) \rightarrow K e^{\lambda(k)t}\, l_k(\bs{y}) r_k(\x) 
\end{equation}
as $t \rightarrow \infty$, with $r_k$ and $l_k$ properly normalized. Using the expressions (\ref{normlk}) and (\ref{normrk}) derived for the eigenfunctions of the antisymmetric current-type observable, we then obtain 
\begin{align}
    G_k^{-}(\x,\bs{y},t) &\rightarrow K e^{\lambda(k)t} \sqrt{\frac{1}{(2\pi)^n\tn{det}C_k}} \exp\left(-\frac{1}{2}\langle\bs{y},\left(C_k^{-1}+2B_k^*\right)\bs{y}\rangle \right)\nonumber \\ &\quad \times\exp\left(\langle \x, B_k^* \x\rangle \right) ,
\end{align}
with $B_k^*$ the appropriate solution of (\ref{ricccur}), $C_k$ satisfying (\ref{ricccovcur}) and $\lambda(k)$ given as in (\ref{curscgf}). \par The generating function (\ref{genint}) now becomes in the long-time limit
\begin{align} \label{longtimegen}
    G_k(\x,t) &\rightarrow K e^{\lambda(k)t}\sqrt{\frac{1}{(2\pi)^n\tn{det}C_k}}  \exp\left(\left\langle \x, \left(B_k^* - \frac{k}{2}\Gamma^{+}\right)\x \right\rangle \right) \nonumber \\ &\quad \times \int_{\mathbb{R}^n} d\bs{y} \, \exp\left(-\frac{1}{2}\langle \bs{y}, \mathcal{B}_k\bs{y}\rangle \right),
\end{align}
where 
\begin{equation} \label{posdef}
    \mathcal{B}_k = C_k^{-1} + 2B_k^* - k\Gamma^{+}.
\end{equation}
The Gaussian integral in the above expression can be evaluated to yield 
\begin{align} \label{gaussint}
    \int_{\mathbb{R}^n} d\bs{y} \, \exp\left(-\frac{1}{2}\langle \bs{y}, \mathcal{B}_k\bs{y}\rangle \right)= \sqrt{(2\pi)^n \tn{det}\mathcal{B}_k^{-1}},
\end{align}
provided that $\mathcal{B}_k$ is positive definite. If this is not the case, then the Gaussian integral is divergent.
\par Continuing, and assuming that $\mathcal{B}_k$ is positive definite, we finally obtain the expression 
\begin{equation}
    G_k(\x,t) \rightarrow K \sqrt{\frac{\tn{det}\mathcal{B}_k^{-1}}{\tn{det}C_k}} \exp\left(\left\langle \x, \left(B_k^* - \frac{k}{2}\Gamma^{+}\right)\x\right\rangle  \right) e^{\lambda(k)t}
\end{equation}
as $t \rightarrow \infty$. This implies that $r_k$ for this observable is proportional to 
\begin{equation}
    \exp\left(\left\langle \x, \left(B_k^* - \frac{k}{2}\Gamma^{+}\right)\x\right\rangle  \right).
\end{equation}
As a result we find at this point that the effective drift associated with $A_T$ is given from (\ref{effdrift}) by 
\begin{align}
    \bs{F}_k(\x) = -M\x + kD\Gamma\x + D\left(2B_k^* - k\Gamma^{+}\right)\x
\end{align}
and, since $\Gamma = \Gamma^{-} + \Gamma^{+}$, this becomes 
\begin{equation} \label{effdriftgenlin}
    \bs{F}_k(\x) = -M\x + 2DB_k^*\x +kD\Gamma^{-} \x,
\end{equation}
which is exactly the effective drift (\ref{effdriftcur}) obtained for the case where $\Gamma$ is purely antisymmetric. This shows that the symmetric part $\Gamma^{+}$ of the observable plays no role in the effective process. However, the symmetric part can still have an effect on the SCGF and rate function because the generating function of the full observable in the long-time limit (\ref{longtimegen}) contains a Gaussian integral that may diverge. 
\par To emphasize this point, we now denote the SCGF of the full observable $A_T$ by $\Lambda(k)$ so that 
\begin{equation}
    \Lambda(k) = \lim_{t\rightarrow \infty} \frac{1}{t} \ln G_k(\x,t).
\end{equation}
For $k$ such that $\mathcal{B}_k$ is positive definite, we have from (\ref{longtimegen}) and (\ref{gaussint}) that 
\begin{align}
    \Lambda(k) = \lim_{t \rightarrow \infty} \frac{1}{t} \ln \left(\sqrt{\frac{\tn{det}\mathcal{B}_k^{-1}}{\tn{det}C_k}} e^{\lambda(k)t} K r_k(\x) \right) = \lambda(k),
\end{align}
given that all other factors vanish in the $t \rightarrow \infty$ limit because they are finite. However, for $k$ such that $\mathcal{B}_k$ is not positive definite the integral (\ref{gaussint}) diverges and we have 
\begin{equation}
    \Lambda(k) = \infty. 
\end{equation}
In other words, we can write 
\begin{equation}
    \Lambda(k) = \begin{cases}\lambda(k) & \mathcal{B}_k \, \, \tn{positive definite} \\ \infty & \mathcal{B}_k \, \, \tn{not positive definite}. \end{cases}
\end{equation}
for the SCGF associated with the full observable $A_T$. The effect of the end-point boundary term associated with the symmetric part $\Gamma^{+}$ is therefore to make the SCGF diverge for a particular range of $k$ values, while for other $k$ the SCGF matches that of (\ref{curscgf}) obtained for the purely antisymmetric observable. 
\par In general there is a specific value of $k$ at which $\mathcal{B}_k$ ceases to be positive definite, which can be either positive or negative. For definiteness we shall assume that this $k$ value is positive and denote it by $k_{+}$ so that 
\begin{equation}
    \Lambda(k) = \begin{cases}\lambda(k) & k < k_{+} \\ \infty & k \geq k_{+}. \end{cases}
\end{equation}
Now, for the fluctuations $a$ associated with $k < k_{+}$ the rate function associated with $A_T$ will also correspond to the rate function associated with $A_T^{-}$. However, for fluctuations associated with $k \geq k_{+}$ the rate function will differ from that of the original. The effect of the SCGF being infinite for $k \geq k_{+}$ is that new rate function will be linear for the fluctuations $a$ corresponding to $k \geq k_{+}$, with the slope and intercept of this linear function determined via the Legendre transform featuring in the G\"{a}rtner-Ellis theorem.  
\par Denoting the rate function for $A_T$ by $\tilde{I}(a)$ and the rate function of $A_T^{-}$ as the usual $I(a)$, we can express the above discussion mathematically as 
\begin{equation} \label{ratelinear}
    \tilde{I}(a) = \begin{cases}I(a) & a < a_{+} \\ k_{+}a - \lambda(k_{+}) & a \geq a_{+}\end{cases},
\end{equation}
where $a_{+}$ is the value of the fluctuations associated with $k_{+}$ and is found via $a_{+} = \lambda'(k_{+})$. We note that it is not known what the appropriate effective process is that is responsible for manifesting the fluctuations $a > a_{+}$ and so the effective process is defined only for those $k$ such that $\mathcal{B}_k$ is positive definite.
\par A similar discussion applies in the event that the $k$ value at which $\mathcal{B}_k$ ceases to be positive definite is negative. In this event we denote this $k$ value by $k_{-}$ and obtain
\begin{equation}
    \Lambda(k) = \begin{cases}\infty & k < k_{-} \\ \lambda(k) & k \geq k_{-} \end{cases}
\end{equation}
for the SCGF of the full observable $A_T$ and 
\begin{equation} \label{ratelinear2}
    \tilde{I}(a) = \begin{cases}k_{-}a - \lambda(k_{-})& a < a_{-} \\ I(a) & a \geq a_{-}  \end{cases}
\end{equation}
for the associated rate function, where $a_{-} = \lambda'(k_{-})$.
\par To close this section we remark that in general, we can consider a diffusion as having a random initial position $\x$, with this initial position distributed according to some density $p_0(\x)$. We then define a generating function 
\begin{equation}
    G_k(t) = \int d\x p_0(\x) G_k(\x,t) = \int d\x \, p_0 (\x) \int d\bs{y} \,G_k(\x,\bs{y},t).
\end{equation}
For the general linear observable considered here we would then typically obtain a boundary term having a singularity at $k_{+} > 0$ and also a boundary term having a singularity at $k_{-} < 0$ due to integration over both the final (stationary) state and the initial density $p_0(\x)$. The above discussions can be easily extended to this case, for which we now obtain two linear tails in the rate function for $a < a_{-}$ and $a> a_{+}$. We will only consider cases where the initial density is a Dirac delta function, for which our previous discussions were sufficient, and so we do not here discuss this further. 
\section{Asymptotic mean and variance}\label{sec:sec42}
The exact results that we have derived before for the three classes of observables considered can be used in principle to find explicit formulas for the moments of these observables. In particular, in the long-time limit, two moments are worth considering. The first is the asymptotic mean, which corresponds to the mean $\mathbb{E}[A_T]$ in the long-time limit and which is linked to the SCGF according to  
\begin{equation}
    \lim_{T\rightarrow\infty} \mathbb{E}\left[A_T\right] = \lambda'(0).
\end{equation}
The second is the variance of $A_T$ which is known to scale as
\begin{equation}
    \tn{var}[A_T] \sim \frac{\sigma_{\tn{asym}}^2}{T},
\end{equation}
when $T \rightarrow \infty$, where $\sigma_{\tn{asym}}^2$ is the so-called asymptotic variance given by 
\begin{equation}
    \sigma_{\tn{asym}}^2 = \lambda''(0).
\end{equation}
In the following we derive explicit expressions for these two quantities for the three observables considered. 
\subsection{Linear additive observable}\label{sec:sec421}
For a linear additive observable $A_T$ having the form (\ref{linadd}) we found that the SCGF is given by (\ref{SCGFlin2}) so that 
\begin{equation} \label{firstder}
    \frac{d\lambda(k)}{dk}=  k\left\langle\left(M^{\mathsf{T}}\right)^{-1}\bs{\eta},D\left(M^{\mathsf{T}}\right)^{-1}\bs{\eta} \right\rangle.
\end{equation}
As a result the asymptotic mean is found to be 
\begin{equation}
    \lambda'(0) = 0. 
\end{equation}
\par Continuing, we obtain from (\ref{firstder}) for the second derivative of the SCGF as
\begin{equation}
    \frac{d^2 \lambda(k)}{dk^2} = \left\langle\left(M^{\mathsf{T}}\right)^{-1}\bs{\eta},D\left(M^{\mathsf{T}}\right)^{-1}\bs{\eta} \right\rangle.
\end{equation}
Given that this expression is independent of $k$ we then obtain 
\begin{equation}
    \lambda''(0) = \left\langle\left(M^{\mathsf{T}}\right)^{-1}\bs{\eta},D\left(M^{\mathsf{T}}\right)^{-1}\bs{\eta} \right\rangle
    \end{equation}
for the asymptotic variance. The same results for the asymptotic mean and variance can be obtained directly from the quadratic expressions of the SCGF and rate function found in (\ref{SCGFlin2}) and (\ref{ratelin}), respectively. 
\subsection{Quadratic additive observable}\label{sec:sec422}
For quadratic additive observables we find from (\ref{quadscgf}) for the associated SCGF that the asymptotic mean has the form 
\begin{equation}
    \lambda'(0) = \tn{Tr}\left(D B_0^{*\prime}\right),
\end{equation}
with $B^*_k$ the solution to the algebraic Riccati equation (\ref{Akquadricc}) and which also satisfies $B_0^* = 0$, and where $B_0^{*\prime}$ represents the derivative of $B^*_k$ with respect to $k$ evaluated at $k = 0$. 
Differentiating equation (\ref{Akquadricc}) with respect to $k$, we obtain
\begin{equation} \label{akquaddiff1}
    2B_k^{*\prime} D B_k^* + 2 B_k^* DB_k^{*\prime} - \left(M^{\mathsf{T}}B_k^{*\prime} + B_k^{*\prime} M\right) + Q = 0,
\end{equation}
which, for $k = 0$, yields the Lyapunov equation
\begin{equation} \label{ricc1}
    M^{\mathsf{T}} B_0^{*\prime} +B_0^{*\prime}M = Q,
\end{equation}
from which the matrix $B_0^{*\prime}$ can be determined. 
\par By using the Lyapunov equation (\ref{ricccov}) defining the stationary covariant matrix $C$ we can write 
\begin{equation}
     \lambda'(0)=\tn{Tr}\left(DB_0^{*\prime}\right) = \tn{Tr}\left((MC + CM^{\mathsf{T}})B_0^{*\prime}\right).
\end{equation}
Using the fact that the trace is invariant under cyclic permutations, this becomes 
\begin{equation} \label{quadmean}
    \lambda'(0)= \tn{Tr}\left(\left[M^{\mathsf{T}}B_0^{*\prime} +B_0^{*\prime}M \right]C \right) = \tn{Tr}\left(QC\right),
\end{equation}
which is the standard known result for the Gaussian expectation of a quadratic form~\cite{seber2012linear}, and which therefore serves as a useful verification of the large deviation results obtained for this type of observable.
\par For the asymptotic variance we need the second derivative of $B_k^*$ at $k = 0$. Therefore, we differentiate (\ref{akquaddiff1}) with respect to $k$, obtaining 
\begin{equation}
    2B_k^{*\prime\prime} D B^*_k + 2 B^*_k DB_k^{*\prime \prime} + 4B_k^{*\prime} DB_k^{*\prime} - \left(M^{\mathsf{T}}B_k^{*\prime\prime} + B_k^{*\prime\prime} M\right)  = 0
\end{equation}
so that
\begin{equation} \label{eqder2}
    M^{\mathsf{T}}B_0^{*\prime\prime} + B_0^{*\prime\prime} M = 4B_0^{*\prime} D B_0^{*\prime}.
\end{equation}
We then have that
\begin{equation}
    \lambda''(0) = \tn{Tr}\left(DB_0^{*\prime\prime}\right) = \tn{Tr}\left((MC + CM^{\mathsf{T}})B_0^{*\prime\prime}\right).
\end{equation}
Using the same logic as before, this becomes
\begin{equation}
    \lambda''(0) =  \tn{Tr}\left(\left[M^{\mathsf{T}} B_0^{*\prime\prime} +B_0^{*\prime\prime}M \right]C \right). 
\end{equation}
Substituting (\ref{eqder2}) into the above then yields
\begin{equation} \label{quadvar}
    \lambda''(0) = 4 \tn{Tr}\left(C B_0^{*\prime} D B_0^{*\prime} \right).
\end{equation}
Using again the Lyapunov equation for the covariance matrix $C$, this becomes
\begin{equation}
    \lambda''(0) = 4 \tn{Tr}\left(CB_0^{*\prime} MC B_0^{*\prime} + C B_0^{*\prime} C M^{\mathsf{T}} B_0^{*\prime}  \right),
\end{equation}
which (using the cyclic invariance property of the trace) can be rearranged to give
\begin{equation}
    \lambda''(0) = 4 \tn{Tr}\left(C B_0^{*\prime} C \left[B_0^{*\prime}M + M^{\mathsf{T}} B_0^{*\prime}\right] \right)
\end{equation}
which reduces to
\begin{equation}
    \lambda''(0) = 4 \tn{Tr}\left(C QC B_0^{*\prime}  \right),
\end{equation}
where we have used the Lyapunov equation (\ref{ricc1}) for $B_0^{*\prime}$. 
\par We note the remarkable property that the formula for the asymptotic variance has been reduced to a quantity involving only the first (and not the second) derivative of $B_k^*$. To find the (asymptotic) mean and variance it is therefore only necessary that we determine the solutions to the Lyapunov equations (\ref{ricccov}) and (\ref{ricc1}) for the matrices $C$ and $B_0^{*\prime}$, respectively. 
These equations can, in principle, be solved without needing to obtain the full solution $B_k^*$ and as such the asymptotic mean and variance for this observable can be found immediately from knowledge of the drift matrix $M$, diffusion matrix $D$ and the particular matrix $Q$ involved in the specific observable under consideration. 
\subsection{Linear current-type observable}\label{sec:sec423}
For (antisymmetric) linear current-type observables, the SCGF $\lambda(k)$ is given by (\ref{curscgf}) and has the same form as that for the quadratic additive observable. As a result, the procedure followed here is identical to that of the previous section. \par The asymptotic mean is given by 
\begin{equation}
    \lambda'(0) = \tn{Tr}(D B_0^{*\prime}),
\end{equation}
where $B_0^{*\prime}$ is the derivative with respect to $k$ (and evaluated at $k = 0$) of the solution to the algebraic Riccati equation (\ref{ricccur}) satisfying $B_0^* = 0$. The derivative of (\ref{ricccur}) with respect to $k$ is given explicitly by
\begin{align} \label{bigasseq}
    &k \Gamma^{\mathsf{T}}D\Gamma - \frac{1}{2}(M^{\mathsf{T}}\Gamma -\Gamma M)\,+ \,(-M + kD\Gamma)^{\mathsf{T}}B_k^{*\prime} + B_k^{*\prime}(-M + kD\Gamma) \nonumber \\ & \quad + 2B_k^{*\prime} D B_k^* + 2 B_k^* D B_k^{*\prime}  + \Gamma^{\mathsf{T}}D B_k^* + B_k^* D\Gamma = 0
\end{align}
which, for $k = 0$, becomes
\begin{equation} \label{bigricc1}
   B_0^{*\prime}M + M^{\mathsf{T}}B_0^{*\prime} = \frac{1}{2}\left(\Gamma M - M^{\mathsf{T}}\Gamma \right).
\end{equation}
Using this result, and the Lyapunov equation for $C$, we then find 
\begin{equation} \label{curmean}
    \lambda'(0) = \tn{Tr}\left(D B_0^{*\prime} \right) = \tn{Tr}\left(\left(MC + CM^{\mathsf{T}}\right)B_0^{*\prime} \right) = \frac{1}{2}\tn{Tr}\left(C \left[\Gamma M - M^\mathsf{T}\Gamma \right] \right)
\end{equation}
for the asymptotic mean, where in the last step we have used the invariance of the trace under cyclic permutations as well as the equation (\ref{bigricc1}). 
\par In order to determine the asymptotic variance we now differentiate (\ref{bigasseq}) with respect to $k$ and set $k = 0$ in the resulting equation, yielding the Riccati equation
\begin{equation}
    M^{\mathsf{T}}B_0^{*\prime\prime} + B_0^{*\prime\prime}M = \Gamma^{\mathsf{T}}D\Gamma + 4B_0^{*\prime}DB_0^{*\prime} + 2\left(\Gamma^{\mathsf{T}}D B_0^{*\prime} + B_0^{*\prime}D\Gamma\right)
\end{equation}
for $B_0^{*\prime\prime}$. We therefore find that 
\begin{align}
    \tn{Tr}(DB_0^{*\prime\prime}) &= \tn{Tr}\left([MC + CM^{\mathsf{T}}]B_0^{*\prime\prime} \right) \nonumber \\ 
    &= \tn{Tr}\left(\left[\Gamma^{\mathsf{T}}D\Gamma + 4B_0^{*\prime}DB_0^{*\prime} + 2\left(\Gamma^{\mathsf{T}}D B_0^{*\prime} + B_0^{*\prime}D\Gamma\right)\right]C \right).
\end{align}
We showed in the previous section that 
\begin{equation}
    4 \tn{Tr}(CB_0^{*\prime}DB_0^{*\prime}) = 4 \tn{Tr}\left(C B_0^{*\prime} C \left[B_0^{*\prime}M + M^{\mathsf{T}} B_0^{*\prime}\right] \right),
\end{equation}
where we use the Lyapunov equation (\ref{ricccov}) for $C$. As such, and using now the relevant Lyapunov equation (\ref{bigricc1}) for $B_0^{*\prime}$, we obtain
\begin{equation}
    4 \tn{Tr}(CB_0^{*\prime}DB_0^{*\prime}) = 2 \tn{Tr}\left(C B_0^{*\prime} C\left[\Gamma M - M^{\mathsf{T}}\Gamma \right]\right)
\end{equation}
so that finally we have for the asymptotic variance the result 
\begin{equation} \label{curvar}
    \lambda''(0) = \tn{Tr}\bigg(C\Gamma^{\mathsf{T}}D\Gamma + 2 C\left[\Gamma M - M^{\mathsf{T}}\Gamma \right]CB_0^{*\prime} + 2 C \left(\Gamma^{\mathsf{T}}D B_0^{*\prime} + B_0^{*\prime}D\Gamma\right)\bigg).
\end{equation}
Again, we note that this is an explicit result in terms of $C$ and $B_0^{*\prime}$, both of which are determined fully via Lyapunov equations. In addition, we note that, as for the quadratic additive observable, the formula for the asymptotic variance for the linear current-type observable reduces to an expression involving only the first derivative of $B_k^*$ at $k = 0$, with determination of the second order derivative not needed.
\chapter{Applications for linear diffusions} \label{chap:chap5}
We here apply the general results obtained in Chap.~\ref{chap:chap4} for a number of specific observables of the three linear systems introduced in Sec.~\ref{sec:sec24}. We focus on quadratic additive and linear current-type observables, as the effective process associated with linear additive observables differs from the original process only in a fairly trivial way. 

\section{Quadratic observable for transverse diffusion}\label{sec:sec51}
For our first illustration, we consider the transverse diffusion in $\mathbb{R}^2$, defined before in Sec.~\ref{sec:sec242}, and study the large deviations of the squared distance 
\begin{equation}
    A_T = \frac{1}{T} \int_0^T \bigg(X_1(t)^2 + X_2(t)^2 \bigg)dt, 
\end{equation}
which corresponds to the choice 
\begin{equation}
    Q = \begin{pmatrix}1 && 0 \\ 0 && 1 \end{pmatrix}
\end{equation}
in the general quadratic observable (\ref{quadadd}). As discussed in Sec.~\ref{sec:sec412}, the generating function of this observable involves the time-dependent matrix $B_k(t)$ satisfying the differential Ricatti equation (\ref{Akquaddiff}). We show in App.~\ref{appendixD} that the solution of this equation is a matrix proportional to the identity matrix: 
\begin{equation} \label{matrix3}
    B_k(t) = \begin{pmatrix}b_k(t) && 0 \\ 0 && b_k(t) \end{pmatrix},
\end{equation}
where the coefficient $b_k(t)$ satisfies the equation 
\begin{equation} \label{scalarricc1}
    \frac{db_k(t)}{dt} = 2\epsilon^2 b_k(t)^2 - 2 \gamma b_k(t) + k, 
\end{equation}
with $b_k(0) = 0$. Of crucial importance in the argument leading to this form for the matrix $B_k(t)$ is the fact that the diffusion matrix $D$ is proportional to the identity and the fact that the drift matrix $M$ has a diagonal symmetric part and a purely antisymmetric part. If the symmetric part of $M$ included non-zero off-diagonal terms then the off-diagonal terms in (\ref{matrix2}) would not cancel and $B_k$ would necessarily also contain off-diagonal terms. 
\par The differential equation (\ref{scalarricc1}) has an exact solution which we can write in a convenient form as 
\begin{equation} \label{timedepsol}
    b_k(t) = \frac{\gamma}{2\epsilon^2}\left(1  -\frac{1 + \frac{\sqrt{\gamma^2 - 2k\epsilon^2}}{\gamma}\tanh\left(t\sqrt{\gamma^2 - 2k\epsilon^2}\right)}{1 + \frac{\gamma}{\sqrt{\gamma^2 - 2k\epsilon^2}}\tanh\left(t\sqrt{\gamma^2 - 2k\epsilon^2}\right)} \right).
\end{equation}
The time-dependent expression for the generating function $G_k$ can now be obtained by substituting (\ref{matrix3}) and (\ref{timedepsol}) into the expression (\ref{Gkquad}) to obtain
\begin{equation}
    G_k(\x,t) = \exp\left(b_k(t)\langle \x, \x\rangle \right)\exp\left(\int_0^t 2\epsilon^2 b_k(s) ds\right).
\end{equation}
The integral in the above expression can be exactly calculated and is given by  
\begin{align}
    \int_0^t 2\epsilon^2 b_k(s) ds &= \gamma t - \tn{arctanh}\left(\frac{\gamma\tanh\left(t\sqrt{\gamma^2 - 2k\epsilon^2}\right)}{\sqrt{\gamma^2 - 2k\epsilon^2}}\right)+ \ln \left(\sqrt{\gamma^2 - 2k\epsilon^2} \right)\nonumber \\ 
    &\quad -\frac{1}{2}\ln \left(\gamma^2 -k\epsilon^2 - k\epsilon^2 \cosh\left(2t \sqrt{\gamma^2 - 2k\epsilon^2}\right)\right).
\end{align}
\par The long-time behavior of $G_k$ is determined, as done before, by obtaining the stationary solution $B_k^*$ of $B_k(t)$ and hence $b_k(t)$ as $t\rightarrow \infty$. This stationary solution $b_k^*$ can  be found either by explicitly taking the limit $t\rightarrow\infty$ in the exact expression (\ref{timedepsol}) for $b_k(t)$ or by solving the algebraic Riccati equation (\ref{Akquadricc}), which reduces in this case to 
\begin{equation}
    2\epsilon^2 {b_k^*}^2 - 2\gamma b_k^* + k = 0,
\end{equation}
choosing the solution which satisfies $b_0^* = 0$. Furthermore it must be verified that this solution gives a modified drift matrix $M_k$ that is positive definite. The two procedures produce identical results, namely
\begin{equation} \label{akstat}
    b_k^* = \frac{\gamma - \sqrt{\gamma^2 - 2k\epsilon^2}}{2\epsilon^2}, \quad k  \in \bigg(-\infty, \frac{\gamma^2}{2\epsilon^2}\bigg]
\end{equation}
so that 
\begin{equation}\label{Aktransquad1}
    B_k^* =\begin{pmatrix}\frac{\gamma - \sqrt{\gamma^2 - 2k\epsilon^2}}{2\epsilon^2} && 0 \\ 0 && \frac{\gamma - \sqrt{\gamma^2 - 2k\epsilon^2}}{2\epsilon^2} \end{pmatrix}.
\end{equation}
Indeed, we now have from (\ref{mkquad}) that 
\begin{equation} \label{mkquadtrans}
    M_k = \begin{pmatrix}\sqrt{\gamma^2 - 2k\epsilon^2} && \xi \\ -\xi && \sqrt{\gamma^2 - 2k\epsilon^2} \end{pmatrix},
\end{equation}
so that $M_k$ has eigenvalues $\sqrt{\gamma^2 - 2k\epsilon^2} \pm i\xi$, which have positive real part for $k$ in the range 
\begin{equation}
    k \in \bigg(-\infty, \frac{\gamma^2}{2\epsilon^2}\bigg).
\end{equation}
Note that we have excluded the endpoint $k = \gamma^2/2\epsilon^2$ since for this value the drift matrix of the effective process has purely imaginary eigenvalues and the effective process is not ergodic. From these results, we find the SCGF directly from (\ref{quadscgf}) as
\begin{equation}\label{transquadSCGF}
    \lambda(k) =2\epsilon^2 b_k^* = \gamma - \sqrt{\gamma^2 - 2k\epsilon^2}, \quad k  \in \bigg(-\infty, \frac{\gamma^2}{2\epsilon^2}\bigg). 
\end{equation}
Taking the Legendre transform of this expression, we then find the rate function $I(a)$ as
\begin{equation}
    I(a) = \frac{\gamma^2 a}{2\epsilon^2} + \frac{\epsilon^2}{2 a} - \gamma, \quad a > 0.
\end{equation}
\begin{figure}[t]
    \centering
    \begin{minipage}{.45\textwidth}
    \includegraphics[width=6cm]{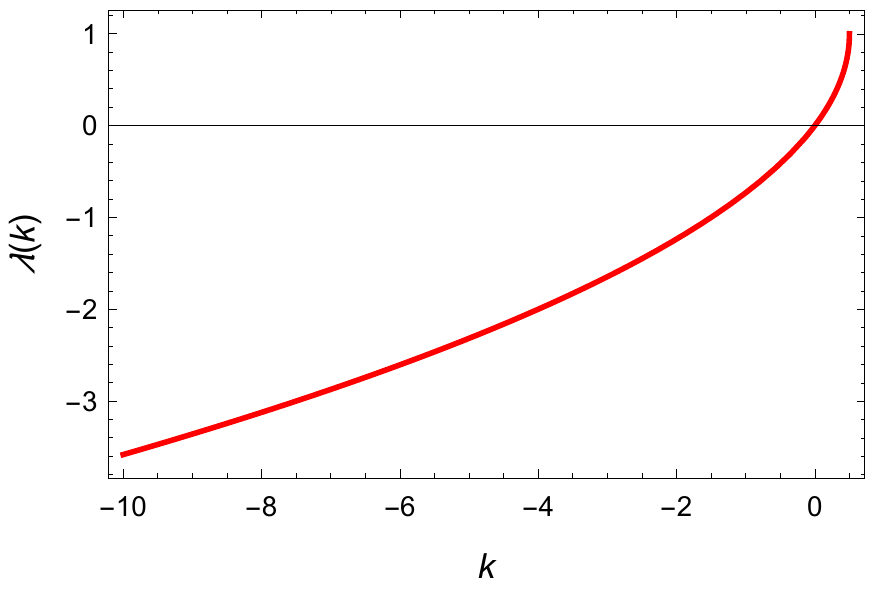}
    \end{minipage}
    \begin{minipage}{.45\textwidth}
    \includegraphics[width=6cm]{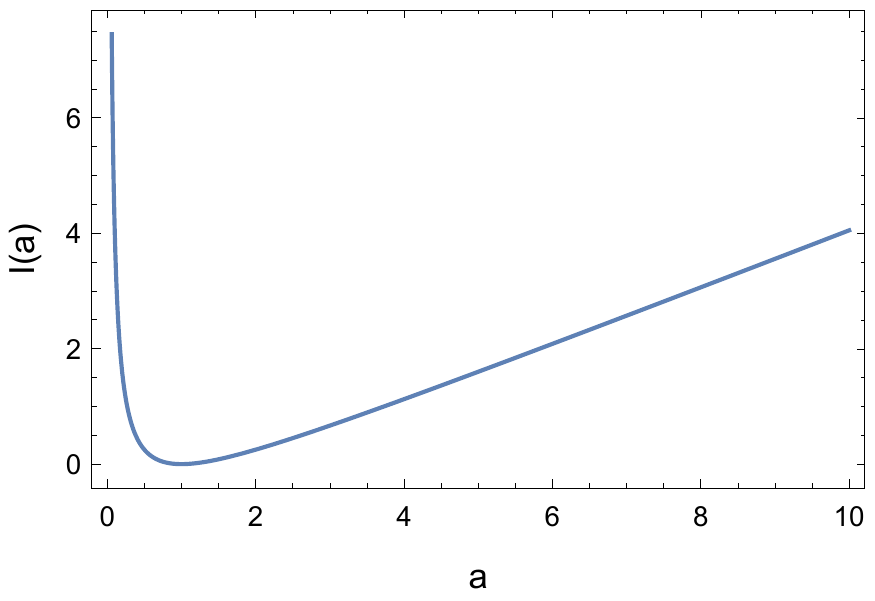}
    \end{minipage}
    \caption{SCGF (left) and rate function (right) for the transverse system with quadratic observable having $Q = \mathbb{I}$ and for parameter values $\gamma = 1,\xi=1$ and $\epsilon=1$. }
    \label{fig:TransquadSCGFrate}
\end{figure}
\par These results are plotted in Fig.~\ref{fig:TransquadSCGFrate} for the values $\gamma =1,\xi =1$ and $\epsilon =1$. We can see that the rate function has a unique minimum $a^* = \epsilon^2/\gamma$ which represents the typical value (mean) of the observable $A_T$ in the limit $T \rightarrow \infty$. This can be found directly from the expression for the rate function or by using the formula (\ref{quadmean}) derived earlier. Similarly, the asymptotic variance $\lambda''(0)$ for this process and observable can be found either directly from (\ref{transquadSCGF}) or by inserting the expression for $B_0^{*\prime}$ found from (\ref{ricc1}) into (\ref{quadvar}) to obtain 
\begin{equation}
    \lambda''(0) = \frac{\epsilon^4}{\gamma^3}.
\end{equation}
\par It is remarkable that both the SCGF and rate function are independent of the nonequilibrium parameter $\xi$. Intuitively, this makes sense: since the observable $A_T$ considered here is radially symmetric, the purely antisymmetric part of the force (responsible for a circulating current and which does not affect the distance of the system from the origin) is of no consequence. 
\par
\begin{figure}[t]
    \centering
            \begin{subfigure}[b]{.45\textwidth}
                \centering
                \includegraphics[width=6cm]{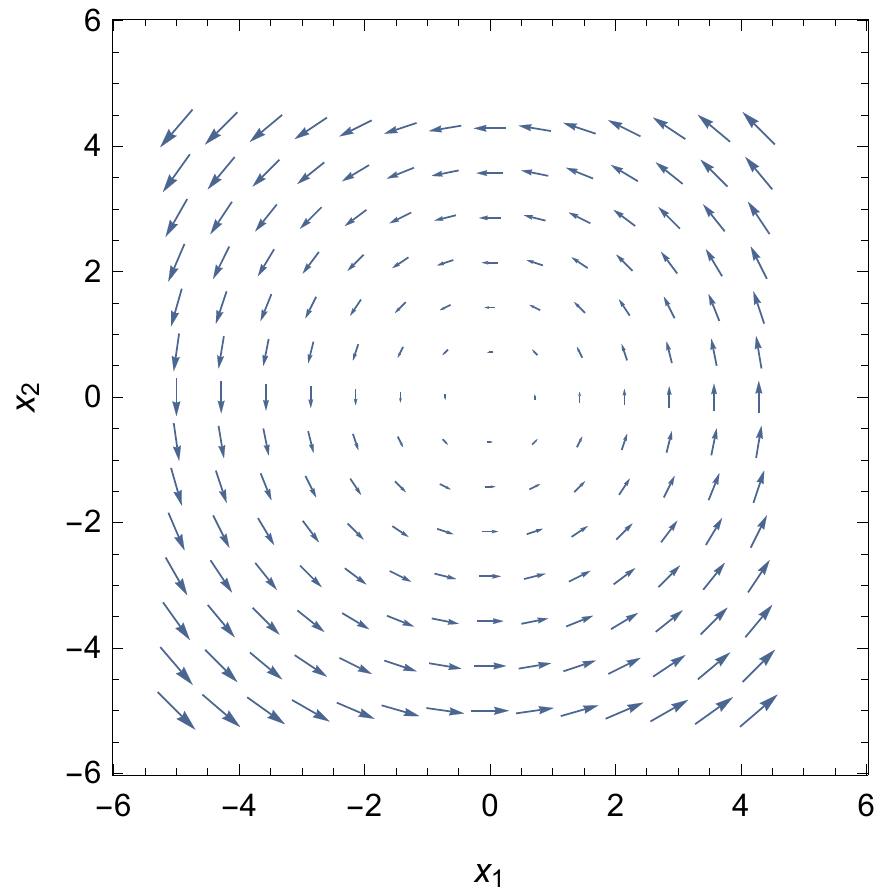}
                \caption{$k \rightarrow \gamma^2/2\epsilon^2= 1/2$}
                
            \end{subfigure}
            \begin{subfigure}[b]{.45\textwidth}
            \centering
                \includegraphics[width=6cm]{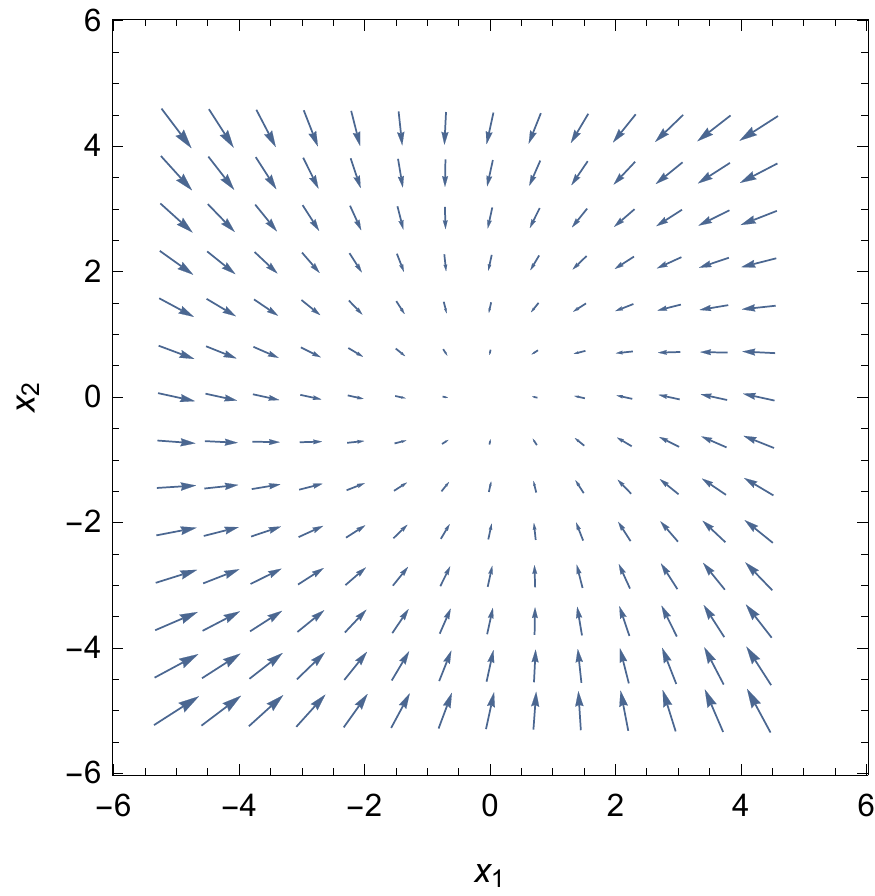}
                \caption{$k = -10$}
            \end{subfigure}
        \caption{Vector field of the effective drift for the transverse system with quadratic observable having $Q = \mathbb{I}$ shown for the case where $k$ approaches the upper limit $\gamma^2/2\epsilon^2$ and for $k = 10$ for the parameters $\gamma=1, \xi = 1$, and $\epsilon =1.$}
        \label{fig:Fktransquad1}
\end{figure}

 A similar independence with respect to $\xi$ can be seen at the level of the effective drift, given from (\ref{effdriftquad}) and (\ref{mkquadtrans}) by 
\begin{equation} \label{Fktransquadform}
    \bs{F}_k(\x) = - \begin{pmatrix}\sqrt{\gamma^2 - 2k\epsilon^2} && \xi \\ -\xi && \sqrt{\gamma^2 - 2k\epsilon^2} \end{pmatrix}\x,
\end{equation}
and modified in a manner which does not depend on $\xi$. We note that the antisymmetric part of the force remains unchanged. Moreover, we observe that the effective drift differs from the original drift in the strength with which the system is attracted to the origin $(0,0)$. Fluctuations $a > a^*$ associated with $k > 0$ are manifested by weakening the diagonal part of the force, in which case the force is dominated by the antisymmetric part, as can be seen in Fig.~\ref{fig:Fktransquad1}, while fluctuations $a < a^*$ associated with $k < 0$ are manifested by strengthening the diagonal part of the force, in which case the force increasingly resembles a purely diagonal linear drift. The stronger the attraction to the origin, the smaller the typical region surrounding $(0,0)$ in which the process spends most of its time, and the smaller the resulting value of $A_T$. The effective drift can therefore be understood as altering the characteristic size of the region surrounding $(0,0)$ in which the process will typically be found. The effective drift is shown for various values of $k$ in Fig.~\ref{fig:Fktransquad1}.
\par As found in Sec.~\ref{sec:sec412}, the stationary density $p_k^*$ associated with $\bs{F}_k$ is centered at $\bs{0}$, as is $p^*$, but has a different covariance matrix $C_k$ satisfying (\ref{ricccovquad}),
and given by 
\begin{equation}
    C_k =  \frac{\epsilon^2}{2 \sqrt{\gamma^2 - 2k \epsilon^2}}\mathbb{I}.
\end{equation}
From (\ref{statcurlin}), we have that the stationary current for the effective process is given by 
\begin{equation}
    \bs{J}_{\bs{F}_k,p^*_k}(\x) = \xi \begin{pmatrix}-x_2 \\ x_1 \end{pmatrix} p_k^*(\x).
\end{equation}
We note that this current differs from the stationary current of the original process only to the extent that $p_k^*$ differs from $p^*$. That is, the effective current is obtained from the original current via the substitution $p^* \rightarrow p_k^*$. For this observable it is therefore clear that fluctuations are realized optimally by altering the density, with the only changes to the current resulting from those same density modifications.
\par This is clear also from the form (\ref{Fktransquadform}) of the expression for the effective drift: given that $D$ here is proportional to the identity matrix and $B_k^*$ is a symmetric matrix, the effective drift can differ from the original drift only in the symmetric part. As a result, the effective process here has a non-zero stationary current for all $k$, given that the original process had a non-zero stationary current. This observation can be extended to include (at minimum) all linear processes having a diffusion matrix proportional to the identity. For such processes, the effective process associated with the fluctuations of a quadratic additive observable has the same equilibrium/nonequilibrium nature as that of the original process. In other words, for such a case we have that if the original process has a non-zero stationary current, then so does the effective  process, and similarly for the case where the stationary current is zero everywhere. 
\section{Linear current-type observables for the gradient diffusion in $\mathbb{R}^2$}\label{sec:sec52}
We now turn our attention to current-type observables having the form (\ref{lincurobs}) and study their fluctuations first for the gradient diffusion in $\mathbb{R}^2$, defined before in Sec.~\ref{sec:sec241}, in order to understand how current fluctuations can arise in a system that does not have a current to begin with. We present next the large deviation calculation for the observable (\ref{lincurobs}) with a general antisymmetric matrix $\Gamma$ having the form 
\begin{equation} \label{generalalpha}
  \Gamma = \begin{pmatrix}0 && -\alpha \\ \alpha && 0 \end{pmatrix}.
\end{equation}
and then consider the special case $\alpha = 1/2$, which corresponds to an interesting observable known as the stochastic area.
\subsection{General $\alpha$}
For the gradient diffusion with $M = \gamma\mathbb{I}$ and $D = \epsilon^2 \mathbb{I}$, the Riccati equation (\ref{riccatidiffcur}) is solved by a matrix $B_k(t)$ that is proportional to the identity matrix, as shown in App.~\ref{appendixE}. Specifically, we have that $B_k(t) = b_k(t) \mathbb{I}$. Substituting this form for $B_k(t)$ into the equation (\ref{riccatidiffcur}) yields the differential equation
\begin{equation} \label{scalarricc2}
    \frac{db_k(t)}{dt} = 2\epsilon^2 b_k(t)^2 - 2\gamma b_k(t) + \frac{k^2 \epsilon^2 \alpha^2}{2}, \quad b_k(0) = 0,
\end{equation}
for the coefficient $b_k(t)$, which has a similar form as that obtained in (\ref{scalarricc1}), but with the constant term now proportional to $k^2$ and not $k$. The exact solution we obtain here for $b_k(t)$ therefore bears a close resemblance to that obtained in (\ref{timedepsol}) and is given explicitly as 
\begin{equation} \label{timedepsol2}
    b_k(t) = \frac{\gamma}{2\epsilon^2}\left(1 - \frac{1 + \frac{\sqrt{\gamma^2 - k^2 \epsilon^4 \alpha^2}}{\gamma}\tanh\left(t\sqrt{\gamma^2 - k^2 \epsilon^4 \alpha^2} \right)}{1 + \frac{\gamma}{\sqrt{\gamma^2 - k^2 \epsilon^4 \alpha^2}}\tanh\left(t\sqrt{\gamma^2 - k^2 \epsilon^4 \alpha^2}\right)} \right).
\end{equation}
\par Using this solution for $b_k(t)$ we can, in principle, obtain the explicit time-dependent expression for the generating function as was done for the quadratic observable for the transverse diffusion. We omit this calculation here and proceed to consider instead the long-time behavior of the solution $b_k(t)$. The stationary solution of (\ref{scalarricc2}) can be obtained either by taking the limit $t \rightarrow \infty$ in the above expression for $b_k(t)$ explicitly or by simply solving for 
\begin{equation}
    2\epsilon^2 {b_k^*}^2 - 2\gamma b_k^* + \frac{k^2 \epsilon^2 \alpha^2}{2} = 0, 
\end{equation}
and then choosing the appropriate solution which satisfies $b_0^* = 0$. The appropriate solution for $b_k^*$ is found to be  
\begin{equation} \label{akalpha}
    b_k^* = \frac{\gamma - \sqrt{\gamma^2 - k^2 \epsilon^4 \alpha^2}}{2\epsilon^2}.
\end{equation}
To determine the range of values of $k$ for which this solution is valid, we note that the drift matrix of the effective process, given by (\ref{mkcur}), is 
\begin{equation} \label{driftmatrix}
    M_k = \begin{pmatrix}\sqrt{\gamma^2 - k^2 \epsilon^4 \alpha^2} && k\epsilon^2 \alpha \\ - k \epsilon^2 \alpha && \sqrt{\gamma^2 - k^2 \epsilon^4 \alpha^2} \end{pmatrix},
\end{equation}
which has eigenvalues $\sqrt{\gamma^2 - k^2 \epsilon^4 \alpha^2} \pm k\epsilon^2 \alpha i$. Therefore, we have that the effective process is ergodic for values of $k$ in the range 
\begin{equation}
    k \in \bigg(-\frac{\gamma}{\epsilon^2 |\alpha|}, \frac{\gamma}{\epsilon^2 |\alpha|} \bigg),
\end{equation}
which means that the the SCGF, given by (\ref{curscgf}), is 
\begin{equation}\label{scgfalpha}
    \lambda(k) = 2\epsilon^2 b_k^* = \gamma - \sqrt{\gamma^2 - k^2 \epsilon^4 \alpha^2}, 
\end{equation}
for $k$ also in that range. The corresponding rate function $I(a)$ can be found by Legendre transform, but we do not show this result explicitly. We can calculate the asymptotic mean and variance directly by using the above expression for the SCGF. For the asymptotic mean we obtain
\begin{equation} \label{meanalpha}
    \lambda'(0) = 0,
\end{equation}
which is an intuitive result: for the current-type observable under consideration, we have from (\ref{ergodicexp}) and given the fact that the gradient process has no stationary current, that $A_T \rightarrow 0$ as $T \rightarrow \infty$. In other words, the stationary mean of this observable is zero for the gradient diffusion with zero stationary current. The asymptotic variance is given by  
\begin{equation} \label{varalpha}
    \lambda''(0) = \frac{\alpha^2\epsilon^4}{\gamma}.
\end{equation}
We note that these results for the asymptotic mean and variance could also be obtained without explicitly solving for the matrix $B_k^*$ and instead by using the results for the mean and variance derived earlier. In particular, the asymptotic mean could be obtained directly from (\ref{curmean}) while the asymptotic variance is found via (\ref{curvar}) along with the Lyapunov equation (\ref{bigricc1}). 
\par Given that the effective drift is linear with drift matrix given as in (\ref{driftmatrix}), we notice by comparison with the SDE (\ref{sdetrans}) governing the evolution of the transverse system that the effective process associated with fluctuations of the current-type observable for the gradient system in fact corresponds to the transverse system with friction coefficient $\sqrt{\gamma^2 - k^2 \epsilon^4 \alpha^2}$ and nonequilibrium parameter $k\epsilon^2 \alpha$. Thus we can obtain the stationary distribution for the effective process and stationary current directly from the expressions (\ref{rhotrans}) and (\ref{Jtrans}) via the substitutions 
\begin{equation} \label{substitutions}
    \gamma \rightarrow  \sqrt{\gamma^2 - k^2 \epsilon^4 \alpha^2},\quad \xi \rightarrow  k\epsilon^2 \alpha.
\end{equation}
With this substitution, we find that the stationary distribution $p^*_k$ associated with the effective process is given by 
\begin{equation}\label{pkalpha}
    p_k^*(\x) =\frac{\sqrt{\gamma^2 - k^2 \epsilon^4 \alpha^2}}{\pi \epsilon^2}\exp\left(-\frac{\sqrt{\gamma^2 - k^2 \epsilon^4 \alpha^2}}{\epsilon^2}\langle \x, \x\rangle \right),
\end{equation}
with the stationary covariance matrix $C_k$ given from (\ref{transcov}) by 
\begin{equation} \label{covalpha}
    C_k = \frac{\epsilon^2}{2\sqrt{\gamma^2 - k^2 \epsilon^4 \alpha^2}}\mathbb{I}. 
\end{equation}
From (\ref{Jtrans}) we similarly find that the stationary current for the effective process is
\begin{equation} \label{curalpha}
    \bs{J}_{\bs{F}_k,p_k^*}(\x) = k\epsilon^2 \alpha \begin{pmatrix}-x_2 \\ x_1\end{pmatrix} p_k^*(\x). 
\end{equation}
These results demonstrate that for the linear current-type observable (and current-type observables more broadly) fluctuations are manifested both by altering the density and the current of the original process in non-trivial ways; unlike in the case of the quadratic additive observable considered previously where the current was modified only in the manner $p^* \rightarrow p^*_k$, we here have that the effective process has non-zero current while the original process has zero current. As such the effective process responsible for manifesting fluctuations is here a nonequilibrium process while the original process is equilibrium. This is intuitive given that a current-type observable can have non-zero ergodic mean only in the case when a stationary current is present. Fluctuations are therefore manifested by simultaneously creating currents and modifying the density.
\subsection{Specific case: Stochastic area}\label{sec:sec521}
The value $\alpha = 1/2$ gives rise to the observable 
\begin{equation}
    A_T = \frac{1}{2T} \int_0^T  \bigg(X_1(t) dX_2(t) - X_2(t) dX_1(t) \bigg),
\end{equation}
which is a (time-normalized) stochastic version of the line integral in $\mathbb{R}^2$ used in Green's theorem to calculate the area
bounded by a closed contour. For this reason, this observable is referred to as the stochastic area or area loop. Because stochastic trajectories are generally not closed, this quantity is interpreted here as the (time-normalized) area enclosed by the stochastic trajectory $(X_1(t), X_2(t))_{t = 0}^{T}$ in $\mathbb{R}^2$ and the chords joining, respectively, $(X_1(0), X_2(0))$ and $(X_1(T), X_2(T))$ to the origin. 
\par Besides this geometric interpretation, the stochastic area is also interesting because it is non-zero only in the presence of a probability current and, therefore, can be used as a measure of the nonreversibility of a process~\cite{gonzalez2019experimental}. A positive value of the stochastic area is associated with an anti-clockwise current whereas a negative value of the stochastic area indicates the presence of a clockwise current. Investigating the fluctuations of the stochastic area and the manner in which they are manifested for the gradient system under consideration therefore allows us to understand the manner in which an equilibrium process can display nonequilibrium behavior, characterized by a non-zero probability current, for large $T$.\par
\begin{figure}[t]
    \centering
    \begin{minipage}{.45\textwidth}
    \includegraphics[width=6cm]{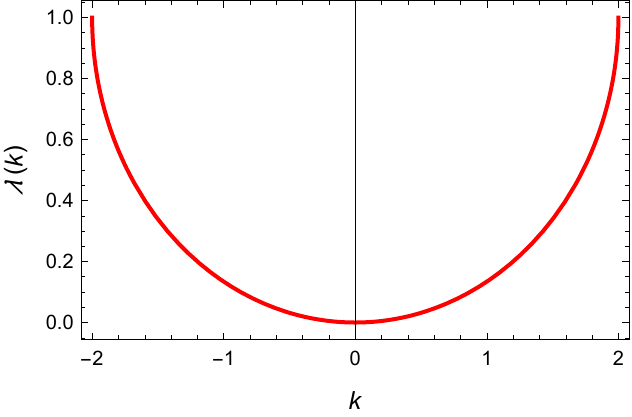}
    \end{minipage}
    \begin{minipage}{.45\textwidth}
    \includegraphics[width=6cm]{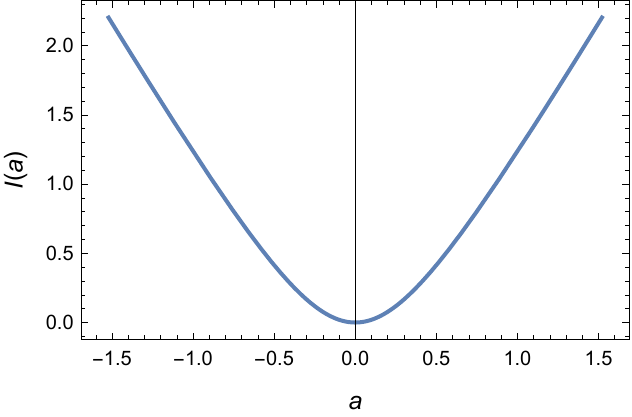}
    \end{minipage}
    \caption{SCGF (left) and rate function (right) for the stochastic area of the gradient system with parameter values $\gamma = 1$ and $\epsilon=1$. }
    \label{fig:gradarealoop1}
\end{figure}

 
The SCGF of the stochastic area is obtained immediately by substituting the value $\alpha = 1/2$ into the expression (\ref{scgfalpha}), yielding 
\begin{equation}
    \lambda(k) = \gamma - \sqrt{\gamma^2 - \frac{k^2\epsilon^4}{4}},\quad k \in \bigg(-\frac{2\gamma}{\epsilon^2}, \frac{2\gamma}{\epsilon^2} \bigg),
\end{equation}
while the rate function $I(a)$ is found as the Legendre transform of $\lambda(k)$. The SCGF and rate function are shown in Fig.~\ref{fig:gradarealoop1} for specific values of $\gamma$ and $\epsilon$
We note that the rate function is perfectly symmetric around the typical value $a^* = 0$ for which $I(a^*)=0$, and which corresponds to the asymptotic mean $\lambda'(0) = 0$. As shown in the preceding section, all antisymmetric linear current-type observables have zero asymptotic mean for the gradient system under consideration, owing to the fact that the system has an equilibrium stationary state with zero stationary current. The asymptotic variance is found from (\ref{varalpha}) for $\alpha = 1/2$ to be \begin{equation}
    \lambda''(0) =\frac{\epsilon^4}{4\gamma}.
\end{equation} 
The fact that the rate function is symmetric around around $a^*$ indicates that positive and negative fluctuations of the area loop are equally likely for this system, reflecting the lack of an inherent rotational bias for gradient drift and $D$ proportional to the identity. We also note that the rate function has asymptotically linear tails. 
\par
\begin{figure}[t]
    \centering
    \begin{minipage}{.45\textwidth}
    \includegraphics[width=6cm]{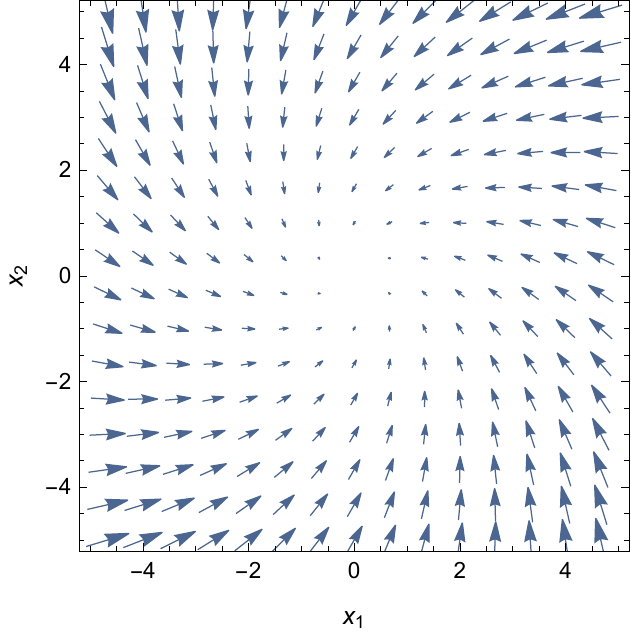}
    \end{minipage}
    \begin{minipage}{.45\textwidth}
    \includegraphics[width=6cm]{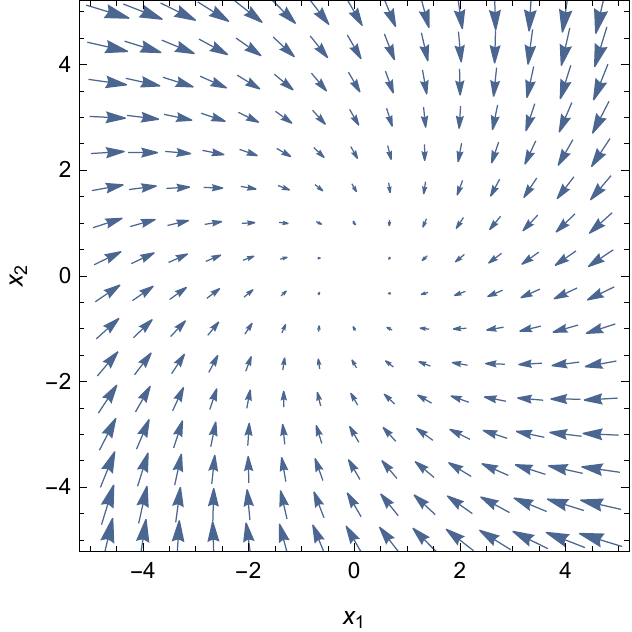}
    \end{minipage}
    \caption{Vector plot of the effective drift associated with the stochastic area of the gradient system for the parameter values $\gamma = 1$ and $\epsilon=1$. Left: $k=-1$. Right: $k=1$.}
    \label{fig:gradarealoop2}
\end{figure}

 To understand how fluctuations arise, we note that the effective drift has from (\ref{effdriftcur}) and (\ref{driftmatrix}) the form
\begin{equation}
    \bs{F}_k(\x) =-\begin{pmatrix}\sqrt{\gamma^2 - \frac{k^2\epsilon^4}{4}} && \frac{k\epsilon^2}{2}\\ -\frac{k\epsilon^2}{2} &&\sqrt{\gamma^2 - \frac{k^2\epsilon^4}{4}} \end{pmatrix} \x.
\end{equation}
We observe an anti-clockwise rotational component in the drift for $k > 0$ and a clockwise rotational component for $k<0$, associated respectively with an anti-clockwise and clockwise stationary current, respectively. This is illustrated Fig.~\ref{fig:gradarealoop2}, where vector plots of the effective drift are shown for various values of $k$. The stationary density and current associated with the effective process are found from (\ref{pkalpha}) and (\ref{curalpha}) as 
\begin{equation}
    p_k^*(\x) = \frac{\sqrt{\gamma^2 - \frac{k^2\epsilon^4}{4}}}{\pi\epsilon^2}\exp\left(-\frac{\sqrt{\gamma^2 - \frac{k^2\epsilon^4}{4}}}{\epsilon^2}\langle\x,\x\rangle \right)
\end{equation}
and 
\begin{equation}
    \bs{J}_{\bs{F}_k,p_k^*}(\x) = \frac{k\epsilon^2}{2}\begin{pmatrix}-x_2 \\ x_1 \end{pmatrix}p_k^*(\x),
\end{equation}
respectively. 
\par We observed in the preceding subsection that the effective drift, as well as the stationary density and stationary current, are here of the same form as that obtained for the transverse system evolving according to (\ref{sdetrans}). Moreover, the effective drift and associated stationary current and its properties has already been discussed in Sec.~\ref{sec:sec242}. For positive values of $k$ the nonequilibrium parameter $\xi_k = k\epsilon^2/2$ is positive and as such the effective drift is associated with an anti-clockwise circulating current. For negative $k$ we have that $\xi_k = k\epsilon^2/2$ is negative and the effective process has a circular clockwise stationary current. This is exactly the expected result: the stochastic area has a positive mean value in the presence of an anti-clockwise current so that fluctuations $a > 0$ (which are associated with values $k > 0$) are manifested via an effective process which has an anti-clockwise stationary current, whereas a negative value of the stochastic area in the long-time limit indicates the presence of a clockwise current so that fluctuations $a < 0$ (associated with $k <0$) are manifested by an effective process with a clockwise stationary current. 
\par Using this logic it is easy to understand why the antisymmetric part of the drift matrix for the effective drift differs from that of the original process: fluctuations of the stochastic area are manifested via the creation of currents and the antisymmetric part of the drift matrix is associated with the presence of a current. It is harder to understand why the symmetric part of the drift matrix for the effective process also differs from that of the original process; it might naively be thought that, since only the antisymmetric part of the drift is here associated with the presence of a current, only the antisymmetric part of the drift need be modified in order to manifest a given fluctuation of a current-type observable. However, it can be seen that the symmetric contribution to the effective drift given here by $2DB_k^*\x$ plays an essential role in manifesting fluctuations. In fact, given that the fluctuations $a$ are related to the parameter $k$ via $a(k) = \lambda'(k)$ and given that the SCGF has the general form (\ref{curscgf}), it is clear that the matrix $B_k^*$ associated with the density modifications is of fundamental importance in determining the particular fluctuation $a$ associated with a particular value of $k$ and, more fundamentally, in the existence of large deviations. For $B_k^* = 0$ the SCGF is zero and large deviations do not exist. The contribution due to $kD\Gamma\x$ in the effective drift can be seen as influencing both the direction and strength of the flow of the probability current while the density is altered in order to affect additional modifications to the magnitude of the current. These modifications together serve to realize a particular fluctuation. \par Notice that the situation is somewhat more complicated in the event that $D$ is not proportional to the identity: in this case the matrix $D\Gamma$ can have a symmetric component and the matrix $DB_k^*$ can have an antisymmetric component. Nevertheless the relation (\ref{curscgf}) continues to hold, emphasizing the importance of this term in determining the large deviations. 

\section{Linear current-type observables for the transverse diffusion}\label{sec:sec53}
In this section, we again consider current-type observables of the form (\ref{lincurobs}) but now consider them for the transverse system described earlier in Sec.~\ref{sec:sec242}, which we know is a nonequilibrium system having a non-zero stationary current. As before, we first consider the matrix $\Gamma$ featuring in the definition of the observable (\ref{lincurobs}) to be a general antisymmetric matrix. We then consider the case $\alpha = 1/2$ corresponding to the stochastic area. In addition, we consider two observables that are relevant in physics, namely,  the nonequilibrium work done on the transverse system and the entropy production. The nonequilibrium work is again related to an antisymmetric matrix $\Gamma$, while the entropy production contains also a non-zero symmetric part. 
\subsection{General $\alpha$} \label{sec:sec531}
We consider here the observable $A_T$ having the form (\ref{lincurobs}), with $\Gamma$ given as in (\ref{generalalpha}). The time-dependent matrix $B_k(t)$ underlying the generating function of $A_T$ also has the diagonal form (\ref{matrix3}) for the transverse system but with the coefficient $b_k(t)$ satisfying now the differential equation
\begin{equation}
    \frac{db_k(t)}{dt} = 2\epsilon^2 b_k(t)^2 - 2\gamma b_k(t) + \frac{k^2\epsilon^2\alpha^2}{2} + k\xi \alpha, \quad b_k(0) = 0. 
\end{equation}
By comparison with (\ref{scalarricc2}) and (\ref{timedepsol}) the solution for the above differential equation can be written as 
\begin{equation}
    b_k(t) = \frac{\gamma}{2\epsilon^2}\left(1 - \frac{1 + \frac{\sqrt{\gamma^2 - k\epsilon^2\alpha(2\xi + k\epsilon^2\alpha)}}{\gamma}\tanh\left(t\sqrt{\gamma^2 - k\epsilon^2\alpha(2\xi + k\epsilon^2\alpha)} \right)}{1 + \frac{\gamma}{\sqrt{\gamma^2 - k\epsilon^2\alpha(2\xi + k\epsilon^2\alpha)}}\tanh\left(t\sqrt{\gamma^2 - k\epsilon^2\alpha(2\xi + k\epsilon^2\alpha)}\right)} \right).
\end{equation}
The stationary solution $b_k^*$ is obtained either as the limit of the above as $t \rightarrow \infty$ or via solution of 
\begin{equation}
    2\epsilon^2 {b_k^*}^2 - 2\gamma b_k^* + \frac{k^2\epsilon^2\alpha^2}{2} + k\xi \alpha = 0 
\end{equation}
and choosing that solution which satisfies $b_0^* = 0$. We find that 
\begin{equation} \label{Aktransgen}
    b_k^* = \frac{\gamma - \sqrt{\gamma^2 - k\epsilon^2\alpha(2\xi + k\epsilon^2\alpha)}}{2\epsilon^2}.
\end{equation}
\par As before, the values of $k$ which are allowed and lead to an ergodic effective process are found by demanding that the eigenvalues of the drift matrix 
\begin{equation}\label{driftalpha}
   M_k = \begin{pmatrix}\sqrt{\gamma^2 - k\epsilon^2\alpha(2\xi + k\epsilon^2\alpha)} && k\epsilon^2 \alpha + \xi \\ - k \epsilon^2 \alpha - \xi && \sqrt{\gamma^2 - k\epsilon^2\alpha(2\xi + k\epsilon^2\alpha)} \end{pmatrix}
\end{equation}
of the effective drift (\ref{effdriftcur}) has strictly positive real part. The eigenvalues $\lambda_{\pm}$ of this matrix are given by 
\begin{equation}
    \lambda_{\pm} = \sqrt{\gamma^2 - k\epsilon^2\alpha(2\xi + k\epsilon^2\alpha)} \pm (k\epsilon^2 \alpha + \xi)i,
\end{equation}
which has positive real part for 
\begin{equation} \label{krange1}
    k \in \bigg(\frac{-\alpha\xi -  \sqrt{\alpha^2(\gamma^2 + \xi^2)}}{\alpha^2\epsilon^2}, \frac{-\alpha\xi +  \sqrt{\alpha^2(\gamma^2 + \xi^2)}}{\alpha^2\epsilon^2} \bigg).
\end{equation}
As a result, the SCGF is found in the usual manner from (\ref{curscgf}) and is given by 
\begin{equation} \label{scgfalpha2}
    \lambda(k) = \gamma - \sqrt{\gamma^2 - k\epsilon^2\alpha(2\xi + k\epsilon^2\alpha)}, 
\end{equation}
for $k$ satisfying (\ref{krange1}). 
\par The result is similar to that obtained previously in (\ref{scgfalpha}) for gradient diffusions. Of particular importance now is the fact that the SCGF satisfies the symmetry 
\begin{equation}\label{scgfsym}
    \lambda(k) = \lambda\bigg(-k - \frac{2\xi}{\alpha \epsilon^2} \bigg).  
\end{equation}
Such a symmetry is an example of a \textit{fluctuation relation}, specifically a Gallavoti-Cohen-type fluctuation relation~\cite{gallavotti1995dynamical,lebowitz1999gallavotti,touchette2018introduction}, and indicates that the probability density $P(A_T = a)$ satisfies, for large $T$, the relation
\begin{equation} \label{FR}
    \frac{P(A_T = a)}{P(A_T = -a)} = \exp\bigg(T\frac{2\xi}{\alpha \epsilon^2}a \bigg).
\end{equation}
For $\xi/\alpha > 0$ this indicates that fluctuations $A_T = a$ with $a > 0$ are exponentially more likely than the corresponding negative fluctuations $A_T = -a$, with the opposite being true for $\xi/\alpha < 0$. This reflects the fact that this system has an inherent rotational behavior controlled by the parameter $\xi$ and associated with a non-zero stationary current. As such the system will be more likely to manifest fluctuations that correspond to a current that resembles the current which is already present in the transverse system than with a current that differs in direction from the original current of the transverse system. 
\par The asymptotic mean and variance of the observable $A_T$ are found either directly from the SCGF (\ref{scgfalpha2}) or by application of the formalism developed in Sec.~\ref{sec:sec423}. It is found, either from (\ref{curmean}) or directly from (\ref{scgfalpha2}) that 
\begin{equation} \label{meanalpha2}
    \lambda'(0) = \frac{\epsilon^2\alpha \xi}{\gamma},
\end{equation}
so that for non-zero $\xi$ the observable $A_T$ has a non-zero stationary mean. This is expected given that the process under consideration has a non-zero stationary current. Note that the sign of the typical value $a^* = \lambda'(0)$ depends on the combination $\xi \alpha$. The asymptotic variance can be obtained via (\ref{curvar}) and (\ref{bigricc1}) or by direct calculation from the expression (\ref{scgfalpha2}) and is given by 
\begin{equation}\label{varalpha2}
    \lambda''(0) = \frac{\alpha^2\epsilon^4(\gamma^2 + \xi^2)}{\gamma^3}.
\end{equation}
\par Finally, the effective process has drift matrix (\ref{driftalpha}) explicitly given by 
\begin{equation} \label{effdriftalpha2}
    \bs{F}_k(\x) =- \begin{pmatrix}\sqrt{\gamma^2 - k\epsilon^2\alpha(2\xi + k\epsilon^2\alpha)} && k\epsilon^2 \alpha + \xi \\ - k \epsilon^2 \alpha - \xi && \sqrt{\gamma^2 - k\epsilon^2\alpha(2\xi + k\epsilon^2\alpha)} \end{pmatrix} \x,
\end{equation}
which corresponds to the transverse system with friction coefficient
\begin{equation}
    \gamma_k = \sqrt{\gamma^2 - k\epsilon^2\alpha(2\xi + k\epsilon^2\alpha)}
\end{equation}
and nonequilibrium parameter 
\begin{equation}
    \xi_k = k\epsilon^2 \alpha + \xi.
\end{equation}
As a result, its stationary density and current can be obtained from (\ref{rhotrans}) and (\ref{Jtrans}), respectively, via the substitution 
\begin{equation}
    \gamma \rightarrow \gamma_k,\quad  \xi \rightarrow \xi_k.
\end{equation}
The result is 
\begin{equation}\label{pkalpha2}
    p_k^*(\x) = \frac{\sqrt{\gamma^2 - k\epsilon^2\alpha(2\xi + k\epsilon^2\alpha)}}{\pi\epsilon^2 } \exp\left(-\frac{\sqrt{\gamma^2 - k\epsilon^2\alpha(2\xi + k\epsilon^2\alpha)}}{\epsilon^2}\langle\x,\x\rangle \right)
\end{equation}
and 
\begin{equation}\label{Jalpha2}
    \bs{J}_{\bs{F}_k, p_k^*}(\x) = \left(k\epsilon^2 \alpha + \xi \right) \begin{pmatrix}-x_2 \\ x_1 \end{pmatrix}p_k^*(\x).
\end{equation}
The stationary current is non-zero for all values of $k$ apart from 
\begin{equation}
    k = -\frac{\xi}{\epsilon^2 \alpha},
\end{equation}
for which the current vanishes. It is therefore clear that this value of $k$ should correspond to the fluctuation $A_T = 0$. This can be verified by solving $a(k) = \lambda'(k) = 0$ for $k$. 
\subsection{Stochastic area}\label{sec:sec532}
For $\alpha = 1/2$, the value corresponding to the stochastic area, the result (\ref{scgfalpha2}) for the SCGF becomes
\begin{equation}
    \lambda(k) = \gamma - \sqrt{\gamma^2 - k\epsilon^2\left(\xi + \frac{k\epsilon^2}{4} \right)}, \quad k \in \bigg(\frac{-2\xi - 2 \sqrt{\gamma^2 + \xi^2}}{\epsilon^2}, \frac{-2\xi + 2 \sqrt{\gamma^2 + \xi^2}}{\epsilon^2} \bigg).
\end{equation}
The SCGF exhibits from (\ref{scgfsym}) the symmetry 
\begin{equation}  \label{scgfsym2}
    \lambda(k) = \lambda\left(-k - \frac{4\xi}{\epsilon^2}\right), 
\end{equation}
which means that the probability density $P(A_T = a)$ for the stochastic area satisfies the fluctuation relation 
\begin{equation} \label{FR2}
    \frac{P(A_T = a)}{P(A_T = -a)} = \exp\left(T\frac{4\xi}{\epsilon^2}a \right) 
\end{equation}
for large $T$. Given that a positive value of the stochastic area is associated with an anti-clockwise current (with the opposite being true for a negative value of the stochastic area), the fluctuation relation expresses the fact that for $\xi > 0$ a positive fluctuation $A_T = a$ is exponentially more likely than the corresponding negative fluctuation for large $T$, while the opposite holds for $\xi < 0$. In this manner the inherent rotation present in the transverse system provides a bias for manifesting fluctuations with a particular sign more easily than fluctuations with the opposite sign. \par The asymptotic mean and variance are given from (\ref{meanalpha2}) and (\ref{varalpha2}) by \begin{equation}
    \lambda'(0) = \frac{\epsilon^2 \xi}{2\gamma}
\end{equation}
and 
\begin{equation}
    \lambda''(0) = \frac{\epsilon^4(\gamma^2 + \xi^2)}{4\gamma^3},
\end{equation}
respectively. The mean value is seen to depend on the sign of $\xi$, with a positive mean value for $\xi > 0$ and a negative mean value for $\xi < 0$. This is expected in light of similar reasoning as that provided for the above analysis regarding the physical meaning of the fluctuation relation (\ref{FR2}). 
\begin{figure}[t]
    \centering
    \begin{minipage}{.45\textwidth}
    \includegraphics[width=6cm]{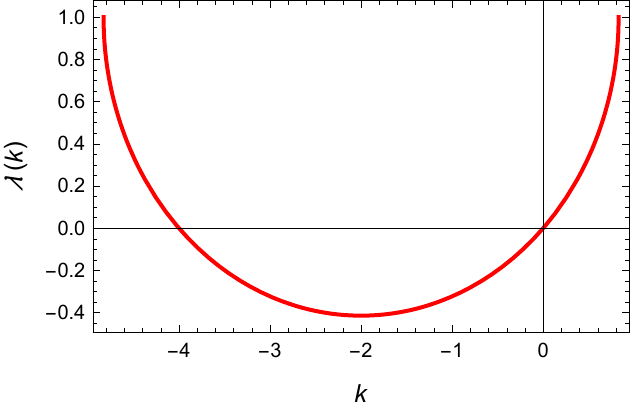}
    \end{minipage}
    \begin{minipage}{.45\textwidth}
    \includegraphics[width=6cm]{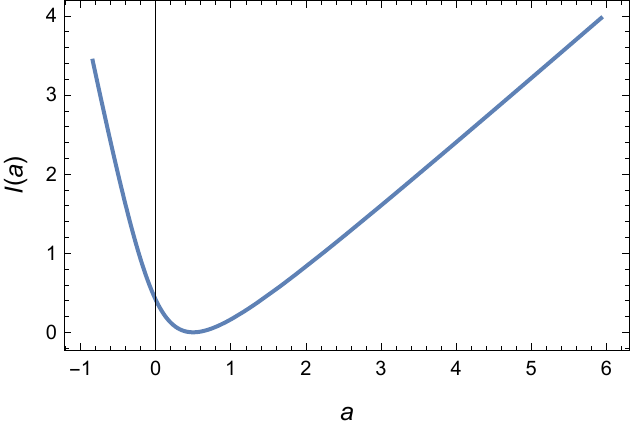}
    \end{minipage}
    \caption{SCGF (left) and rate function (right) for the stochastic area of the transverse system for the parameter values $\gamma = 1,\xi=1$ and $\epsilon=1$. }
    \label{fig:transarealoop1}
\end{figure}
\par The SCGF and rate function are shown for specific parameter values in Fig.~\ref{fig:transarealoop1}. The SCGF is seen to be symmetric around $k = - 2\xi/\epsilon^2$, in correspondence with (\ref{scgfsym2}). Furthermore, the rate function is seen to be asymmetric around the typical value $a^* = \epsilon^2\xi/2\gamma$, with the rate function being steeper for $a < a^*$ than $a > a^*$, indicating the presence of non-Gaussian fluctuations. 
\par The effective drift is given from (\ref{effdriftalpha2}) by 
\begin{equation}
    \bs{F}_k(\x) = - \begin{pmatrix}\sqrt{\gamma^2 - k\epsilon^2\left(\xi + \frac{k\epsilon^2}{4} \right)}&& \xi + \frac{k\epsilon^2}{2} \\ -\xi - \frac{k\epsilon^2}{2}&& \sqrt{\gamma^2 - k\epsilon^2\left(\xi + \frac{k\epsilon^2}{4} \right)} \end{pmatrix}\x,
\end{equation}
with the associated stationary density and current obtained from (\ref{pkalpha2}) and (\ref{Jalpha2}) and given as 
\begin{equation}
    p^*_k(\x) = \frac{\sqrt{\gamma^2 - k\epsilon^2\left(\xi + \frac{k\epsilon^2}{4} \right)}}{\pi\epsilon^2}\exp \left(-\frac{\sqrt{\gamma^2 - k\epsilon^2\left(\xi + \frac{k\epsilon^2}{4} \right)}}{\epsilon^2}\langle\x, \x\rangle \right)
\end{equation}
and 
\begin{equation} \label{effcurtransstoc}
    \bs{J}_{\bs{F}_k,p_k^*}(\x) = \left( \xi + \frac{k\epsilon^2}{2}\right) \begin{pmatrix}-x_2 \\ x_1 \end{pmatrix}p_k^*(\x),
\end{equation}
respectively. 
\begin{figure}[t]\centering
\subfloat[$k = 0.5$ ($k > 0$)]{\includegraphics[width=.45\linewidth]{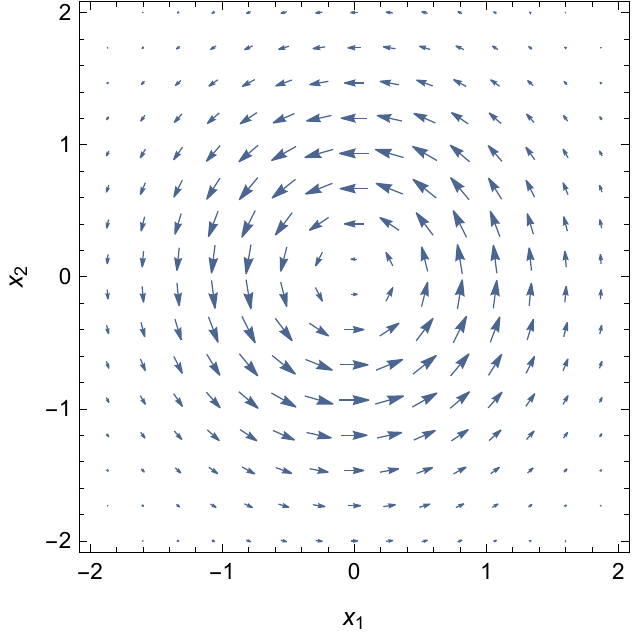}}\hfill
\subfloat[$k = -1$ ($-2\xi/\epsilon^2 < k < 0$)]{\includegraphics[width=.45\linewidth]{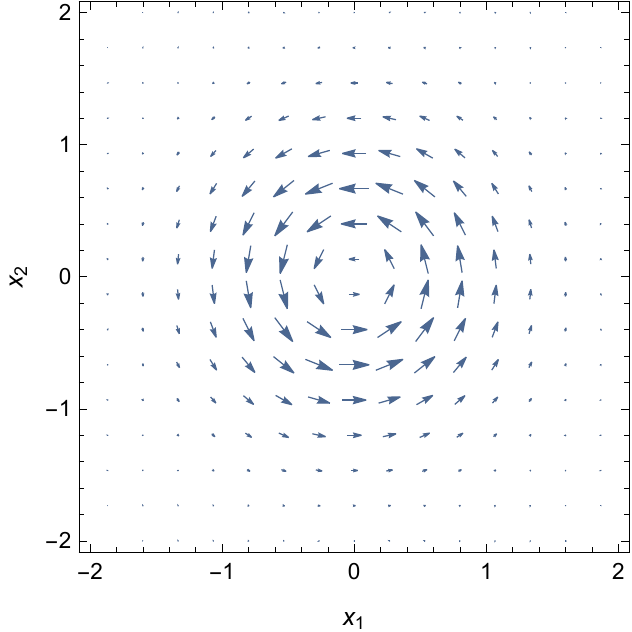}}\par 
\subfloat[$k=-3$ ($k < -2\xi/\epsilon^2$)]{\includegraphics[width=.45\linewidth]{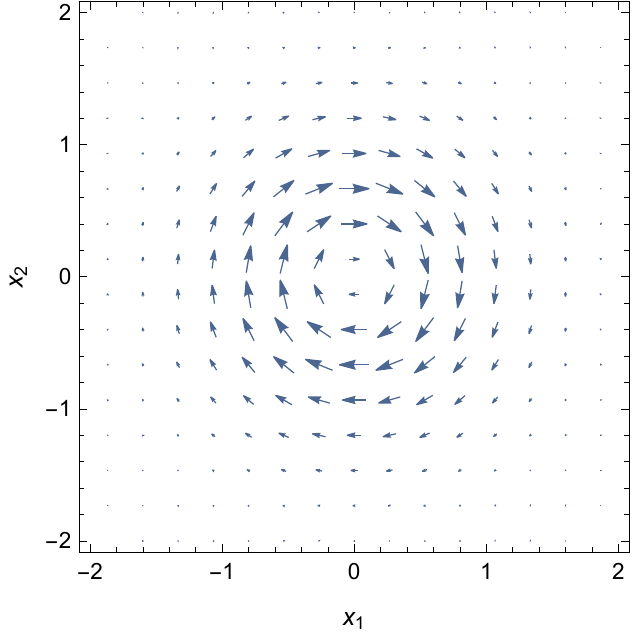}}
\caption{Vector plot of the stationary current for the effective process associated with the transverse diffusion and stochastic area for various values of $k$ for the parameters $\xi =1,\gamma =1$, and $\epsilon=1$.}
\label{fig:Jkstocarea}
\end{figure}
Vector plots of the stationary current associated with the effective process are shown in Fig.~\ref{fig:Jkstocarea} for the three cases $k > 0$, $-2\xi/\epsilon^2 < k < 0$ and $k < -2\xi/\epsilon^2$ for the case where $\xi = 1, \gamma = 1$ and $\epsilon = 1$. A consistent scale relating the size of the arrows and the magnitude of the vector field is used for these 3 figures so as to allow easy comparison of both the direction and magnitude of the vector fields for different values of $k$.
\par For $\xi > 0$ we see from Fig.~\ref{fig:Jkstocarea} as well as the expression (\ref{effcurtransstoc}) that fluctuations $a > a^*$, associated with $k > 0$, are manifested by increasing the magnitude of the anti-clockwise rotational component in the effective drift in comparison to the original drift. Positive fluctuations $0 < a < a^*$ associated with $-2\xi/\epsilon^2 < k < 0$ are manifested by decreasing the magnitude of the rotational component in the effective drift, but with the direction of the rotation still seen to be anti-clockwise. For $k = -2\xi/\epsilon^2$ the anti-symmetric component of the drift vanishes and the stochastic area is zero. This represents an equilibrium fluctuation of a nonequilibrium process and is therefore manifested via an equilibrium effective process. Finally, fluctuations $a < 0$ correspond to $k < -2\xi/\epsilon^2$, for which the anti-symmetric component of the effective drift has the opposite sign as that of $\xi > 0$ and for which the effective process has a clockwise stationary current. A similar analysis holds (with opposite results) for $\xi < 0$.

\subsubsection{Nonequilibrium work}\label{sec:sec5321}
For linear diffusions there are many quantities of physical interest in nonequilibrium statistical physics that have the form of the linear current-type observable (\ref{lincurobs}) introduced previously. The study of such quantities forms part of the formalism of stochastic thermodynamics or stochastic energetics~\cite{seifert2012stochastic, sekimoto1998langevin,sekimoto2010stochastic}, which is concerned with extending the traditional notions and laws of thermodynamics to individual realizations of the stochastic dynamics governing a system. In stochastic thermodynamics, quantities such as work, heat and entropy take the form of time-integrated functionals of the system's state. Defined in this manner, these thermodynamic quantities can be shown to satisfy the first and second laws of thermodynamics. For a review of stochastic energetics we refer to~\cite{seifert2012stochastic}.
\par Two of the most important quantities in stochastic thermodynamics are the nonequilibrium work done on the system and defined for a linear system (in its time-averaged form) as
\begin{equation} \label{neqwork}
    \mathcal{W}_T = \frac{1}{T} \int_0^T \left(M^{\mathsf{T}}D^{-1}\bs{X}(t) - D^{-1}M \bs{X}(t)\right) \circ \, d\bs{X}(t),
\end{equation}
and the entropy production defined as 
\begin{equation} \label{entropyproduction}
    \mathcal{E}_T = -\frac{1}{T} \int_0^T 2 D^{-1} M \bs{X}(t) \circ d\bs{X}(t).
\end{equation}
It is important to note that the nonequilibrium work constitutes the antisymmetric part of the entropy production, and as such constitutes a linear current-type observable (\ref{lincurobs}) with purely antisymmetric $\Gamma$. For the particular system under consideration, we have that the expression for the nonequilibrium work is
\begin{equation} \label{neqworktrans}
    \mathcal{W}_T = \frac{1}{T} \int_0^T \begin{pmatrix}0 && -2\xi/\epsilon^2 \\ 2\xi/\epsilon^2 && 0 \end{pmatrix}\bs{X}(t)\circ d\bs{X}(t),
\end{equation}
which corresponds to the observable (\ref{lincurobs}) with $\Gamma$ as in (\ref{generalalpha}) for
\begin{equation} \label{alphaneq}
    \alpha = \frac{2\xi}{\epsilon^2}.
\end{equation}
\par In contrast to the stochastic area, which has a definition independent of the system under consideration and always has the same form, the nonequilibrium work done on a system depends on the particular SDE satisfied by that system. For a linear system the nonequilibrium work is therefore defined in terms of the drift matrix $M$ and diffusion matrix $D$ associated with that linear diffusion. This fact will be of fundamental importance in understanding the differences in the large deviations associated with the stochastic area versus those of the nonequilibrium work.
\par We observed previously that a positive value of the stochastic area is associated with an anti-clockwise current while a negative value of the stochastic area is associated with a clockwise current. This is no longer the case for the nonequilibrium work, as can be seen from the definition (\ref{neqworktrans}): for $\xi > 0$ an anti-clockwise current is associated with a positive value of the nonequilibrium work, while for $\xi < 0$ a clockwise current is associated with a positive value of the nonequilibrium work. Given that the transverse system has an anti-clockwise stationary current for $\xi > 0$ and a clockwise stationary current for $\xi < 0$ this immediately implies that the asymptotic mean of the nonequilibrium work is positive regardless of the value of $\xi$ (assuming $\xi$ is non-zero) whereas for the stochastic area the sign of the asymptotic mean depended on the sign of $\xi$. 
\par The SCGF for $\mathcal{W}_T$ can be obtained by substituting (\ref{alphaneq}) into the expression (\ref{scgfalpha2}). We find that
\begin{equation} \label{scgftranswork}
    \lambda(k) = \gamma - \sqrt{\gamma^2 - 4k(1+k)\xi^2},
\end{equation}
which is immediately seen to be independent of the noise strength $\epsilon$, due to the presence of $D^{-1}$ in the definition (\ref{neqwork}). The SCGF satisfies here the symmetry 
\begin{equation} \label{lambdasym}
    \lambda(k) = \lambda(-k - 1)
\end{equation}
so that the SCGF is symmetric around $k = -1/2$ and the density $P(\mathcal{W}_T=w)$ satisfies the fluctuation relation 
\begin{equation} \label{FR3}
    \frac{P(\mathcal{W}_T = w)}{P(\mathcal{W}_T = -w)} = e^{Tw}
\end{equation}
for large $T$, so that positive values $w$ of the nonequilibrium work are always exponentially more likely than the corresponding negative value $-w$. 
\par The asymptotic mean and variance for the nonequilibrium work done on the transverse system can be found either directly via explicit differentiation of $\lambda(k)$ or via (\ref{curmean}) and (\ref{curvar}). The result is
\begin{equation}
    w^* = \lambda'(0) = \frac{2\xi^2}{\gamma}
\end{equation}
for the asymptotic mean and
\begin{equation}
    \lambda''(0) = \frac{4\xi^2(\gamma^2 + \xi^2)}{\gamma^3}
\end{equation}
for the asymptotic variance. The typical value $w^*$ is seen to be proportional to $\xi^2$ and is always positive. This is the expected result in light of our previous discussions regarding the nature of the nonequilibrium work and the form the fluctuation relation (\ref{FR3}) takes. Furthermore, this result is expected also from thermodynamic considerations. It was mentioned previously that the nonequilibrium work constitutes the antisymmetric part of the entropy production. From the second law of thermodynamics we know that the average entropy production is always greater than or equal to zero and as such the typical value for the nonequilibrium work is also expected to be greater than or equal to zero. 
\par
\begin{figure}[t]
    \centering
    \begin{minipage}{.45\textwidth}
    \includegraphics[width=6cm]{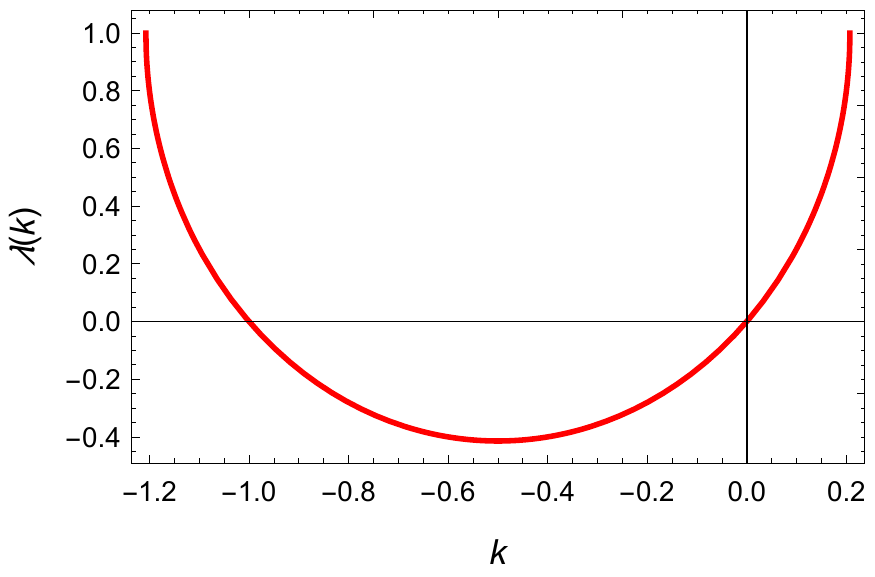}
    \end{minipage}
    \begin{minipage}{.45\textwidth}
    \includegraphics[width=6cm]{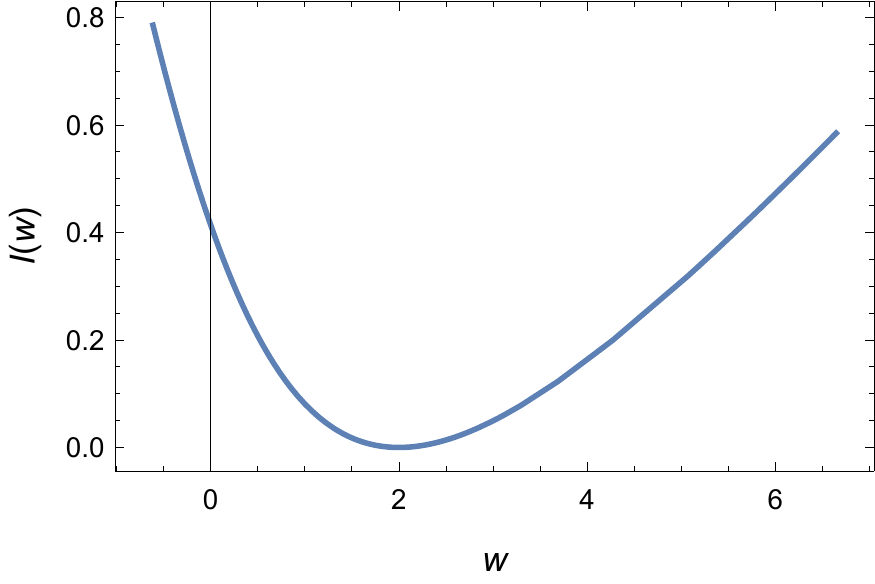}
    \end{minipage}
    \caption{SCGF (left) and rate function (right) for the nonequilibrium work done on the transverse system for the parameter values $\gamma = 1,\xi=1$ and $\epsilon=1$. }
    \label{fig:transwork1}
\end{figure}

 The SCGF and rate function $I(w)$ obtained as the Legendre transform of $\lambda(k)$ are shown in Fig.~\ref{fig:transwork1}. We note the asymmetry of the rate function around the typical value $w^*$, which serves as a visual confirmation of the fluctuation relation (\ref{FR3}). We note also that the same plot would be obtained if a negative value were used for $\xi$: this can be seen explicitly also from the expression (\ref{scgftranswork}) for the SCGF, which depends only on the magnitude of $\xi$ and not on its sign. It can also be seen explicitly that the SCGF is symmetric around $k = -1/2$. 
\par
\begin{figure}[t]
    \centering
            \begin{subfigure}[b]{.45\textwidth}
                \centering
                \includegraphics[width=6cm]{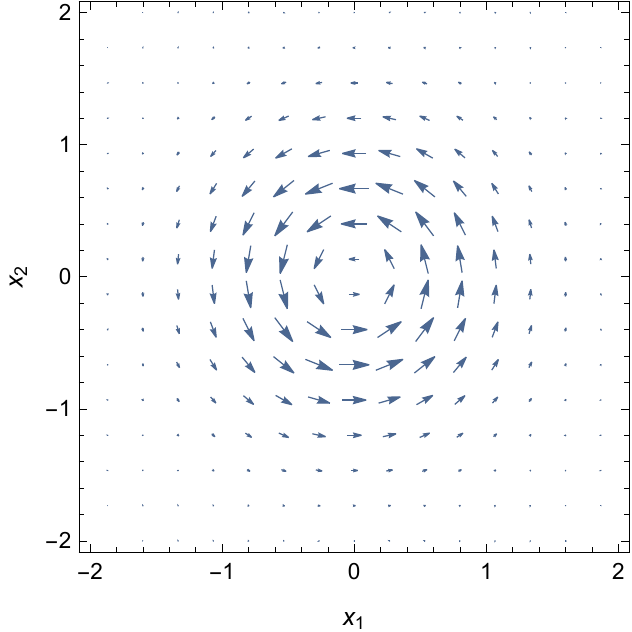}
                \caption{$k = -\frac{1}{4}$}
                
            \end{subfigure}
            \begin{subfigure}[b]{.45\textwidth}
            \centering
                \includegraphics[width=6cm]{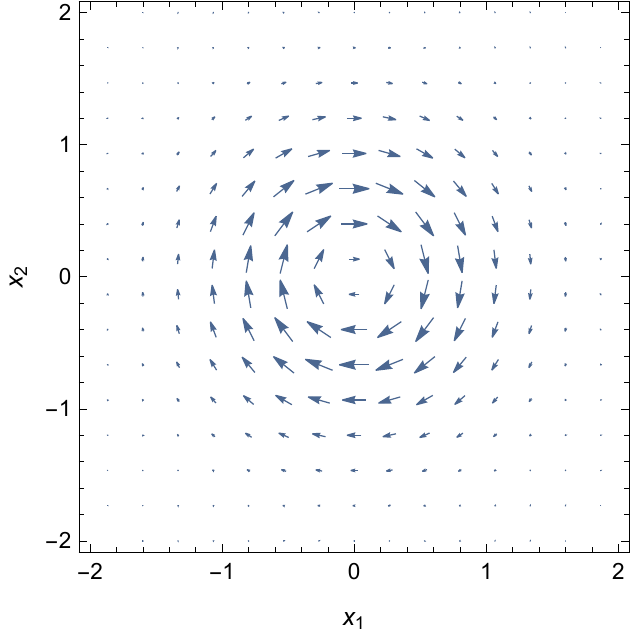}
                \caption{$k = -\frac{3}{4}$}
            \end{subfigure}
        \caption{Vector plot of the stationary current of the effective process associated with the nonequilibrium work done on the transverse system with $\xi =1$ for various values of $k$ and $\gamma =1, \epsilon =1$.}
        \label{fig:Jkwork1}
\end{figure}

\begin{figure}[h]
    \centering
            \begin{subfigure}[b]{.45\textwidth}
                \centering
                \includegraphics[width=6cm]{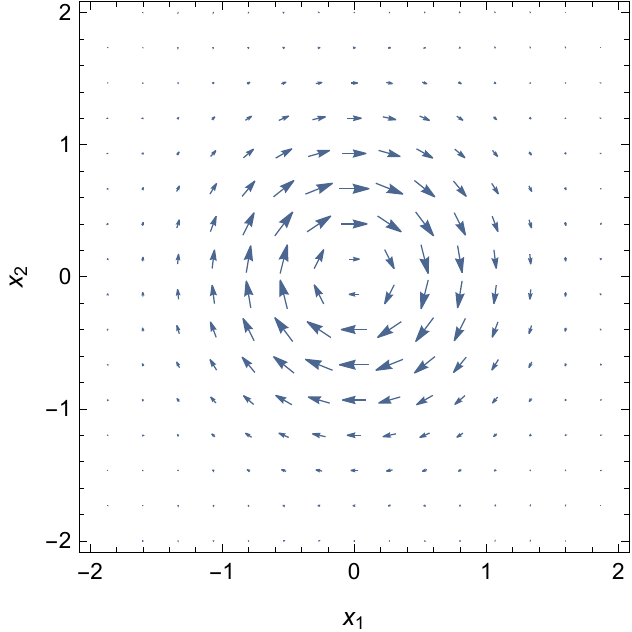}
                \caption{$k = -\frac{1}{4}$}
                
            \end{subfigure}
            \begin{subfigure}[b]{.45\textwidth}
            \centering
                \includegraphics[width=6cm]{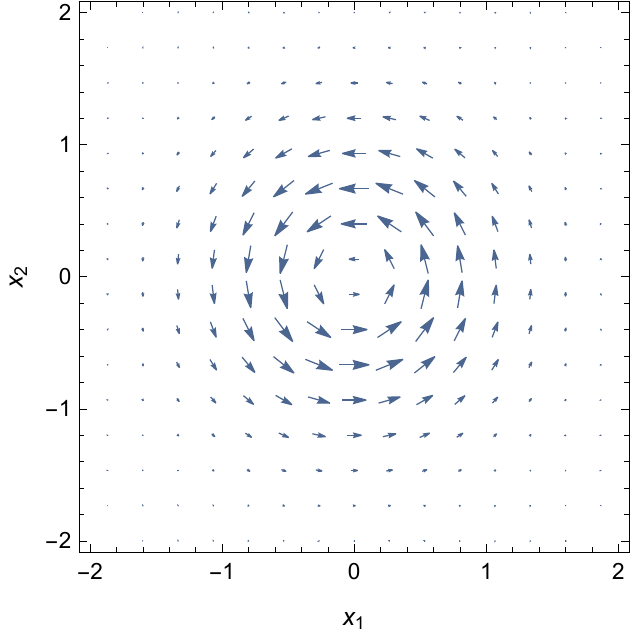}
               \caption{$k = -\frac{3}{4}$}
            \end{subfigure}
        \caption{Vector plot of the stationary current of the effective process associated with the nonequilibrium work done on the transverse system with $\xi = -1$ for various values of $k$ and $\gamma =1, \epsilon =1$.}
        \label{fig:Jkwork2}
\end{figure}

 The effective drift $\bs{F}_k$ is found by substituting the relevant value for $\alpha$ as in (\ref{alphaneq}) into (\ref{effdriftalpha2}) to obtain 
\begin{equation}
    \bs{F}_k(\x) = -\begin{pmatrix}\sqrt{\gamma^2 - 4k(1+k) \xi^2} && \xi(1 + 2k) \\ -\xi(1 + 2k) && \sqrt{\gamma^2 - 4k(1+k)\xi^2} \end{pmatrix} \x.
\end{equation}
This drift is associated with an ergodic process for 
\begin{equation} \label{krange}
    k \in \bigg(\frac{-\xi^2 - \sqrt{\gamma^2 \xi^2+ \xi^4}}{2\xi^2}, \frac{-\xi^2 + \sqrt{\gamma^2 \xi^2+ \xi^4}}{2\xi^2} \bigg).
\end{equation}Finally, we show the stationary density and current associated with the effective process. We have from (\ref{pkalpha2}) that 
\begin{equation}
    p_k^*(\x) = \frac{\sqrt{\gamma^2 - 4k (1+k)\xi^2}}{\pi\epsilon^2} \exp\left(-\frac{\sqrt{\gamma^2 - 4k (1+k)\xi^2}}{\epsilon^2}\langle \x, \x\rangle \right)
\end{equation}
for the stationary density with stationary covariance matrix given by
\begin{equation} \label{Cktrans}
    C_k = \frac{\epsilon^2}{2 \sqrt{\gamma^2 - 4k(1+k)\xi^2}}\mathbb{I},
\end{equation}
and from (\ref{Jalpha2}) we obtain 
\begin{equation} \label{workcurrent1}
    J_{\bs{F}_k, p_k^*}(\x) = \xi(1 + 2k) \begin{pmatrix} -x_2 \\ x_1 \end{pmatrix}p_k^*(\x)
\end{equation}
for the associated stationary density. The stationary current is shown for various values of $k$ for $\xi > 0$ in Fig.~\ref{fig:Jkwork1} and for $\xi < 0$ in Fig.~\ref{fig:Jkwork2}.
\par For $\xi > 0$ it is clear from the expression for the stationary current of the effective process that fluctuations $w > w^*$ associated with $k>0$ will be manifested by a current having greater anti-clockwise magnitude than that of the original process. For $-1/2 < k < 0$ we find that the current is anti-clockwise but with decreased magnitude compared to the original process. For $k = -1/2$ which is associated with the fluctuation $w = 0$ which occurs only in the presence of zero stationary current we observe that the stationary current of the effective process vanishes. Finally, for the fluctutations $w < 0$ associated with $k < -1/2$ we find that the stationary current is clockwise with increasing magnitude as $k$ decreases. 
\par For $\xi < 0$, for which the transverse process has a clockwise circulating current, we can repeat the above analysis. In this case values $w > 0$ of the nonequilibrium work are now associated with a clockwise stationary current and as such for $k > 0$ fluctuations $w> w^*$ are manifested by increasing the magnitude of the clockwise current. For fluctuations $0 < w < w^*$ associated with $-1/2 < k < 0$ the magnitude of the current is decreased, but the direction of the current remains clockwise. Finally fluctuations $w < 0$ are manifested by an anti-clockwise current, with the magnitude of this current increasing for decreasing $k$ in $k < -1/2$. 

\subsection{Entropy production}\label{sec:sec534}
We complete our study of the transverse system by considering the large deviations of the entropy production defined earlier in (\ref{entropyproduction}). It was noted there that the nonequilibrium work constitutes the antisymmetric part of the entropy production, so that we can write 
\begin{equation}
    \mathcal{E}_T = \mathcal{W}_T + \frac{1}{T}\int_0^T \Gamma^{+} \bs{X}(t) \circ d\bs{X}(t)
\end{equation}
with 
\begin{equation}
    \Gamma^{+} = -\frac{2\gamma}{\epsilon^2}\mathbb{I}. 
\end{equation}
Following our discussion regarding general linear current-type observables in Sec.~\ref{sec:sec413}, we now must find the value of $k$ for which the matrix $\mathcal{B}_k$, as given in (\ref{posdef}), is positive definite, where $C_k$
is as given in (\ref{Cktrans}) and $B_k^*$ is found by substituting $\alpha = 2\xi /\epsilon^2$ in (\ref{Aktransgen}) to be 
\begin{equation}
    B_k^* = \frac{\gamma - \sqrt{\gamma^2 - 4k(1+k)\xi^2}}{2\epsilon^2} \, \mathbb{I}. 
\end{equation}
The relevant value of $k$ for which $\mathcal{B}_k$ ceases to be positive definite is found to be negative and is given by 
\begin{equation}
    k_{-} = -1,
\end{equation}
independent of the values of the parameters $\gamma, \xi$ and $\epsilon$. 
\par
\begin{figure}[t]
    \centering
    \includegraphics[width=0.5\textwidth]{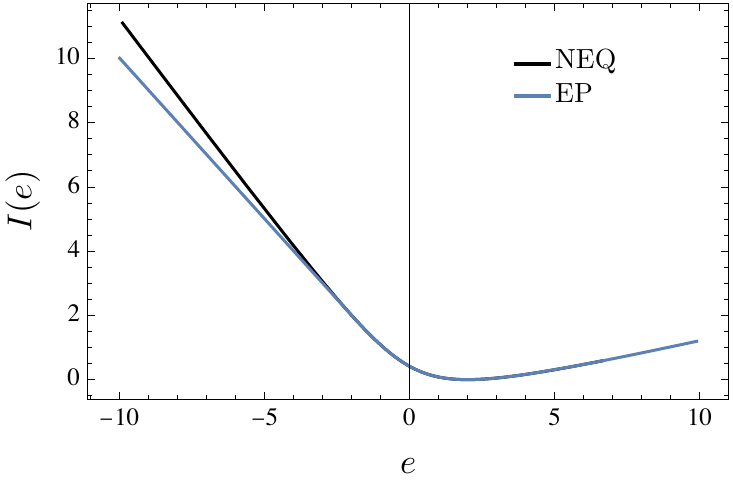}
    \caption{Rate functions for the nonequilibrium (NEQ) work and entropy production (EP) are shown for $\gamma = 1$, $\xi = 1$ and $\epsilon = 1$.}
    \label{fig:ratecomp1}
\end{figure}

 Following the discussion regarding general linear current type observables in Sec.~\ref{sec:sec413}, the SCGF for the entropy production matches that obtained for the nonequilibrium work only for 
\begin{equation}
    k \in \bigg[-1,\frac{-\xi^2 + \sqrt{\gamma^2 \xi^2+ \xi^4}}{2\xi^2}\bigg),
\end{equation}
while the rate function $\tilde{I}(e)$ for the entropy production now satisfies from (\ref{ratelinear2})
\begin{equation}
    \tilde{I}(e) = \begin{cases}-e & e < \frac{-2\xi^2}{\gamma} \\ I(e) & e \geq \frac{-2\xi^2}{\gamma} \end{cases},
\end{equation}
with the crossover value $e_{-}$ found from 
\begin{equation}
    e_{-} = \lambda'(-1) = -\frac{2\xi^2}{\gamma}
\end{equation}
and given that $\lambda(-1) = \lambda(0) = 0$ following from the Gallavoti-Cohen symmetry (\ref{lambdasym}).
The rate function for the entropy production and the nonequilibrium work are compared for a specific set of parameters in Fig.~\ref{fig:ratecomp1}. We observe that the rate functions are identical for $e \geq e_{-}$ while the entropy production has a linear rate function for $e < e_{-}$ as a result of the fact that the SCGF is divergent for all $k \leq k_{-}$. 
\par We conclude by mentioning that the SCGF and rate function of the nonequilibrium work and entropy production were obtained previously by Noh \cite{noh2014fluctuations} for the transverse system using path integral methods. Here we have obtained these results using different methods and have studied in addition the effective process associated with the fluctuations of these observables, allowing us to study the manner in which fluctuations are created. 
\section{Nonequilibrium work for the spring coupled system with two temperatures} \label{sec:sec54}
As a final application we study the fluctuations of the nonequilibrium work done on the spring system introduced in Sec.~\ref{sec:sec243} involving two heat baths at different temperatures. From the SDE (\ref{sdespring}) and the definition (\ref{neqwork}) we have that the nonequilibrium work done on this system takes the form 
\begin{equation}
    \mathcal{W}_T = \frac{1}{T} \int_0^T \begin{pmatrix}0 && -\frac{\kappa(\epsilon_1^2 - \epsilon_2^2)}{\epsilon_1^2 \epsilon_2^2} \\ \frac{\kappa(\epsilon_1^2 - \epsilon_2^2)}{\epsilon_1^2 \epsilon_2^2} && 0\end{pmatrix} \bs{X}(t) \circ d\bs{X}(t),
\end{equation}
so that it corresponds to 
\begin{equation}
    \alpha = \frac{\kappa(\epsilon_1^2 - \epsilon_2^2)}{\epsilon_1^2 \epsilon_2^2}
\end{equation}
in the notation used before. 
\par The generating function of the nonequilibrium work cannot be obtained analytically for this system, contrary to the transverse system, since $B_k(t)$ in the Riccati equation (\ref{riccatidiffcur}) does not here have a diagonal form, due to the off-diagonal symmetric part of the drift matrix $M$. As far as we know, an exact solution for $G_k(\x,t)$ for all times $t$ is not possible. Nevertheless, the large deviations can be found in the same manner as before by obtaining the appropriate stationary solution $B_k^*$ to the algebraic Riccati equation (\ref{ricccur}). 
\par The result for $B_k^*$ is too long to show, but it can be shown that the corresponding $M_k$ given as in (\ref{mkcur}) is positive definite for $k$ in the range
\begin{equation} \label{krange3}
    k \in (k^{-},k^{+})
\end{equation}
where 
\begin{equation} \label{krange4}
    k^{\pm} = \frac{-\kappa  \text{$\epsilon_1$}^2+\kappa  \text{$\epsilon_2$}^2\pm\sqrt{4 \gamma ^2 \text{$\epsilon_1 $}^2 \text{$\epsilon_2$}^2+8 \gamma  \kappa  \text{$\epsilon_1$}^2 \text{$\epsilon_2$}^2+\kappa ^2 \left(\text{$\epsilon_1$}^2+\text{$\epsilon_2$}^2\right)^2}}{2 \kappa  \left(\text{$\epsilon_1$}^2-\text{$\epsilon_2$}^2\right)},
\end{equation}
so that the associated $r_k$ and $l_k$ as in (\ref{normrk}) and (\ref{normlk}) are normalizeable for these $k$ and therefore constitute valid eigenfunctions corresponding to the eigenvalue (\ref{curscgf}) given by 
\begin{equation} \label{scgfspring}
    \lambda(k) = \gamma + \kappa - \sqrt{\gamma^2 + 2\gamma \kappa - \kappa^2 \frac{\left((1+k)\epsilon_1^2 - k\epsilon_2^2 \right) \left(k \epsilon_1^2 - (1+k)\epsilon_2^2\right)}{\epsilon_1^2 \epsilon_2^2} },
\end{equation}
which is continuous in $k$ and has $\lambda(0) = 0$. Therefore this eigenvalue represents the SCGF, and the large deviation eigenfunctions $r_k$ and $l_k$ are well defined and lead to an ergodic effective process for $k$ as in (\ref{krange3}) and (\ref{krange4}). 
\par The SCGF satisfies the Gallavoti-Cohen symmetry 
\begin{equation} \label{sym}
    \lambda(k) = \lambda(-k -1),
\end{equation}
which was obtained also for the nonequilibrium work done on the transverse system. As a result we have again a fluctuation relation of the form 
\begin{equation} \label{FR4}
    \frac{P(\mathcal{W}_T = w)}{P(\mathcal{W}_T = -w)} = e^{Tw}
\end{equation}
for the nonequilibrium work done on the spring system, which has a similar interpretation as that which was obtained for the transverse system. 
\par The asymptotic mean (typical value $w^*$ for the nonequilibrium work) is obtained in the usual manner from the SCGF (\ref{scgfspring}) 
and is found to be 
\begin{equation}
    w^* = \frac{\kappa^2(\epsilon_1^2 - \epsilon_2^2)^2}{2(\gamma + \kappa)\epsilon_1^2 \epsilon_2^2},
\end{equation}
which is seen to be positive for all allowed values of the parameters. This expresses again the fact that the expected entropy production due to the nonequilibrium work is always greater than or equal to zero. The asymptotic variance is not shown here because the expression is too long. 
\begin{figure}[t]
    \centering
    \begin{minipage}{.45\textwidth}
    \includegraphics[width=6cm]{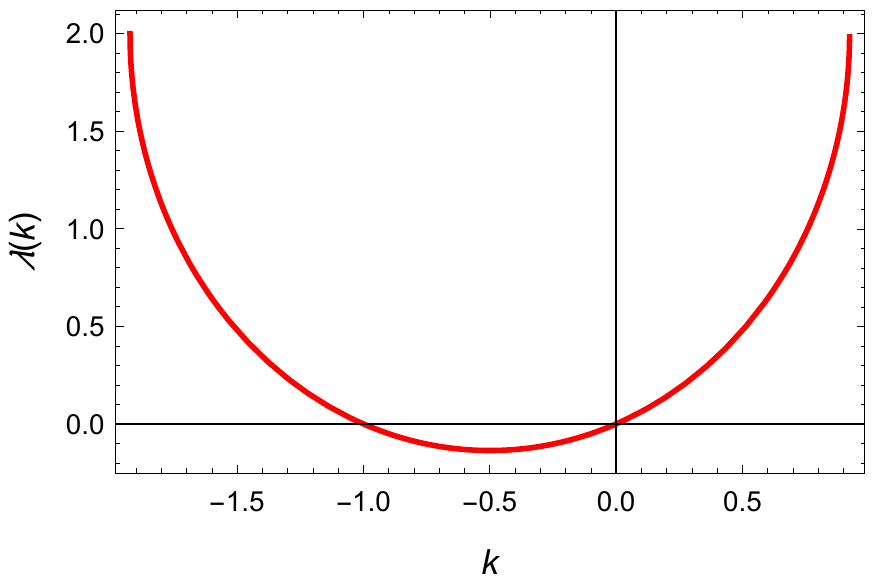}
    \end{minipage}
    \begin{minipage}{.45\textwidth}
    \includegraphics[width=6cm]{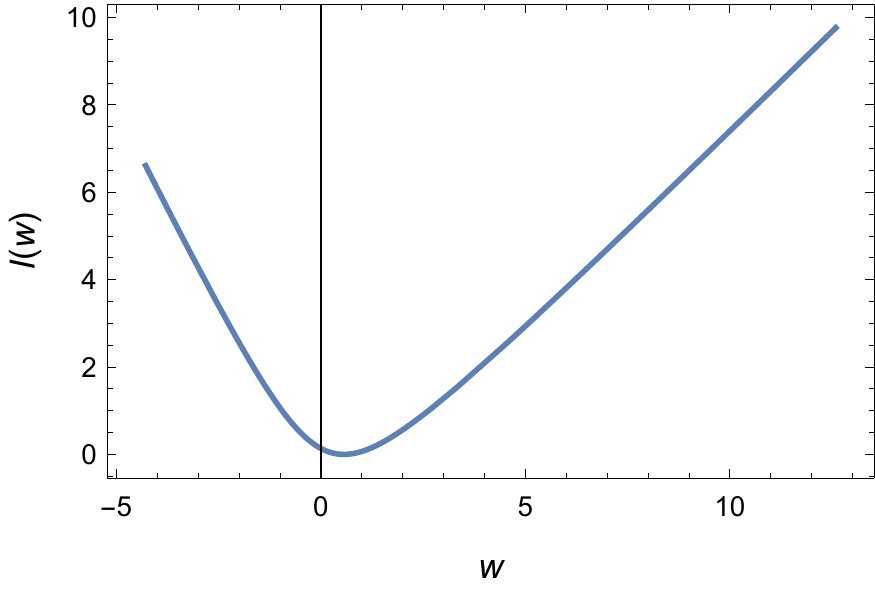}
    \end{minipage}
    \caption{SCGF (left) and rate function (right) for the nonequilibrium work done on the spring system for the parameter values $\gamma = 1,\kappa=1$ and noise strengths $\epsilon_1 = 1$ and $\epsilon_2 =2$.}
    \label{fig:springwork1}
\end{figure}
\par The SCGF and rate function are shown in Fig.~\ref{fig:springwork1}. The SCGF is seen to be symmetric around $k = -1/2$ in accordance with the symmetry (\ref{sym}) and the rate function is asymmetric around its typical value $w^*$, with fluctuations $w > w^*$ being more likely than fluctuations $w < w^*$, indicating again the presence of non-Gaussian fluctuations. The SCGF and rate function remain invariant under the transformation $\epsilon_1 \rightarrow \epsilon_2$, $\epsilon_2 \rightarrow \epsilon_1$, indicating that only the magnitude of the difference $|\epsilon_1 - \epsilon_2|$ in noise strengths and not the sign of the difference $\epsilon_1 - \epsilon_2$ determines the large deviations. As for the transverse system, this is explained by the fact that what constitutes positive and negative values of the nonequilibrium work is determined by the sign of $\epsilon_1 - \epsilon_2$ and hence by the direction of the current already present in the spring system, in contrast to the stochastic area which has a definition independent of the system under consideration. 
\begin{figure}[t]\centering
\subfloat[$k = 1/4$ ($k > 0$)]{\includegraphics[width=.45\linewidth]{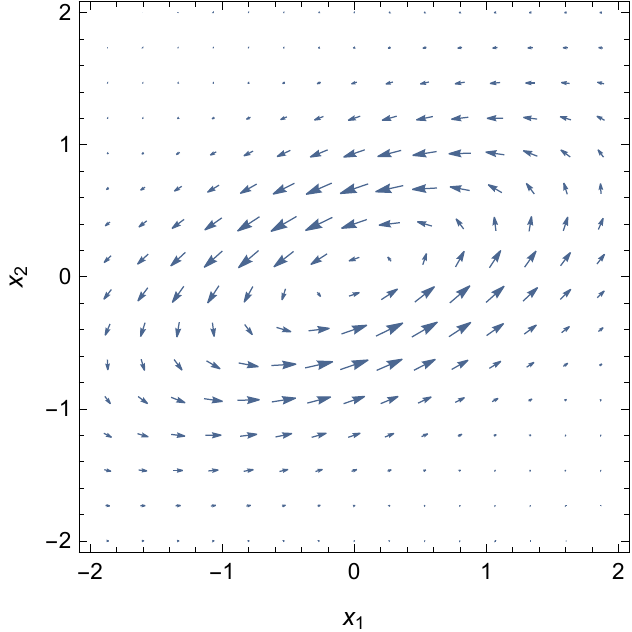}}\hfill
\subfloat[$k = -1/4$ ($-1/2 < k < 0$)]{\includegraphics[width=.45\linewidth]{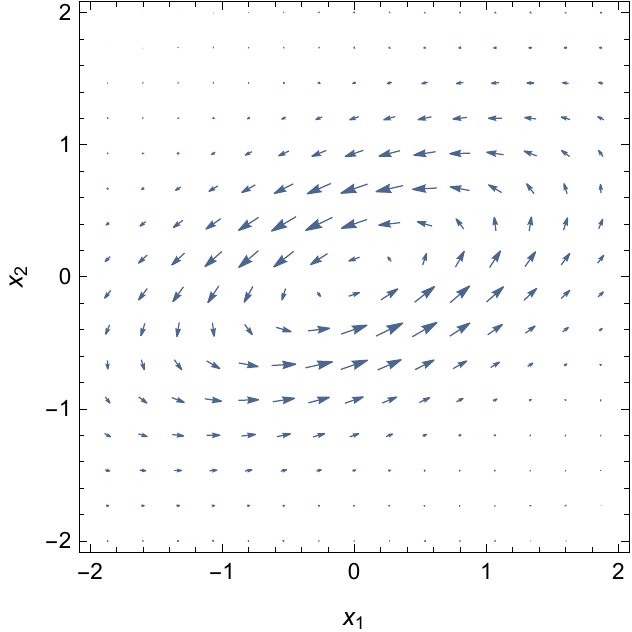}}\par 
\subfloat[$k=-3/4$ ($k < -1/2$)]{\includegraphics[width=.45\linewidth]{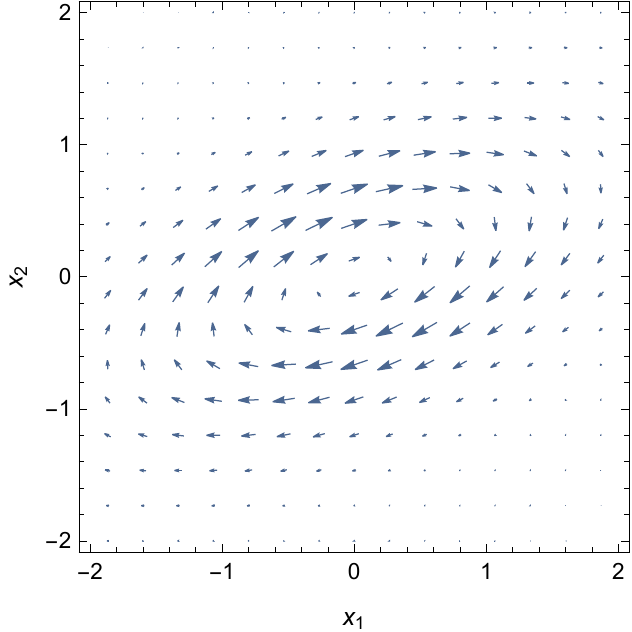}}
\caption{Vector plot of the stationary current for the effective process associated with the nonequilibrium work done on the spring system for the parameters $\gamma =1, \kappa = 1$ and for noise strengths $\epsilon_1 = 2, \epsilon_2 = 1$.}
\label{fig:Jkspringwork}
\end{figure}
\par Finally, it can be shown that the stationary current associated with the effective process has the form 
\begin{equation}
    \bs{J}_{\bs{F}_k,p_k^*}(\x) = (1+2k)H \x \, p_k^*(\x),
\end{equation}
where $H$ is the matrix given in (\ref{Hspring}). As such, it can be seen by comparison with the expression (\ref{Jspring}) for the stationary current of the original system that the effective current $\bs{J}_{\bs{F}_k,p_k^*}$ can be obtained from the stationary current $\bs{J}_{\bs{F},p^*}$ present in the original system by multiplying $\bs{J}_{\bs{F},p^*}$ by $(1+2k)$ and performing the substitution $p^* \rightarrow p_k^*$. Importantly, we note that the stationary current (\ref{workcurrent1}) for the effective process associated with the nonequilibrium work for the transverse diffusion is obtained from the stationary current for the transverse diffusion (\ref{Jtrans}) in exactly the same manner: via multiplication by $(1 + 2k)$ and the substitution $p^* \rightarrow p_k^*$. This might indicate a general structure in the current modifications associated with manifesting the fluctuations of linear current type observables for linear diffusions. 
\par Consequently, we see that the manner in which the stationary current is modified in order to manifest fluctuations here can be understood in exactly the same manner as for the nonequilibrium work of the transverse system, with $\epsilon_1 - \epsilon_2$ (or the temperature difference $T_1 - T_2$) playing here the role of the nonequilibrium parameter $\xi$. For $\epsilon_1 - \epsilon_2 > 0$ the spring system has an anti-clockwise stationary current and fluctuations $w > w^*$ associated with $k > 0$ are manifested by increasing the strength of this anti-clockwise current. For $0 < w < w^*$ associated with $-1/2 < k < 0$ fluctuations are manifested by decreasing the strength of the anti-clockwise current while for $k < -1/2$ the direction of the current is reversed in order to manifest negative fluctuations $w < 0$. For $k = -1/2$ the current vanishes, as it must, to achieve the fluctuation $w = 0$, which is an equilibrium fluctuation. A similar analysis holds for $\epsilon_1 - \epsilon_2 < 0$, for which the system has a clockwise current. This discussion is illustrated graphically for $\epsilon_1 - \epsilon_2 > 0$ in Fig.~\ref{fig:Jkspringwork}. 

\chapter{Conclusions and open problems}\label{chap:chap6}
We considered in this dissertation the large deviations of both reflected diffusions and linear diffusions. We provide here a summary of the dissertation and the results we obtained, with particular emphasis on those results that are novel. Finally we mention some open problems and possible directions for further study. 
\par In Chap.~\ref{chap:chap2} we introduced the theory of Markov diffusions and illustrated this theory for linear diffusions, a class of diffusions studied extensively in this dissertation. We also provided an introduction to the theory of dynamical large deviations, introducing the generating function that is central to much of our work, the spectral problem associated with calculating the long-time limit of this generating function, and the effective process, which describes the manner in which fluctuations are manifested dynamically in time. 
\par Chapter~\ref{chap:chap3} was concerned with the large deviations of reflected diffusions evolving in a subset of $\mathbb{R}^n$. We provided in that chapter a summary of the results previously obtained~\cite{dubuissonmasters, Buisson2020} for the large deviations of additive observables, discussing the argument used to find the appropriate boundary conditions for the spectral problem associated with the SCGF. We then explained why the arguments employed for these observables fail to generalize to the case of current-type observables and then proceeded to present a new argument~\cite{Mallmin2021}, based on the local time formulation and the Feynman-Kac formula, to obtain the proper boundary conditions on $\mathcal{L}_k$ and $\mathcal{L}_k^{\dagger}$. An interesting consequence of these boundary conditions that we have derived is that they imply that the effective process responsible for manifesting fluctuations of a current-type observable has a stationary current satisfying a zero current condition at the reflecting boundary, which means that it is again a reflected diffusion. Furthermore, the component of the effective drift normal to the boundary was shown to be identical to that of the original drift. Both of these results were obtained~\cite{dubuissonmasters, Buisson2020} also for additive observables, and as such these results are now shown to be true regardless of the type of dynamical observable considered. Finally, we demonstrated our results by discussing the large deviations of a current-type observable for the heterogeneous single-file diffusion, published recently in~\cite{Mallmin2021}. The correct boundary conditions on the tilted generator $\mathcal{L}_k$ proved crucial in obtaining both the rate function and the effective process. 
\par Turning to the study of linear diffusions in Chap.~\ref{chap:chap4}, we derived, using the Feynman-Kac formula, an explicit expression for the generating function $G_k$ associated with linear additive, quadratic additive and linear current-type observables. By studying the long-time limit of this exact result, we obtained the SCGF $\lambda(k)$ and the effective process responsible for manifesting the fluctuations for these classes of observables in the long time limit. 
\par Remarkably, it was shown that for all three classes of observables considered, the effective process remains a linear process. This is an important result which enables us to study the manner in which linear diffusions manifest fluctuations for a variety of physically important observables such as the nonequilibrium work and entropy production. In particular, it allowed for an investigation into the manner in which equilibrium processes manifest nonequilibrium fluctuations (or vice versa) via the creation or modification of probability currents and the alteration of probability densities. 
\par Finally, we obtained explicit expressions for the asymptotic mean and variance associated with all the observables studied in the context of linear diffusions. This extends much work in the literature, where often only the first moment is obtained.
\par To illustrate our results, we studied in Chap.~\ref{chap:chap5} a number of systems and observables. We found for the transverse system with quadratic observable that fluctuations were manifested primarily via the modification of the stationary density, while the stationary current was modified only in trivial ways. In this sense we can view the quadratic additive observable as an `equilibrium observable', for which the effective process has the same equilibrium or nonequilibrium nature as that of the original process for all fluctuations. 
\par Next we studied general anti-symmetric linear current-type observables, and in particular the stochastic area, for a gradient equilibrium system, showing that here fluctuations are manifested instead via the creation of a probability current and hence by a nonequilibrium effective process. As such the linear current-type observable constitutes a `nonequilibrium observable' for which the associated effective process can have different reversibility properties than that of the original process. 
\par For the transverse process, we contrasted the manner in which fluctuations are manifested for the stochastic area and the nonequilibrium work, emphasizing the fact that the nonequilibrium work is defined in terms of the parameters defining the transverse system while the stochastic area has a definition independent of the system. The nonequilibrium work is found to have a typical value that is always positive, in accordance with the second law of thermodynamics, while the typical value of the stochastic area can be either positive or negative depending on the sign of the nonequilibrium parameter. For both of these observables (and in fact for any antisymmetric linear current-type observable) we obtained Gallavoti-Cohen type fluctuation relations constraining the associated probability density and indicating that fluctuations with a particular sign are exponentially more likely than the equivalent fluctuation with opposite sign. In addition, we illustrated the effect of the stationary state of the transverse system on fluctuations of the entropy production, producing a linear tail in the rate function for fluctuations larger than a particular value. 
\par Finally we considered the fluctuations of the nonequilibrium work for a spring system, obtaining results that can be understood in much the same way as those of the transverse diffusion, but with the noise difference now playing the role of the nonequilibrium parameter.
\par To conclude the dissertation we provide some possible directions for further study. For diffusions in bounded domains, these include: 
\begin{itemize}
    \item The study of diffusions with non-smooth reflecting boundaries $\po$. It is not clear at this point how to accommodate kinks in boundaries in the local time framework used to study reflected diffusions in this thesis. In particular, what is the appropriate direction in which the local time acts at a kink in a boundary? 
    \item The simplicity of the heterogeneous single-file diffusion makes an exact solution of the large deviation problem associated with a simple class of current-type observables possible. Can we find more complicated systems (such as systems with a drift $\bs{F}$ depending on the state of the process) with reflecting boundaries that also allows for an exact solution? 
    \item Investigating the large deviations of diffusions with boundary behavior other than reflection. An interesting type of boundary behavior that might have non-trivial large deviations is that of sticky boundaries~\cite{harrison1981sticky,bou2020sticky,engelbert2014stochastic}, where the process remains at the boundary for a random amount of time upon reaching the boundary. It is not clear at this time whether a generating function approach similar to the one employed for reflecting boundaries could be used to obtain the appropriate large deviation boundary conditions.  
\end{itemize}
\par Potentially interesting directions for further work related to the large deviations of linear diffusions are: 
\begin{itemize}
    \item Studying time-dependent linear diffusions, and particularly time-periodic linear diffusions. A framework for the large deviations of time-periodic systems has been developed~\cite{barato2018current} and application of this framework, along with the general results obtained for the generating function of linear diffusions could prove fruitful. 
    \item It is known~\cite{chetrite2015variational} that the effective process associated with an observable can be formulated as an optimal control process which minimizes a particular control cost. For the observables considered in this dissertation the associated control problem is of linear-quadratic type~\cite{stengel1986stochastic} and the Riccati equations (\ref{Akquaddiff}) and (\ref{riccatidiffcur}) arise also~\cite{stengel1986stochastic,abou2012matrix} in the control formulation in determining the drift matrix for the appropriate linear control which minimizes the relevant control cost. Exploring this link with control theory more fully constitutes an interesting direction for further study.
    \item Linear diffusions provide analytically tractable approximations to non-linear systems having a stable fixed point. Using the framework developed in  this dissertation it might be possible to obtain useful approximate results for the large deviations of observables such as the nonequilibrium work or entropy production of interesting non-linear systems. Using the results obtained for the effective process associated with linear diffusions it might also be possible to understand, in an approximate way, how non-linear systems manifest fluctuations for these physically interesting observables. 
\end{itemize}
\par Finally, the study of reflected linear diffusions could prove interesting. It is already known that many of the results obtained in this dissertation for linear diffusions in unbounded domains, such as the fact that the effective process associated with a large class of observables is again a linear process, does not hold for reflected linear diffusions. This was illustrated explicitly by Du Buisson and Touchette~\cite{dubuissonmasters, Buisson2020}, where the dynamical large deviations associated with a linear additive observable of the reflected Ornstein-Uhlenbeck process was obtained. It is therefore natural to ask which of the results obtained for linear diffusions in unbounded domains remain valid for reflected diffusions and whether similar methods used here to examine linear diffusions could prove useful also for reflected linear diffusions. In particular, are there general results for the effective process associated with the fluctuations of such processes? What form does the effective drift take and what are the general principles involved? 

\appendix

\chapter[Duality for Markov operators]{Duality relation for Markov operators}\label{appendixA}
We show here the calculation leading to the duality relation 
\begin{equation}
    \langle p, \mathcal{L}h\rangle = \langle \mathcal{L}^{\dagger}p, h\rangle - \int_{\po} h(\x) \bs{J}_{\bs{F}, p}(\x) \cdot \hat{\bs{n}}(\x)d\x  - \frac{1}{2} \int_{\po}  p(\x) D \bs{\nabla} h(\x) \cdot \hat{\bs{n}}(\x)d\x,
\end{equation}
for the Markov operators of a process constrained to a region $\om \in \mathbb{R}^n$. \par To obtain this result, we use a mathematical identity, which amounts to integration by parts in higher dimensions. For a scalar field $u$ and vector field $\bs{V}$ we have \begin{equation} \label{eq:int-by-parts}
    \int_{\om}  u(\x) \bs{\nabla} \cdot \bs{V}(\x)d\x = -\int_{\po} u(\x) \bs{V}(\x)\cdot \hat{\bs{n}}(\x)d\x  - \int_{\om} \bs{V}(\x) \cdot \bs{\nabla} u(\x)d\x ,
\end{equation}
where $\hat{\bs{n}}(\x)$ is the inward normal vector at $\x \in \po$. 
\par Starting from the inner product   
\begin{equation} \label{eq:appendA0}
    \langle  p, \mathcal{L} h\rangle = \int_{\om} p(\x) \left[\bs{F}(\x) \cdot \nabla + \frac{1}{2} \nabla \cdot D \nabla \right] h(\x) d\x 
\end{equation}
and using \eqref{eq:int-by-parts}, we have 
\begin{align} \label{eq:appendA1}
    \int_{\om} p(\x) \bs{F}(\x) \cdot \nabla h(\x)d\x  &= - \int_{\po} p(\x) h(\x) \bs{F}(\x) \cdot \hat{\bs{n}}(\x)d\x   \nonumber \\ & \quad \quad 
    - \int_{\om} \bs{\nabla} \cdot \left[\bs{F}(\x) p(\x) \right] h(\x) d\x
\end{align}
and 
\begin{align} \label{eq:appendA2}
    \int_{\om}  p(\x) \left[\frac{1}{2} \nabla \cdot D \bs{\nabla} \right] h(\x)d\x  &= - \frac{1}{2}\int_{\po}   p(\x) D \bs{\nabla}h(\x) \cdot \hat{\bs{n}}(\x) d\x \nonumber \\ & \quad \quad - \frac{1}{2} \int_{\om}  \bs{\nabla} p(\x) \cdot D \nabla h(\x)d\x.
\end{align}
Given that the diffusion matrix $D$ is symmetric, we have 
\begin{equation}
    \frac{1}{2}\int_{\om}  \bs{\nabla}p(\x) \cdot D \bs{\nabla} h(\x) d\x= \frac{1}{2}\int_{\om}  D\bs{\nabla} p(\x) \cdot  \bs{\nabla} h(\x)d\x 
\end{equation}
and applying \eqref{eq:int-by-parts} to this last expression, we obtain 
\begin{align} \label{eq:appendA3}
    \frac{1}{2}\int_{\om}  D\nabla p(\x) \cdot  \nabla h(\x) d\x &= -\frac{1}{2} \int_{\po} h(\x) D \nabla p(\x) \cdot \hat{\bs{n}}(\x) d\x  \nonumber \\ &\quad \quad - \int_{\om} h(\x) \left[\frac{1}{2} \nabla \cdot D \nabla \right]p(\x)d\x . 
\end{align}
Substituting \eqref{eq:appendA1}, \eqref{eq:appendA2} and \eqref{eq:appendA3} into \eqref{eq:appendA0}, we obtain 
\begin{align}
    \avg{p, \mathcal{L} h} =
    \begin{multlined}[t]
         \langle\mathcal{L}^{\dagger} p, h\rangle - \int_{\po}d\x h(\x) \left[p(\x) \bs{F}(\x) - \frac{1}{2} D \bs{\nabla} p(\x) \right]\cdot \hat{\bs{n}}(\x)  \\ - \frac{1}{2} \int_{\po} d\x\,  p(\x) D \bs{\nabla} h(\x) \cdot \hat{\bs{n}}(\x) ,
      \end{multlined}
\end{align}
where $\mathcal{L}^{\dagger}$ is defined in (\ref{FPO}). Using the definition (\ref{current}) of the current $\bs{J}_{\bs{F},p}$ associated with the drift $\bs{F}$ and density $p$ to write
\begin{equation}
    \bs{J}_{\bs{F},p}=p(\x) \bs{F}(\x) - \frac{1}{2} D \bs{\nabla} p(\x),
\end{equation}
we then obtain
\begin{equation} \label{surface2}
    \langle p, \mathcal{L}h\rangle = \langle \mathcal{L}^{\dagger}p, h\rangle - \int_{\po} h(\x) \bs{J}_{\bs{F}, p}(\x) \cdot \hat{\bs{n}}(\x)d\x  - \frac{1}{2} \int_{\po}  p(\x) D \bs{\nabla} h(\x) \cdot \hat{\bs{n}}(\x)d\x.
\end{equation}

\chapter[Duality for large deviation operators]{Duality relation for large deviation operators}\label{appendixB}
We here show the derivation leading to the duality relation (\ref{tiltedcurrentduality}) for the large deviation operators associated with a current-type observable (\ref{currentobs}). 
We start again from the inner product $\langle l, \mathcal{L}_k r\rangle$, which we can write explicitly as 
\begin{equation}\label{startingpoint}
    \avg{l, \mathcal{L}_k r} = \int_{\om}  l(\x) \left[\bs{F}(\x)\cdot (\nabla + k \bs{g}(\x)) + \frac{1}{2} (\nabla + k \bs{g}(\x)) \cdot D (\nabla + k \bs{g}(\x))  \right] r(\x)d\x,
\end{equation}
using the expression (\ref{tiltedcurrent}) for the tilted generator. For the first term on the right hand side of the above we have, using integration by parts as in \eqref{eq:int-by-parts}, that 
\begin{align} \label{eq:appendB2}
    \int_{\om}  l(\x) \bigg[\bs{F}(\x)\cdot (\bs{\nabla} + k \bs{g}(\x))\bigg] r(\x)d\x =
    \begin{multlined}[t]
    - \int_{\po}  l(\x)r(\x) \bs{F}(\x)  \cdot \hat{\bs{n}}(\x) d\x \\ + \int_{\om} r(\x) \bigg[(-\bs{\nabla} + k \bs{g}(\x)) \cdot \left(\bs{F}(\x) l(\x)\right) \bigg] d\x.
    \end{multlined}
\end{align}
For the second term in the RHS of (\ref{startingpoint}), we first note that 
\begin{equation} \label{eq:appendB1}
    \frac{1}{2} (\bs{\nabla} + k \bs{g}(\x)) \cdot D (\bs{\nabla} + k \bs{g}(\x)) = \frac{1}{2} (\bs{\nabla} \cdot D \bs{\nabla} + k \bs{g}(\x) \cdot D \bs{\nabla} + \bs{\nabla} \cdot k D \bs{g}(\x) + k^2 \bs{g}(\x)\cdot D \bs{g}(\x)).
\end{equation}
The last term in the above contains no derivatives and produces no boundary terms, while the first term has already been dealt with in App.~\ref{appendixA} in \eqref{eq:appendA2} and \eqref{eq:appendA3}, with the understanding that $l$ and $r$ are to replace the $p$ and $h$ used there, respectively. We can therefore write 
\begin{align} \label{eq:appendB3}
    \int_{\om} l(\x) \left[\frac{1}{2} \bs{\nabla} \cdot D \bs{\nabla}  \right]r(\x) d\x&=
         \int_{\om}  r(\x) \left[\frac{1}{2} \bs{\nabla} \cdot D \bs{\nabla}  \right]l(\x)d\x \nonumber  \\ &\quad - \frac{1}{2} \int_{\po} l(\x) D\bs{\nabla} r(\x) \cdot \hat{\bs{n}}(\x)d\x \nonumber \\ &\quad + \frac{1}{2} \int_{\po} r(\x) D\bs{\nabla} l(\x) \cdot \hat{\bs{n}}(\x)d\x.
\end{align}
\par For the remaining two terms in \eqref{eq:appendB1}, we have 
\begin{align} \label{eq:appendB4}
    \int_{\om}  l(\x) \left[\frac{1}{2} k \bs{g}(\x) \cdot D \bs{\nabla} \right]r(\x)d\x &=
     \int_{\om}  l(\x) \left[\frac{1}{2} k D \bs{g}(\x) \cdot  \bs{\nabla} \right]r(\x)d\x \nonumber\\
    &=-\frac{1}{2} \int_{\po} l(\x) r(\x) k D \bs{g}(\x) \cdot \hat{\bs{n}}(\x)d\x \nonumber\\ & \quad- \frac{1}{2}\int_{\om}  r(\x) \bs{\nabla} \cdot l(\x) k D \bs{g}(\x)d\x,
\end{align}
where we have used the symmetry of $D$ in the first line and integration by parts in the second, and 
\begin{align} \label{eq:appendB5}
    \int_{\om}  l(\x) \left[\frac{1}{2} \bs{\nabla} \cdot k D \bs{g}(\x) \right]r(\x)d\x = 
    \begin{multlined}[t]
     -\frac{1}{2}\int_{\po}  l(\x)r(\x) k D \bs{g}(\x) d\x\\  - \int_{\om}  \left[\frac{1}{2} k D \bs{g}(\x) \cdot \bs{\nabla} l(\x) \right] r(\x)d\x.
     \end{multlined}
\end{align}
Combining \eqref{eq:appendB2}, \eqref{eq:appendB3}, \eqref{eq:appendB4} and \eqref{eq:appendB5} we then obtain
\begin{align} \label{tiltedcurrentduality3}
    \avg{l, \mathcal{L}_k r} =
    \begin{multlined}[t]
         \int_{\om}  r(\x) \bigg[(-\bs{\nabla} + k \bs{g}(\x))\cdot (\bs{F}(\x) l(\x)) + \frac{1}{2} \bigg(\bs{\nabla} \cdot D \bs{\nabla} l(\x)  \\  - kD \bs{g}(\x) \cdot \bs{\nabla} l(\x) - \bs{\nabla} \cdot (kD \bs{g}(\x) l(\x)) + k^2 \bs{g}(\x) \cdot D \bs{g}(\x) l(\x)  \bigg)  \bigg]d\x  \\ 
    	 - \int_{\po}  \bigg\{l(\x) r(\x)\bigg(\bs{F}(\x) + k D \bs{g}(\x) \bigg) + \frac{1}{2} l(\x) D \bs{\nabla} r(\x) \\ - \frac{1}{2} r(\x) D \bs{\nabla} l(\x) \bigg\}\cdot \hat{\bs{n}}(\x)d\x.
    \end{multlined}
\end{align}
The final result for the duality relation relating the large deviation operators can therefore be written as 
\begin{align}  \label{tiltedcurrentduality2}
    \avg{l, \mathcal{L}_k r} &= \langle \mathcal{L}_k^{\dagger} l, r\rangle  - \int_{\po}  \bigg\{l(\x) r(\x)\bigg(\bs{F}(\x) + k D \bs{g}(\x) \bigg) + \frac{1}{2} l(\x) D \bs{\nabla} r(\x) \nonumber \\ &\quad - \frac{1}{2} r(\x) D \bs{\nabla} l(\x) \bigg\}\cdot \hat{\bs{n}}(\x)d\x,
\end{align}
where the adjoint operator $\mathcal{L}_k^{\dagger}$ is seen to act on a density $l$ in the manner 
\begin{align} 
    \mathcal{L}_k^{\dagger}l(\x) &= (-\bs{\nabla} + k \bs{g}(\x))\cdot (\bs{F}(\x) l(\x)) + \frac{1}{2} \bigg(\bs{\nabla} \cdot D \bs{\nabla} l(\x) \nonumber \\  &- kD \bs{g}(\x) \cdot \bs{\nabla} l(\x) - \bs{\nabla} \cdot (kD \bs{g}(\x) l(\x)) + k^2 \bs{g}(\x) \cdot D \bs{g}(\x) l(\x)  \bigg),
\end{align}
from which it follows that $\mathcal{L}_k^{\dagger}$ has the form  
\begin{equation}
    \mathcal{L}_k^{\dagger} =(-\bs{\nabla} + k \bs{g})\cdot \bs{F} + \frac{1}{2}\left(-\bs{\nabla}+k\g\right)\cdot D\left(-\bs{\nabla}+k\g\right).
\end{equation}
This could be inferred directly from the form of the tilted generator (\ref{tiltedcurrent}) by observing that $\bs{\nabla}^{\dagger}$ = $-\bs{\nabla}$ (leaving out the boundary term produced, since all boundary terms must collectively add up to zero in the end), given that the adjoint is obtained via integration by parts. 
\par 

\chapter[Induction argument for current-type observables]{Induction argument for linear current-type observable}\label{appendixC}
We show here the calculation leading to the exact result for the generating function associated with a purely antisymmetric linear current-type observable $A_T$. Using again a time-discretization argument, our induction hypothesis is that 
\begin{equation} \label{inductionmain2}
    G_k(\x,m\Delta t) = \exp(\langle\x, B^{(m)}_k\x\rangle) \exp\left(\sum_{i=0}^{m-1} \tn{Tr}\left(DB^{(i)}_k\right)\right),
\end{equation}
which has the same form as (\ref{inductionmain1}) for the quadratic additive observable, but with the matrix $B^{(m)}_k$ now satisfying the recursion relation
\begin{align} \label{induction2}
    B^{(m)}_k &= B^{(m-1)}_k + \Delta t\bigg(\frac{k^2}{2} \Gamma^{\mathsf{T}}D\Gamma \,-\, \frac{k}{2}(M^{\mathsf{T}}\Gamma-\Gamma M) \,+ \,(-M + kD\Gamma)^{\mathsf{T}}B^{(m-1)}_k \nonumber \\ &\quad + B^{(m-1)}_k(-M + kD\Gamma) +  2B^{(m-1)}_kDB^{(m-1)}_k \bigg)
\end{align}
with initial condition $B^{(0)}_k = 0$. As for the quadratic additive observable, the induction hypothesis (\ref{inductionmain2}) and (\ref{induction2}) is only claimed to hold up to first order in $\Delta t$. It is clear, by similar reasoning as that employed for the quadratic additive observable following (\ref{induction1}), that $B^{(i)}_k$ given by (\ref{induction2}) is symmetric for all $i$. Now we prove that the induction hypothesis holds for $m=1$. We have, using the relevant expression (\ref{tiltedcurrentoperator}) for the tilted generator $\mathcal{L}_k$, that
\begin{align}
    G_k(\x,\Delta t) &= (1+\Delta t\mathcal{L}_k) 1 \nonumber \\ &= \bigg(1 + \Delta t\bigg(-\frac{k}{2}\langle\x, (M^{\mathsf{T}}\Gamma - \Gamma M)\x\rangle + \frac{k^2}{2} \langle \x, \Gamma^{\mathsf{T}}D\Gamma \x\rangle\bigg)\bigg)1 \nonumber \\ 
    &= e^{\langle\x, B^{(1)}_k\x\rangle)},
\end{align}
with 
\begin{equation}
    B^{(1)}_k = \Delta t \bigg(-\frac{k}{2}(M^{\mathsf{T}}\Gamma - \Gamma M)+\frac{k^2}{2} \Gamma^{\mathsf{T}}D\Gamma\bigg),
\end{equation}
which satisfies (\ref{induction2}) upon recognizing that $B^{(0)}_k = 0$. For the induction step, we assume that (\ref{inductionmain2}) and (\ref{induction2}) hold for $m = j$. We will show that it then holds also for $m = j +1$. We have 
\begin{align} \label{eq5}
    \mathcal{L}_k &\exp(\langle \x, B^{(j)}_k \x\rangle) \nonumber \\ &= \bigg(-\frac{k}{2}\langle\x, (M^{\mathsf{T}}\Gamma - \Gamma M)\x\rangle +  (-M + kD\Gamma)\x\cdot\bs{\nabla} + \frac{1}{2} \bs{\nabla}\cdot D \bs{\nabla} \nonumber \\ &\quad \quad + \frac{k^2}{2} \langle \x, \Gamma^{\mathsf{T}}D\Gamma\x\rangle \bigg)
     \exp(\langle \x, B^{(j)}_k \x\rangle)\nonumber \\
    &= \bigg(-\frac{k}{2}\langle\x, (M^{\mathsf{T}}\Gamma - \Gamma M)\x\rangle + \frac{k^2}{2} \langle \x, \Gamma^{\mathsf{T}}D\Gamma\x\rangle + 2 \langle \x, (-M + kD\Gamma)^{\mathsf{T}}B^{(m)}_k \x\rangle  \nonumber \\ &\quad + \frac{k^2}{2} \langle \x, \Gamma^{\mathsf{T}}D\Gamma \x\rangle + \tn{Tr}(DB^{(m)}_k)\bigg) \exp(\langle \x, B^{(m)}_k \x\rangle),
\end{align}
where we have used (\ref{eq3}) and (\ref{eq4}) for the derivatives of the exponential. 
Using the fact that only the symmetric part of a matrix is relevant inside the inner product, we can write
\begin{equation} \label{eq6}
    2\langle \x, (-M + kD\Gamma)^{\mathsf{T}}B^{(m)}_k \x\rangle = \bigg\langle \x, \bigg((-M+kD\Gamma)^{\mathsf{T}}B^{(m)}_k + B^{(m)}_k(-M+kD\Gamma)\bigg)\x\bigg\rangle.
\end{equation}
Given that 
\begin{equation}
    G_k(\x,(j+1)\Delta t) = (1 +\Delta t \mathcal{L}_k)\exp(\langle\x, B^{(j)}_k\x\rangle) \exp\left(\sum_{i=0}^{j-1} \tn{Tr}(DB^{(i)}_k)\right)
\end{equation}
we now insert the results of (\ref{eq5}) and (\ref{eq6}) into the above and use the fact that for a matrix $A$ the expression
\begin{equation}
    \bigg[1 + \Delta t A\bigg] = e^{\Delta tA}
\end{equation}
holds up to first order in $\Delta t$. We obtain then
\begin{equation}
    G_k(\x,(j+1)\Delta t)\exp(\langle\x, B^{(j+1)}_k\x\rangle) \exp\left(\sum_{i=0}^{j} \tn{Tr}(DB^{(i)}_k)\right)
\end{equation}
where 
\begin{align}
    B^{(j+1)}_k &= B^{(j)}_k + \Delta t\bigg(\frac{k^2}{2} \Gamma^{\mathsf{T}}D\Gamma \,-\, \frac{k}{2}(M^{\mathsf{T}}\Gamma-\Gamma M) \,+ \,(-M + kD\Gamma)^{\mathsf{T}}B^{(j)}_k \nonumber \\&\quad + B^{(j)}_k(-M + kD\Gamma) + 2B^{(j)}_kDB^{(j)}_k\bigg),
\end{align}
thereby completing the proof by induction. Taking now the limit $n\rightarrow \infty$ so that $\Delta t = t/n \rightarrow 0$ in $G_k(\x,n \Delta t)$, we obtain
\begin{equation}
    G_k(\x,t) = \exp(\langle \x, B_k(t) \x\rangle) \exp\left(\int_0^t \tn{Tr}(DB_k(s)) ds\right),
\end{equation}
with $B_k(t)$ satisfying the differential Riccati equation
\begin{align} 
    \frac{dB_k(t)}{dt} &= \frac{k^2}{2} \Gamma^{\mathsf{T}}D\Gamma \,-\, \frac{k}{2}(M^{\mathsf{T}}\Gamma-\Gamma M) \,+ \,(-M + kD\Gamma)^{\mathsf{T}}B_k(t) \nonumber \\ &\quad + B_k(t)(-M + kD\Gamma) + 2B_k(t) DB_k(t),
\end{align}
with initial condition $B_k(0) = 0$ and with stationary solutions $B_k^*$ satisfying 
\begin{align} 
    0 &= \frac{k^2}{2} \Gamma^{\mathsf{T}}D\Gamma \,-\, \frac{k}{2}(M^{\mathsf{T}}\Gamma-\Gamma M) \,+ \,(-M + kD\Gamma)^{\mathsf{T}}B_k^* + B_k^*(-M + kD\Gamma) \nonumber \\ &\quad + 2B_k^* DB_k^* .
\end{align}
\chapter[$B_k(t)$ diagonal for the transverse diffusion]{Derivation of $B_k(t)$ for quadratic observables for the transverse diffusion} \label{appendixD}
We show here the derivation of the result (\ref{matrix3}). We have that the matrix $B_k(t)$ satisfies the differential Riccati equation (\ref{Akquaddiff}) with the initial condition $B_k(0) = 0$ and which therefore clearly satisfies $B_0(t) = 0$ for all time $t$. In order to obtain the analytic solution for $B_k(t)$ we will first show, using a time-discretized version of the differential equation (\ref{Akquaddiff}), that $B_k(t)$ is proportional to the identity matrix for all times $t$. 
\par Consider a time step $\Delta t$. We have, from (\ref{Akquaddiff}), that 
\begin{equation}
    B_k(t + \Delta t) = B_k(t) + \Delta t \bigg[2B_k(t) D B_k(t) -\left(M^{\mathsf{T}}B_k(t)  +  B_k(t)M \right) + kQ\bigg]
\end{equation}
up to first order in $\Delta t$. It is to be understood that all expressions for the remainder of the discrete-time calculation are valid up to first order in $\Delta t$. Continuing, we have in particular for $t = 0$ that 
\begin{equation}
    B_k(\Delta t) = \Delta t \, k Q,
\end{equation}
where we have used the fact that $B_k(0) = 0$. 
\par For the second step in the time evolution we now have 
\begin{equation} \label{timediscAkquad2}
    B_k(2\Delta t) = B_k(\Delta t) + \Delta t \bigg[2B_k(\Delta t) D B_k(\Delta t) -\left(M^{\mathsf{T}}B_k(\Delta t)  +  B_k(\Delta t)M \right) + kQ\bigg].
\end{equation}
Given that $Q = \mathbb{I}$ so that $B_k(\Delta t) = \Delta t k Q$ is proportional to the identity we then have 
\begin{equation}
    2 B_k(\Delta t) D B_k(\Delta t) = 2 \Delta t^2 \, k^2 \epsilon^2 \mathbb{I},
\end{equation}
since $D = \epsilon^2 \mathbb{I}$. Furthermore, we have
\begin{equation} \label{matrix2}
    M^{\mathsf{T}}B_k(\Delta t)  +  B_k(\Delta t)M = \Delta t \, k\begin{pmatrix}\gamma && -\xi \\ \xi &&\gamma \end{pmatrix} + \Delta t \, k \begin{pmatrix}\gamma && \xi \\ -\xi  && \gamma \end{pmatrix} = 2\Delta t \, k \gamma \mathbb{I}
\end{equation}
so that all terms on the RHS of (\ref{timediscAkquad2}) are seen to be proportional to the identity, implying that $B_k(2\Delta t)$ is also proportional to the identity. But similar reasoning can be used also for every following time step so that we must conclude that $B_k(m\Delta t)$ remains proportional to the identity for all $m \in \mathbb{N}$. Taking then the $\Delta t \rightarrow 0$ limit we can write for all times $t$ 
\begin{equation}\label{matrix4}
    B_k(t) = \begin{pmatrix}b_k(t) && 0 \\ 0 && b_k(t) \end{pmatrix}.
\end{equation}
It can easily be shown by substituting the expression (\ref{matrix4}) for $B_k(t)$ into (\ref{Akquaddiff}) that $b_k(t)$ satisfies the scalar differential Riccati equation 
\begin{equation} 
    \frac{db_k(t)}{dt} = 2\epsilon^2 b_k(t)^2 - 2 \gamma b_k(t) + k, \quad \tn{with} \quad b_k(0) =  0. 
\end{equation}

\chapter[$B_k(t)$ diagonal for the gradient diffusion]{Derivation of diagonal form of $B_k(t)$ for current-type observable of the gradient diffusion}\label{appendixE}
We show here that the time-dependent solution $B_k(t)$ of the Riccati equation (\ref{riccatidiffcur}) for a linear current-type observable of the gradient diffusion introduced in Sec.~\ref{sec:sec241} has the form of a diagonal matrix for all times $t$. 
\par For this system, with $M = \gamma\mathbb{I}$ and $D = \epsilon^2\mathbb{I}$, we have that 
\begin{equation}
    M^{\mathsf{T}}\Gamma - \Gamma M = 0
\end{equation}
and 
\begin{equation}
    \Gamma^{\mathsf{T}}D\Gamma = \epsilon^2 \alpha^2 \mathbb{I},
\end{equation} 
so that we can write the differential Riccati equation (\ref{riccatidiffcur}) explicitly as 
\begin{equation}\label{akcurdiffgrad}
    \frac{dB_k(t)}{dt} = 2\epsilon^2 B_k(t)^2 - \begin{pmatrix}\gamma && -k\epsilon^2\alpha \\ k\epsilon^2\alpha&& \gamma \end{pmatrix}B_k(t) - B_k(t) \begin{pmatrix}\gamma && k\epsilon^2 \alpha \\ -k\epsilon^2 \alpha && \gamma \end{pmatrix} + \frac{k^2\epsilon^2\alpha^2}{2} \mathbb{I}. 
\end{equation}
Using a similar time-discretization argument as that employed for the transverse diffusion and quadratic observable in App.~\ref{appendixD}, the evolution equation (\ref{akcurdiffgrad}) takes the form 
\begin{align}
    B_k(t + \Delta t) &= B_k(t) + \Delta t \bigg[- \begin{pmatrix}\gamma && -k\epsilon^2\alpha \\ k\epsilon^2\alpha&& \gamma \end{pmatrix}B_k(t) - B_k(t) \begin{pmatrix}\gamma && k\epsilon^2 \alpha \\ -k\epsilon^2 \alpha && \gamma \end{pmatrix}\nonumber \\  &\quad + 2 \epsilon^2 B_k(t)^2 + \frac{k^2\epsilon^2 \alpha^2}{2}\mathbb{I}\bigg] + \mathcal{O}(\Delta t^2).
\end{align}
We therefore have, using the initial condition $B_k(0) = 0$, that
\begin{equation}
    B_k(\Delta t) = \Delta t \frac{k^2\epsilon^2 \alpha^2}{2}\mathbb{I}
\end{equation}
for the first step in the time evolution, up to first order in $\Delta t$. For the second step we note that, since $B_k(\Delta t) \propto \mathbb{I}$, then 
\begin{equation}
    \begin{pmatrix}\gamma && -k\epsilon^2\alpha \\ k\epsilon^2\alpha&& \gamma \end{pmatrix}B_k(\Delta t) + B_k(\Delta t) \begin{pmatrix}\gamma && k\epsilon^2 \alpha \\ -k\epsilon^2 \alpha && \gamma \end{pmatrix} = 2\gamma B_k(\Delta t) 
\end{equation}
so that 
\begin{equation}
    B_k(2\Delta t) = B_k(\Delta t) + \Delta t \bigg[-2\gamma B_k(\Delta t) + 2\epsilon^2 B_k(\Delta t)^2 + \frac{k^2 \epsilon^2 \alpha^2}{2} \mathbb{I} \bigg]
\end{equation}
and since $B_k(\Delta t) \propto \mathbb{I}$ we clearly have that $B_k(2\Delta t) \propto \mathbb{I}$. The same reasoning holds for all time steps and so, taking the continuum limit $\Delta t \rightarrow \infty$, we obtain 
\begin{equation}
    B_k(t) = \begin{pmatrix}b_k(t) && 0 \\ 0 && b_k(t) \end{pmatrix},
\end{equation}
completing the derivation. 
\backmatter

\bibliography{phdbib.bib}

\end{document}